\documentclass[twocolumn]{aastex631}

\usepackage{amsmath}	
\usepackage{breakurl}
\usepackage{xcolor}

\newcommand{\lam}[1]{\,$\lambda$#1}

\shorttitle{Catalog of PNe detected by GALEX and corollary optical surveys}
\shortauthors{G{\'o}mez-Mu{\~n}oz et al.}

\begin{document}

\title{Catalog of Planetary Nebulae detected by GALEX and corollary optical surveys}

\correspondingauthor{M. A. G{\'o}mez-Mu{\~n}oz}
\email{magm@iac.es}

\author[0000-0002-3938-4211]{M. A. G{\'o}mez-Mu{\~n}oz}
\affiliation{Instituto de Astrof{\'i}sica de Canarias, E-38205 La Laguna, Tenerife, Spain}
\affiliation{Departamento de Astrof{\'i}sica, Universidad de La Laguna, E-38206 La Laguna, Tenerife, Spain}
\altaffiliation{Visiting student from the Instituto de Astrof\'isica de Canarias 
in the Dept. of Physics and Astronomy of the Johns Hopkins 
University (from September 15th 2018 to December 10th 2018).}

\author[0000-0001-7746-5461]{L. Bianchi}
\affiliation{Department of Physics and Astronomy, The Johns Hopkins University, 3400 N. Charles Street, Baltimore, MD 21218, USA}

\author[0000-0002-3011-686X]{A. Manchado}
\affiliation{Instituto de Astrof{\'i}sica de Canarias, E-38205 La Laguna, Tenerife, Spain}
\affiliation{Departamento de Astrof{\'i}sica, Universidad de La Laguna, E-38206 La Laguna, Tenerife, Spain}
\affiliation{Consejo Superior de Investigaciones Científicas, Spain}

\begin{abstract}
Planetary nebulae (PNe) consist of an ionized envelope surrounding 
a hot central star (CSPN) that emits mostly at ultraviolet (UV) wavelengths.
Ultraviolet observations, therefore, provide important information on both the CSPN and the nebula.
We have matched the PNe in The Hong Kong/AAO/Strasbourg 
H$\alpha$ (HASH) catalog with the \textit{Galaxy Evolution Explorer} 
(GALEX) UV sky surveys, the Sloan Digital Sky Survey data 
release 16 (SDSS), and the Panoramic Survey Telescope and Rapid
 Response System (Pan-STARRS) PS1 second release.
A total of 671 PNe were observed by GALEX with the far-UV 
(FUV; 1344-1786{\AA}) and/or the near-UV (NUV; 1771-2831{\AA}) 
detector on (GUVPNcat); 83 were observed by SDSS (PNcatxSDSSDR16) 
and 1819 by Pan-STARRS (PNcatxPS1MDS). We merged a distilled version of these matched catalogs into 
GUVPNcatxSDSSDR16xPS1MDS, which contains a total of 375 PNe with 
both UV and optical photometry over a total spectral coverage of $\sim$1540--9610{\AA}.
We analyzed separately 170 PNe resolved in GALEX images and 
determined their UV radius by applying a flux profile analysis. 
The CSPN flux could be extracted separately from the PN emission 
for 8 and 50 objects with SDSS and Pan-STARRS counterparts respectively.
The multiband photometry was used to distinguish between 
compact and extended PNe and CSPNe (binary CSPNe) by
 color--color diagram analysis. We found that compact PNe candidates 
could be identified by using the  $r-i < -0.4$ and $-$1$<$FUV$-$NUV$<$1 colors,
whereas binary CSPNe candidates in given $T_\mathrm{eff}$ 
ranges (all with color r$-$i$>-$0.4) can be identified in the 
color region (FUV$-$NUV)$\leq$6(r$-$i)+1.3, $-$0.8$<$FUV$-$NUV$<$0.4 and r$-$i$<$0.75.
\end{abstract}




\section{Introduction} \label{sec:introduction}

Planetary Nebulae (PNe) are the late evolutionary products 
of low- and intermediate-mass stars ($\sim$0.8--8.0~M$_{\sun}$);
they consist of an ionized envelope surrounding a stellar nucleus. 
The removal of the external shell  (H envelope) occurs through
mass loss experienced in the asymptotic giant branch (AGB) phase.
Subsequently, during a brief PN evolutionary phase ($\sim10^4$~yr,
depending on the star mass) the central star of the PN (CSPN) 
increases its effective temperature, $T_\text{eff}$, from 
$\sim$30\,000~K up to $\sim$150\,000~K, becoming bright at
ultraviolet (UV) wavelengths with a low optical luminosity
owing to its small radii (down to Earth-like sizes).
The CSPN is hot enough to ionize the expelled shell 
($\ga$30\,000~K). As the age of the PN increases,
the ionized shell of ejected and swept up gas expands
and fades, blending into the interstellar medium,
 the CSPN then entering the white dwarf (WD) cooling phase 
\citep{vassiliadis1994,bloecker1995,miller2016}.
The most important nucleosynthesis enrichment of the outer layer 
\citep[both light and heavy neutron-rich elements;][]{karakas2014} 
takes place during the AGB phase and marks the 
evolutionary phases that follow; during the AGB phase the so-called 
``third dredge-up'' can add sufficient carbon to the envelope 
and transform an O-rich star into a C-rich star.

The spectrum of a PN at UV wavelengths includes the both stellar
continuum emission from the CSPN and nebular emission (both 
emission lines and continuum emission). Nebular emission lines, 
such  as \ion{He}{2}\lam{1640}, \ion{C}{4}\lam{1549}, and 
\ion{C}{3}]\lam{1909}, and P-Cygni profiles of stellar wind, 
lines such as \ion{C}{4}, \ion{P}{4}, and \ion{N}{5}, are useful 
in the study of the hottest highly ionized regions in PNe and 
for the characterization of the ionizing star respectively 
\citep[e.g.,][]{feibelman2000,gauba2001,hoogerwerf2007,herald2011}.
PNe UV spectra have been taken with the \textit{International 
Ultraviolet Explorer} (\textit{IUE\/}) and with the \textit{Hubble Space 
Telescope} (\textit{HST\/}) spectrographs, mostly of the CSPN. In addition, the 
\textit{Far Ultraviolet Spectroscopic Explorer} (\textit{FUSE\/}; 
wavelength range of 905{\AA} to 1195{\AA}) observed 80 PNe and 
this data has led to important discoveries, such as highly ionized neon 
in the wind of CSPNe \citep{herald2011,keller2011}, whose lines 
are a crucial diagnostic for the hottest ($T_\text{eff}\geq85\,000$~K) 
CSPN, and P-Cygni profiles in highly ionized ions 
(e.g., \ion{P}{4}\lam{1118,1128}, \ion{C}{3}\lam{1175}, and 
\ion{S}{4}\lam{1073}) with stellar wind velocities between 
200~km\,s$^{-1}$ and 4300~km\,s$^{-1}$ \citet{guerrero2010}.
\citet{bianchi2018} recently presented broad-band UV imaging 
of PNe from the \textit{Galaxy Evolution Explorer} (GALEX) and 
showed that UV morphology and colors reflect the ionization 
structure of the PNe studied.
\citet{rao2018a,rao2018b} investigated the UV structure of 
NGC\,6302 and NGC\,40 using observations from the 
\textit{Ultraviolet Imaging Telescope} (\textit{UVIT\/}) on board 
ASTROSAT, mapping the \ion{C}{4}\lam{1549} emission line in 
order to study the shock interaction between the nebula and the ISM.

The origin of the morphology of PNe has been the subject of debate for
many years. Although the interacting stellar wind model
and its generalization \citep{kwok1978,balick1987} provide
a good explanation for the simplest PN morphologies (round
or slightly elliptical), roughly $\sim$80\% of PNe
present asymmetries in a sample
of 225, 900, 119, and up to 2699 true PNe in
\textit{The Hong Kong/AAO/Strasbourg H$\alpha$ catalog}
(HASH) database \citep[][respectively]{manchado2004,Parker2006,Sahai2011,parker2016}.
Recent studies suggest that the mechanism for producing
more complex morphologies is related to binary interaction
of the CSPN \citep[see][for a review]{jones2017}.
Many studies have been devoted to identifying binary companions
of CSPN by employing methods such as infrared (IR)
excess \citep{douchin2015,barker2018}, photometric variability
\citep{bond2009,guerrero2018} and radial velocity variations
\citep{jones2017a,jones2019,jones2019b}. Most of these studies
are based on optical or IR surveys, or long-term monitoring
of optical spectra. Hot WDs in binaries, however, are extremely
hard to identify unless UV data are available
\citep[e.g.,][]{bianchi2011b,bianchi2011a,bianchi2018z}.

GALEX \citep{martin2005} performed imaging surveys of the
sky in two UV bands: far-UV (FUV; 1344--1786{\AA}) and
near-UV (NUV;1771--2831{\AA}). GALEX UV catalogs matched with optical
surveys have proven to be an important tool in the detection
of hot-WD and main-sequence companions
\citep{bianchi2011b,bianchi2018z,bianchi2020arxiv}, making the
GALEX database a unique resource not only for the study of the
nebular gas and the CSPN properties, but also in the search for
binary CSPNe \citep[e.g.,][]{miszalski2012}.

In this paper we examine a sample of PNe from the HASH database
within the footprint of the GALEX surveys (Section~\ref{sec:constructing_uv_catalog}).
We also matched the PNe sample with the \textit{Sloan Digital Sky Survey}
\citep[SDSS;][]{york2000} and the \textit{Panoramic Survey Telescope
and Rapid Response System} \citep[Pan-STARRS;][]{chambers2016} sky surveys.
We then merged the matched catalogs to construct a catalog
of PNe with photometry covering the UV--optical spectral range
(Section~\ref{subsec:comprehensive_catalog}).
In Section~\ref{sec:analysis_pne_galex} we examine the effect 
of nebular emission lines and nebular continuum emission on 
the broad-band photometry considered here within 
the wavelength range of the \textit{GALEX}, SDSS, and Pan-STARRS data.
An analysis of the most extended PNe (more extended than a 
GALEX resolution element), using color--color diagrams with 
different UV--optical color combinations, is presented 
in Section~\ref{subsec:pne_cc_diagram}.
A summary and conclusions are given in 
Section~\ref{sec:summary_conclusions}.

\section{Constructing a UV-optical photometric catalog of Planetary Nebulae}
\label{sec:constructing_uv_catalog}

In this section, we match all 
known confirmed and presumed PNe from the HASH database with
UV imaging data from GALEX, as well as 
optical imaging from the SDSS and Pan-STARRS databases. Finally,
we distill these matched
catalogs to extract a PN sample with observations in the UV and optical ranges.

\subsection{The reference catalog}
\label{subsec:assemblinh_pn_sample}

The HASH database is the most up-to-date catalog 
of Galactic PNe; it includes a compilation of all previous 
PN catalogs 
\citep[e.g., ][]{Perek1967,kerber2003,stanghellini2010,Parker2006} 
and is actively ingesting new candidate (or confirmations of) PNe \citep[e.g., ][]{LeDu2022}.
We used the coordinates from the HASH 
database,\footnote{\url{http://hashpn.space}} 
which gives the position of 3865 objects with a reported 
positional accuracy of $\simeq$1{\arcsec}; 
2700, 459 and 706 are classified as true (T; spectroscopically 
confirmed), likely (L) and probable (P) PNe respectively.
For the purpose of this paper, we included all HASH PNe, even 
those that are classified as L or P. We use the latest (December 2022) update of 
the HASH database, which includes an additional 
209 spectroscopically confirmed Galactic PNe \citep{LeDu2022}.

For the following matches we preserved the HASH \texttt{status} 
column, which describes the PN's status (T, L, or P), and the 
\texttt{MajDiam} and \texttt{MinDiam} columns, which give the major 
and minor axes of the PN, and the \texttt{Catalog} column,
 which gives the source catalog.
We also included the \texttt{mainClass} and \texttt{subClass} columns 
that describe the main morphological type and the sub-morphological type respectively.
We refer to this extracted catalog as PNcat.

\subsection{Matching PNcat to the GALEX database: GUVPNcat}
\label{subsec:galex_match}

The \textit{Galaxy Evolution Explorer} (GALEX) imaged the sky in the far-UV
(FUV, 1344--1786{\AA}, $\lambda_\text{eff}=1538.6${\AA}) and near-UV
(NUV, 1771--2831{\AA}, $\lambda_\text{eff}=2315.7${\AA}) simultaneously,
with a field of view of 1{\fdg}2 diameter and a spatial resolution of
4{\farcs}2 and 5{\farcs}3 respectively \citep{morrissey2007}.
The images, reconstructed from photon-counting recordings, are sampled
with virtual pixels of size 1{\farcs}5 . The widest sky
coverage is provided by the All-Sky Image Survey (AIS) and the Medium
(depth) Imaging Sky Survey (MIS), which reach typical depths of 19.9 and
20.8\,mag (FUV/NUV), and 22.6/22.7\,mag (FUV/NUV) respectively
in the AB magnitude system
(see \citealt{bianchi2009b,bianchi2011a,bianchi2014,bianchi2017}
for a review and sky coverage).

In this paper we use data from the GALEX sixth and seventh releases (GR6/GR7), which
provides a sky
coverage of 24\,790~deg$^2$ and 2251~deg$^2$ for AIS and MIS respectively \citep{bianchi2019}.
The data were extracted from the \textit{Space Telescope Science Institute}
\textit{Mikulski Archive for Space Telescope} (MAST\footnote{\url{http://archive.stsci.edu/}})
at the CASJobs SQL interface.\footnote{\url{http://galex.stsci.edu/casjobs/}}. PNcat
was matched to the GALEX \texttt{visitphotoobjall}\footnote{The \texttt{visitphotoobjall} table,
includes
all the observations. The source catalog is compiled from individual visits. 
Therefore, there may be repeated observations of the same source.}
table using a match radius of 5{\arcsec}. This
radius value was chosen to include objects with positional error $\leq$5{\arcsec}.
Of the 3865 PNe in PNcat, 1605 matches of 671 unique PNe were found
in the \texttt{visitphotoobjall}  table.
The extracted matches constitute the matched catalog, which we name as GUVPNcat.

The GALEX database photometry table \texttt{visitphotoobjall}
also contains information
related to the \texttt{artifacts} in the images. According to the GALEX GR6
documentation,\footnote{\url{http://galex.stsci.edu/GR6/?page=ddfaq\#6}} the only artifact
flags causing real concern are \textit{Dichroic reflection} (\texttt{artifact} = 4 or 64) and
\textit{Window reflection} (NUV only; \texttt{nuv\_artifact} = 2). It is also recommended to
remove objects that are near the edge of the field of view as these sources could
have \textit{edge reflection} or the \textit{rim artifact} (\texttt{artifact} = 32) set
\citep{bianchi2011a,bianchi2014,bianchi2017}. Sources with 
\texttt{fov\_radius}\footnote{\texttt{fov\_radius} is the distance of 
the source from the center of the field.} $>$ 0{\fdg}55 have to be
examined carefully because of the poor astrometry and photometry 
near the edge of the field \citep[as explained in ][]{morrissey2007,bianchi2017}.
A total of 240 measurements have \texttt{fov\_radius}$>$0{\fdg}55, 
of which only 31 have \texttt{artifact} = 32.
These tags are included in GUVPNcat.

The \texttt{visitphotoobjall} table gives several types of 
magnitudes for each source
obtained by different procedures (source profile fits or aperture 
magnitudes), such as Kron elliptical
magnitude (MAG\_AUTO), isophotal (MAG\_ISO), and circular aperture 
magnitudes (\texttt{FUV\_MAG\_APER\_}\textit{\#n} and
\texttt{NUV\_MAG\_APER\_}\textit{\#n}; these are seven magnitude measurements 
with different aperture radius, as described by
\citet[][their Figure~4]{morrissey2007}. We extracted the 
\texttt{fuv\_mag} and \texttt{nuv\_mag} tags, which are the
``best'' measurement for each source as determined by the GALEX 
pipeline.  We also include in GUVPNcat
other measurements (with their respective errors), including the 
size (semi-major and semi-minor axes with
tags \texttt{A\_IMAGE} and \texttt{B\_IMAGE} respectively) and 
ellipticity (ELLIPTICITY = $1-$B/A) of the source which
correspond to the area of integration for the \texttt{MAG\_AUTO} magnitude.
GALEX magnitudes are in the AB magnitude system \citep{oke1983}.

Each row resulting from the match with \texttt{visitphotoobj} 
corresponds to a GALEX observation, typically taken 
with both detectors on; therefore, the row for one observation 
includes resulting positions and magnitudes in the FUV and NUV. In some 
observations the FUV detector was off \citep[see][]{bianchi2017} 
so that only an NUV image was taken.
The \texttt{nuv\_weight} and \texttt{fuv\_weight} (from visitphotoobj) tags
indicate effective exposure time for NUV and FUV in each observation
and are reported in GUVPNcat.
A total of 739 out of 1605 observations were taken with both 
detectors on; of these,  138 do not have an FUV magnitude because 
the source flux is fainter than the detection threshold 
in the FUV frame,  while 90 are not detected in NUV. We enter magnitude values 
of $-$999 for non detection in GUVPNcat to distinguish them from the cases 
when FUV detection was off.
Out of the 1605 GALEX observations, 856 were taken with only the
NUV detector on, and ten were taken with only the
FUV detector on.
When one detector was off, the photometry tags relative to that 
detector are set to $-$888 in GUVPNcat.
Exposure times range from $\sim$31 to $\sim$1705~s; 
one should be aware that the magnitude limit is not constant in this 
serendipitous compilation.

\begin{figure}
\centering
\includegraphics[width=0.5\textwidth]{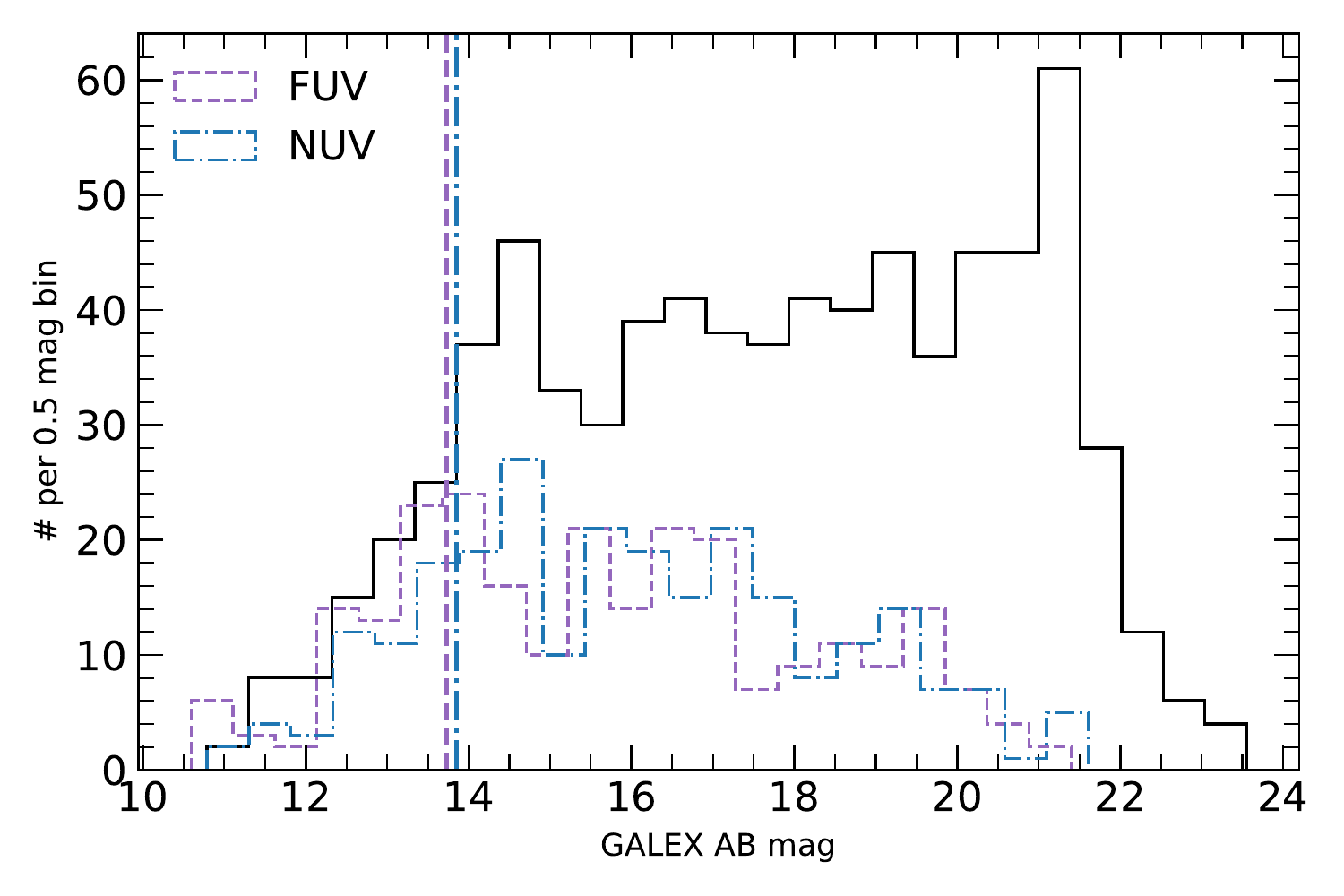} \\
\includegraphics[width=0.5\textwidth]{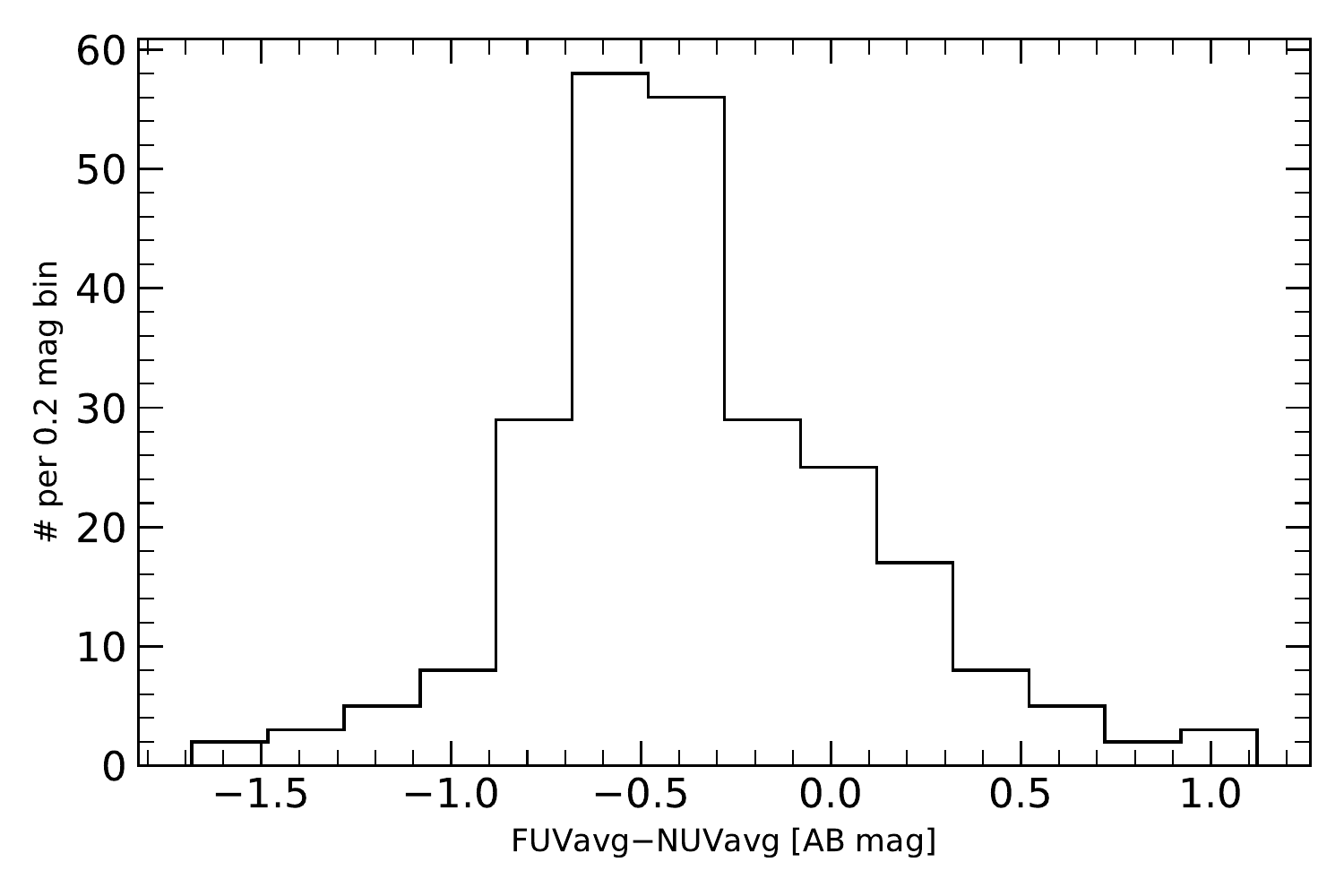}
\caption{Distribution of GALEX  FUVavg and NUVavg magnitudes 
(top), and FUV$-$NUV (bottom) in GUVPNcat. The vertical lines
mark the saturation limit of each band in the upper panel.}
\label{im:fuvnuv_stair}
\end{figure}

The GALEX \texttt{visitphotoobjall} table contains all existing measurements.
Following
\citet{bianchi2017}, we added tags to identify multiple measurements of the same source.
Sources from \texttt{visitpohotoobjall}  within 2.5{\arcsec}  
of each other and measured in different observations are considered
duplicate measurements of the same source. The best measurement, 
using the criteria of \citet{bianchi2017}, is chosen as  primary
(and \texttt{primgid} is the primary's object id) and given 
\texttt{grankdist} = 1, while the other measurements of the same source are
given \texttt{grankdist} $>$ 1 in order of distance from the primary.
For a given GALEX source, \texttt{grankdist} = 0 implies that 
the object has one GALEX observation.
The \texttt{ngrankdist} tag gives the number of measurements
associated with the source, and the
\texttt{primggroupid} tag gives the concatenation of the \texttt{objid}s of these measurements; 
\texttt{objid}, a unique GALEX identifier for the source,
is extracted from the \texttt{visitphotoobjall} table and is included in
GUVPNcat, as well as an IAU-style identifier constructed from the coordinates.
Averaged magnitudes, \texttt{FUVavg} and \texttt{NUVavg}, were 
calculated for each source as the mean values of their 
corresponding repeated measurements weighted by the errors.
Of the 1605 matches in GUVPNcat, 392 have only one observation 
(\texttt{grankdist} = 0);  370 have multiple observations 
of the same source (\texttt{grankdist} = 1); 1202 have 
\texttt{grankdist} $\geq$ 1, from 2 (e.g., GALEX J174232.4$-$180943) 
up to 48  (GALEX J125927.8$+$273811), and only 11 have 
\texttt{grankdist} = $-$1.
The distribution of \texttt{FUVavg}, \texttt{NUVavg} and 
\texttt{FUVavg}$-$\texttt{NUVavg} of the sources is shown in 
Figure~\ref{im:fuvnuv_stair}.
Vertical dashed lines correspond to the non-linearity limit for FUV and NUV
\citep[13.73 and 13.85\,mag respectively; see][for details 
on non-linearity limits]{bianchi2018}.
Selecting entries with \texttt{grankdist} = 0,1 or $-1$ gives a 
list of unique GALEX sources found within 5{\arcsec} of the PN coordinates.

We can now examine whether there are multiple matches,  
i.e., more than one GALEX source within the match radius for each PN.
We counted multiple matches for observations with 
\texttt{grankdist}=1, 0, or $-1$, as those with \texttt{grankdist}$>$1 
are deemed repeated measurements of the same match.
To track the multiple matches we followed the flagging system of the
 GUVmatch catalogs given in Table~1 of \citet{bianchi2020arxiv}. We 
assign \texttt{distancerank} = 0 to the GALEX match if the PN has only 
one GALEX match; otherwise, we rank
the multiple matches based on the distance, with the closest source
defined as the primary match (\texttt{distancerank}=1), the additional
GALEX matches being assigned \texttt{distancerank}$>$1 in order of 
distance \citep[as defined in][]{bianchi2020arxiv}.
We examined the 85 PNe that have multiple matches and found that some of them
correspond to a part of the nebular gas that was detected as an 
independent source (e.g., GALEX J190432.4$+$175710), a few of 
them correspond to a nearby star (e.g., GALEX J183533.4$-$313543), 
and the majority of them correspond to artifacts due to a saturated 
central star (e.g., GALEX J222938.5$-$205014, GALEX J155159.8$+$325658 
and GALEX J201508.8$+$124217), as the non-linearity affects not 
only the count rate of a bright source but also its shape and position 
\citep{morrissey2007}. All multiple matches are included in GUVPNcat. 
However, they can be filtered out by selecting objects with 
\texttt{distancerank}=0 or 1 (671 observations).

The columns of GUVPNcat are described in Table~\ref{tab:gpncat} 
in Appendix~\ref{ap:description_catalog}.

\subsection{Matching PNcat to SDSS: PNcatxSDSSDR16}
\label{subsec:sdss_match}

The \textit{Sloan Digital Sky Survey} \citep[SDSS;][]{york2000} has mapped
the sky  in five  broad bands using a dedicated 2.5\,m 
telescope located at the Apache Point
Observatory (APO) in New Mexico. The telescope used a 
wide field-of-view camera to acquire
the images from 3048{\AA} to 10\,833{\AA}, in five pass bands:
$u$ (3048--1028{\AA}, $\lambda_\text{eff} = 3594.9${\AA}), 
$g$ (3783--5549{\AA}, $\lambda_\text{eff} = 4640.4${\AA}),
$r$ (5415--6989{\AA}, $\lambda_\text{eff} = 6122.3${\AA}), 
$i$ (6689--8389{\AA}, $\lambda_\text{eff} = 7439.5${\AA}),
and $z$ (7960--10\,830{\AA}, $\lambda_\text{eff} = 8897.1${\AA}), 
with a spatial resolution of $\sim$1{\farcs}4.

We used SDSS data release 16 
\citep[DR16;][]{Ahumada2020}, which covers a unique footprint of 14\,724~deg$^2$
of sky \citep[from 
AREACat.\footnote{\url{http://dolomiti.pha.jhu.edu/uvsky/area/AREAcat.php}};][]{bianchi2019}
The match to PNcat was performed with the STScI 
MAST database at the CASJobs SQL 
interface,\footnote{\url{https://skyserver.sdss.org/casjobs/}}
with the SDSS DR16 \texttt{photobojall} table, using a 
match radius of 5{\arcsec}. We removed objects with the ``edge''
flag set, as suggested in the SDSS web 
page,\footnote{\url{https://www.sdss4.org/dr17/algorithms/photo_flags_recommend/}}
and kept only SDSS matches
with \texttt{mode} $=$ 1 set; only primary sources.
The DR16 \texttt{photoobjall} table contains mostly unique sources 
(found by selecting \texttt{mode} = 1), except 
that a few primaries have duplicate entries due to duplicate spectra. We checked and  found that
there were no duplicate entries.

We found 108 SDSS DR16 sources within 5{\arcsec} of 83 PNcat objects.
To track the multiple matches we again 
followed the flagging system of the GUVmatch catalogs 
given in Table~1 of \citet{bianchi2020arxiv}.
Of the 83 PNe in SDSS, 66 have only one SDSS 
counterpart (\texttt{distancerank}=0) and 17 have more 
than one SDSS counterpart (\texttt{distancerank}=1) 
within the match radius. The resulting 
matched catalog, PNcatxSDSSDR16, is available in electronic 
form only; the SDSS tags included in PNcatxSDSSDR16 are 
described in Table~\ref{tab:pncat_sdss} in Appendix~\ref{ap:description_catalog}.

\subsection{Matching PNcat to Pan-STARRS: PNcatxPS1MDS}
\label{subsec:panstarrs_match}

The \textit{Panoramic Telescope and Rapid Response System} 
\citep[Pan-STARRS;][]{chambers2016} is a system for wide-field 
astronomical imaging in the northern hemisphere 
(Dec.\ $> -$30{\degr}). Pan-STARRS1 PS1 $3\pi$ is the first 
part of the Pan-STARRS sky surveys to be
completed and comprises the current data releases DR1 (3$\pi$ 
survey) and DR2 (Medium Deep Survey; hereafter MDS).
The PS1 survey used a 1.8\,m ground-based telescope, located 
at Haleakala Observatory in Hawaii and its 1.4 gigapixel camera
to image the sky in five broad bands: $g_\text{PS1}$ 
(3943$-$5593{\AA}, $\lambda_\text{eff}=4775.6${\AA}),
$r_\text{PS1}$ (5386--7036{\AA}, $\lambda_\text{eff}=6129.5${\AA}), 
$i_\text{PS1}$ (6778--8304{\AA}, $\lambda_\text{eff}=7484.6${\AA}),
$z_\text{PS1}$ (8028--9346{\AA}, $\lambda_\text{eff}=8657.8${\AA}), 
$y_\text{PS1}$ (9100--10\,838{\AA}, $\lambda_\text{eff}=9603.1${\AA}),
with a field of view of 3{\degr} and a single epoch depth 
(5$\sigma$) of 22.0, 21.8, 21.5, 20.9 and 19.7\,mag respectively.

We matched PNcat with PS1 MDS using the CasJobs SQL interface at 
MAST\footnote{\url{https://panstarrs.stsci.edu}}
with a match radius of 5{\arcsec}. Sources were extracted from 
the \texttt{MeanObjectView} joined to the \texttt{StackObjectAttributes}
table.

The \texttt{MeanObjectView} table contains the mean photometry 
for objects based on single-epoch data
(i.e., we do not expect multiple measurements of the same object 
included in this table).
As for the match with the GALEX and SDSS catalogs, we flag 
(and retain) multiple matches within the match radius, with the 
\texttt{distancerank} tag \citep[see Table~1 of][for definitions]{bianchi2020arxiv}.

Table~13 of \citet{flewelling2020} lists flags relative to the 
quality of the extracted photometry contained in the
\texttt{MeanObjectView} table. We restricted our results to 
sources with \texttt{objInfoFlag} containing the `GOOD'
flag set (Good quality measurement in the PS1 data; e.g., PS) 
and sources which have 
\texttt{nDetections} $\geq$ 3 in
at least the $g_\text{PS1}$, $r_\text{PS1}$, and $i_\text{PS1}$ bands 
(where the majority of pixels were not masked;
\texttt{gQfPerfect}, \texttt{rQfPerfect}, and \texttt{iQfPerfect} $>$ 0.85).

Out of the 3865 PNe in PNcat, we found 3301 matches within 
5{\arcsec} of 1819 unique PNe in the PS1 MDS \texttt{MeanObjectView} 
table. The tags of the resulting matched catalog, 
PNcatxPS1MDS, are described in Appendix~\ref{ap:description_catalog} 
(Table~\ref{tab:pncat_ps1}).
Of the 1819 PNcat objects in PS1 MDS, 927 have only 
one counterpart (\texttt{distancerank}=0) and 892 PNcat 
objects have more than one PS1 MDS counterpart 
(\texttt{distancerank} = 1) within the match radius.

The columns of PNcatxPS1MDS are described in Table~\ref{tab:pncat_sdss} 
in Appendix~\ref{ap:description_catalog}.

\subsection{Matched GUVPNcat, PNcatxSDSSDR16, and PNcatxPS1MDS: GUVPNcatxSDSSDR16xPS1MDS}
\label{subsec:comprehensive_catalog}

Here we distill and compile all the above matched catalogs into a single catalog.
Whereas in the matches with each database we 
kept multiple measurements and multiple matches with 
flags to identify them, in the merged 
match catalog we keep only the ``primary'' match from 
each database, but we preserve the information on existing 
additional matches by including  ad hoc tags.

The comprehensive matched catalog of GALEX, SDSS and Pan-STARRS data,
GUVPNcatxSDSSDR16xPS1MDS, is a concatenation
of GUVPNcat, PNcatxSDSSDR16, and PNcatxPS1MDS, including only 
matches with \texttt{distancerank} = 0 or 1 which implicitly 
includes only \texttt{grankdist} = 0, 1, or $-$1 for GUVPNcat, resulting
in a catalog of unique PNe and matched sources.
A total of 36 and 362 PNe from GUVPNcat have a match in SDSS 
and Pan-STARRS (the 36 SDSS objects are also in Pan-STARRS),
respectively. The GUVPNcatxSDSSDR16xPS1MDS catalog 
contains a total 375 PNe with spectral coverage from $\sim$1540 to
$\sim$9610{\AA}, including 13 PNe observed by us; see Section~\ref{sec:observational_data}.
The GUVPNcatxSDSSDR16xPS1MDS columns are described in Table~\ref{tab:gpncat_unique}.

\begin{figure}
    \centering
    \includegraphics[width=1.0\columnwidth]{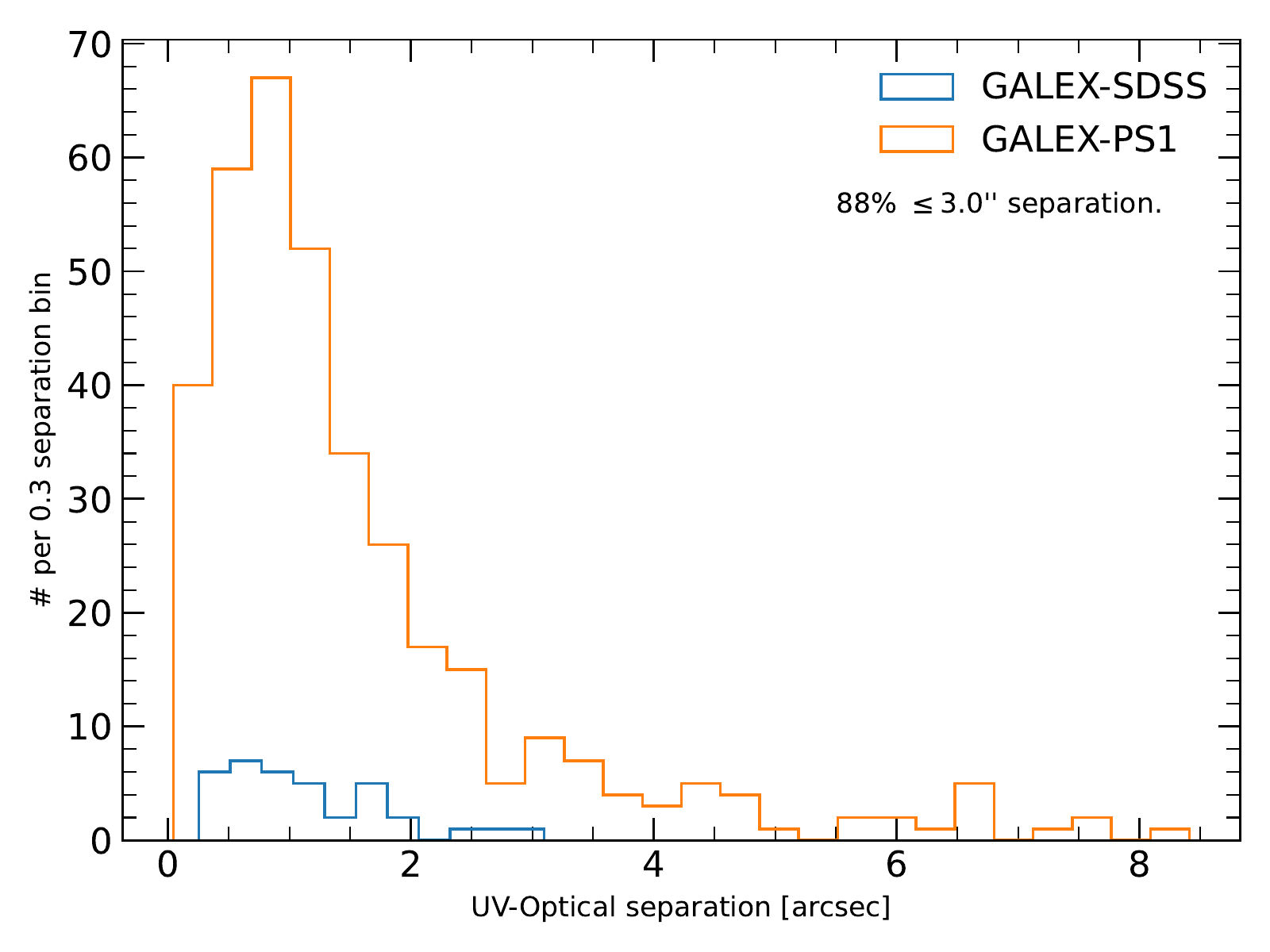}
    \caption{Distribution of GALEX-SDSS (blue) and GALEX-PS1 (orange) best match coordinates separation.
    About 88\% of the separations are smaller than the GALEX resolution (5{\arcsec}).\label{fig:hist_uvopt}}
\end{figure}

Figure~\ref{fig:hist_uvopt} shows the distribution of the 
separation between the GALEX and the optical counterpart 
coordinates (SDSS and Pan-STARRS in blue and orange, respectively).
 While the separation between the UV and optical coordinates is 
usually good (88\% are $\leq$3{\arcsec} separated), special 
attention should be given when analyzing objects 
with \texttt{distancerank} = 1. Although the GALEX 
astrometric accuracy is better than 2{\farcs}5 
($\sim$1{\arcsec}), deblending sources closer than this separation 
is not always robust because of the instrument resolution 
\citep[5{\arcsec}; see][]{morrissey2007,bianchi2017}. Therefore,
when analyzing individual objects, it is important to 
examine multiple matches (those with \texttt{distancerank} $\geq$ 1), 
as the closest optical source may not necessarily be the 
correct match and because the GALEX source may include flux 
from multiple sources unresolved in the GALEX images, making the 
UV-optical color biased. We analyzed the objects with UV-optical 
separation larger than 5{\arcsec} and found that the majority of
 them correspond to GALEX magnitudes $>$ 19~mag and are usually 
low-surface brightness and unresolved objects in UV in which the
 CSPN is not visible. Tags giving the separation between GALEX 
and SDSS (\texttt{sep\_GS}) and GALEX and Pan-STARRS 
(\texttt{sep\_GP}) positions are included in GUVPNcatxSDSSDR16xPS1MDS.

The GUVPNcatxSDSSDR16xPS1MDS catalog includes a total of 362
 PNe with measurements in the GALEX, SDSS DR16
and/or PS1 MDS surveys
from a total of 3865 PNe in PNcat. In addition, it includes 13
PNe observed by us.
The catalog includes columns (tags) with measurement 
of sizes \citep[diameters; ][]{parker2016},
Gaia EDR3 distances from \citet{gonzalessantamaria2021} and \citet{Chornay2021},
and the surface brightness-radius relation distances from \citet{Frew2016}.
The extracted CSPNe photometry is also included for the selected
PNe (see sections~\ref{subsec:cspne_photmetry} and \ref{sec:observational_data}).

A description of the tags is given in Apendix~\ref{ap:description_catalog}.

\section{Analysis: Effect of Nebular emission on the CSPN photometry}
\label{sec:analysis_pne_galex}

In this section we investigate the effect of the nebular 
emission in the CSPNe photometry
on the filters included in the GUVPNcatxSDSSDR16xPS1MDS 
catalog. We analyze separately those PNe larger than the GALEX
resolution element, and those unresolved in GALEX imaging. 
For the resolved PNe
we isolated the CSPNe flux from the nebular emission by 
aperture photometry.

It has been shown that the combination of UV and 
optical colors is sensitive to the ionization structure,
 and hence to the effective temperature and age, of the CSPN 
\citep[e.g., ][who folded synthetic PNe model spectra 
with the LSST photometric system]{vejar2019}. 
In general, the GALEX, SDSS, and PS1 measurements may contain
flux from both the CSPN continuum and PN nebular emission (lines and continuum).

\begin{figure*}
\centering
\includegraphics[width=0.95\columnwidth]{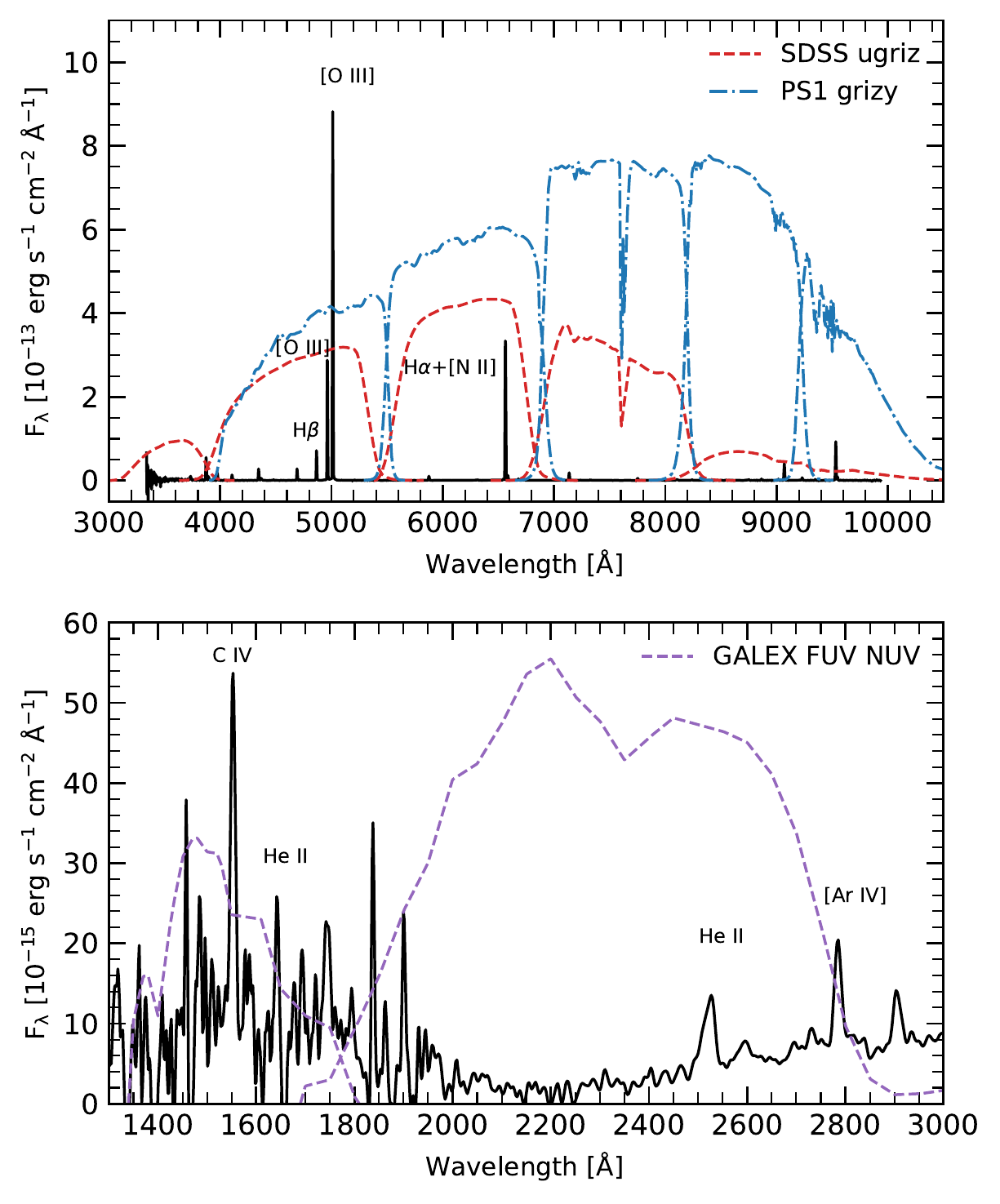}	
\includegraphics[width=0.95\columnwidth]{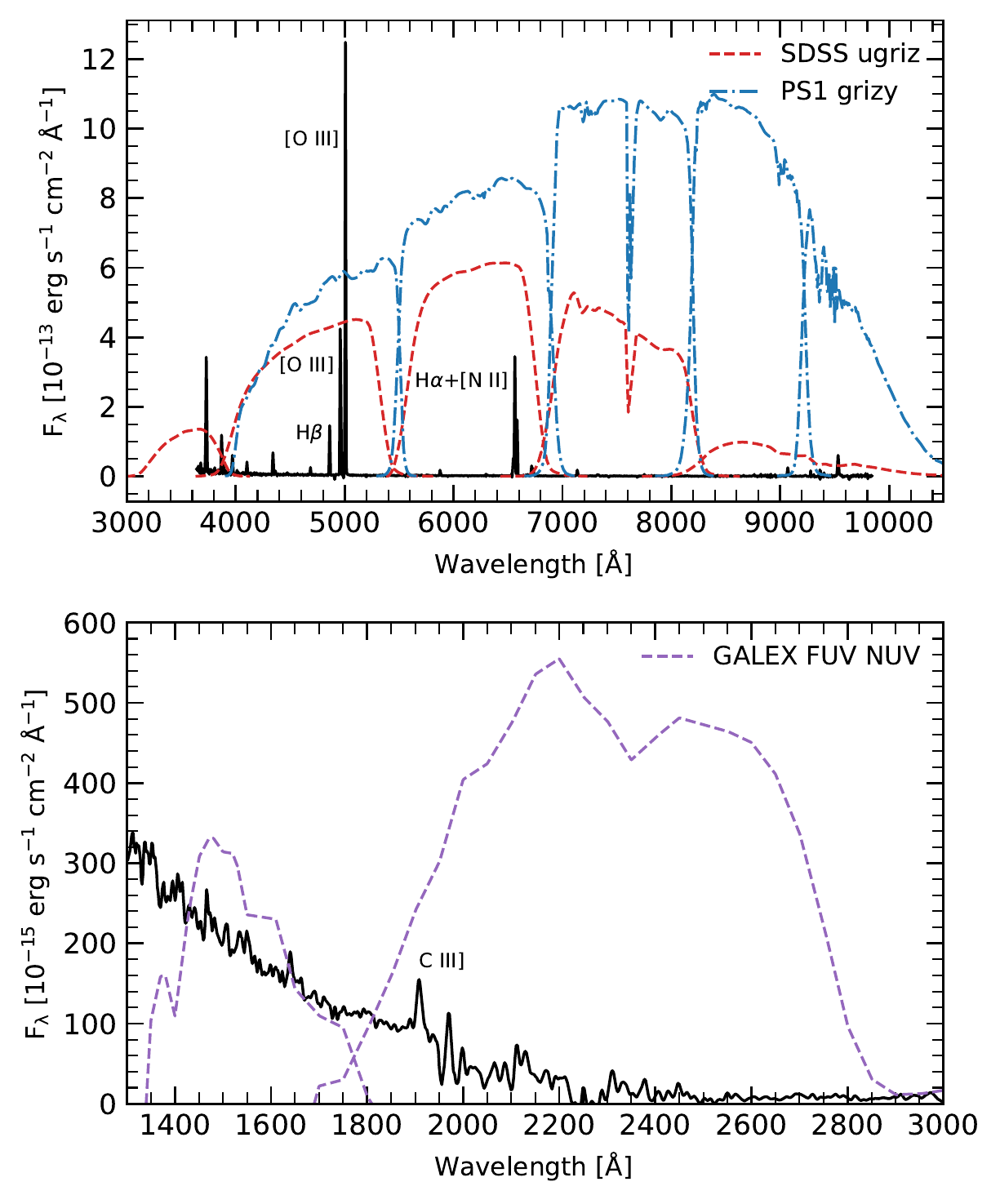}	
\caption{Optical (top panels) and UV (bottom panels) spectra of 
PN NGC~1501 (left) and NGC\,3587 (right).
The SDSS and PS1 (top), and GALEX (bottom) transmission curves 
are overlaid.
\label{im:opt_neb_example}}
\end{figure*}

Figure~\ref{im:opt_neb_example} shows
UV and optical spectra of two PNe with 
different ionization structure, revealing the rich
information contained in the UV and optical ranges.
The figure shows the optical and UV
spectrum of NGC\,1501\footnote{Taken from the Gallery of
PNe spectra:
\url{https://web.williams.edu/Astronomy/research/PN/nebulae/} \label{foot:8}}
(left panels), with the
transmission curves (top panel) of SDSS $u$, $g$, 
$r$, $i$ and $z$ (red dashed line), 
PS1 $g$, $r$, $i$,$z$, $y$ (blue dot-dashed line),
and the IUE SWP28952 and LWP08948 UV spectra with the
transmission curves of GALEX FUV and NUV overlaid 
(bottom panel, purple dashed line).
In the optical range, an [\ion{O}{3}] emission line is 
prominent in the \textit{g} band, and H$\alpha$ and
[\ion{N}{2}] emission lines in the \textit{r} band. In the 
UV range, \ion{C}{4} and
\ion{He}{2} emission lines are present in FUV, and \ion{He}{2} and 
[\ion{Ar}{4}] lines in NUV.
Note,  however,  that the \ion{He}{2}~1641{\AA} and 
\ion{C}{4}~1551{\AA} nebular emission lines
are a characteristic signature of high-ionization PNe 
and, in the case of low-resolution spectra,
may be a composite of stellar and nebular emission lines.
Figure~\ref{im:opt_neb_example} also shows the optical and UV spectrum of
NGC~3587$^{\footnotesize \,\ref{foot:8}}$ (right panels),
again with the transmission curves 
of SDSS (red dashed line)
and PS1 (blue dot-dashed line),  and the IUE SWP04920 and 
LWR04251 UV spectra with the GALEX FUV and NUV transmission 
curves overlaid.
Similar to NGC ~1501, optical emission lines of [\ion{O}{3}],  
[\ion{N}{2}],  H$\alpha$, and H$\beta$ are seen in 
the SDSS~$g$ (PS1~$g$) and SDSS~$r$ (PS1~$r$) 
bands; the [\ion{N}{2}] 6548{\AA} and 6584{\AA}
emission lines are stronger in 
low-ionization PNe. In the UV range, strong emission 
lines of \ion{C}{4} and \ion{He}{2}  are not present
in low-ionization PNe (e.g., NGC~3587).
More examples showing a variety of cases are given by \citet{bianchi2018}.

Nebular emission complicates measurements
of the CSPN flux, particularly for those PNe that have bright central emission
and for compact PNe.
In the case of GALEX, the ``best'' magnitude measurement reported by the pipeline
is obtained with a Kron elliptical aperture of the size of the nebular extent. The 
best magnitude reported by the GALEX pipeline contains the CSPN flux and 
the surrounding nebular flux.
Therefore, for extended PNe (compared with the GALEX instrumental PSF), we measured
the CSPN flux with
an aperture the size of the instrumental PSF and corrected
for the nebular contribution estimated in an annulus
(of the size of the extended emission)
to subtract the PN emission plus local background flux.
The accuracy of the correction depends on the radial 
profiles of the flux in the
emission lines contributing to each filter. If the 
correction is small with respect
to the CSPN flux, the estimated error will be small.
\citet{bianchi2018}, for example, show seven PN radial profiles 
from GALEX; in some
of them the nebular emission is several orders of magnitude 
fainter than the CSPN flux.
However, for hot CSPNe ($T_\text{eff} >$ 45\,000~K),
the extracted CSPN flux could be affected by the PN's 
contribution to the \ion{He}{2}~$\lambda$1640 emission line, which is
strong in the inner parts of the PN close to the CSPN.

\subsection{PNe resolved by GALEX}
\label{subsec:extended_pne_galex}

The size and morphology of a PN at given wavelengths 
depends on the ionization structure, projection effects,
the sensitivity of the instrument and depth of the exposure.
A number of surveys have measured the sizes
 of PNe at different wavelengths 
\citep[e.g., ][]{acker1992,stanghellini2010,frew2013}.
As a starting point, we used the size reported in 
PNcat (measured in optical images),
which is included in all the matched catalogs constructed in this work,
to separate PNe potentially resolved in GALEX imaging.
As a first approximation we assume a PN to be resolved 
in GALEX imaging if its size is at least twice the
resolution of GALEX in the NUV band (2$\times$5{\farcs}2).
A total of 252 PNe were found with diameter $>$ 10{\arcsec} 
(using \texttt{MajDiam} when available).
However, PN sizes from PNcat were measured in optical imaging.
We estimate PN sizes in the GALEX UV imaging.

\begin{figure*}
\centering
\includegraphics[width=0.35\textwidth]{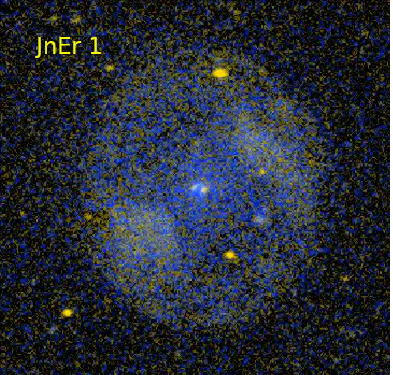}
\includegraphics[width=0.35\textwidth]{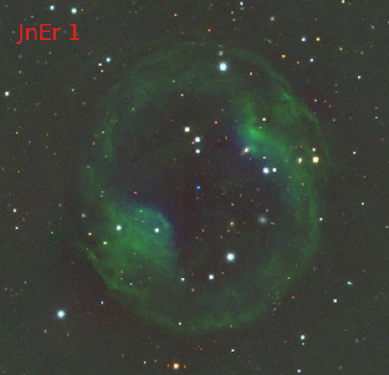}
\includegraphics[width=0.65\textwidth]{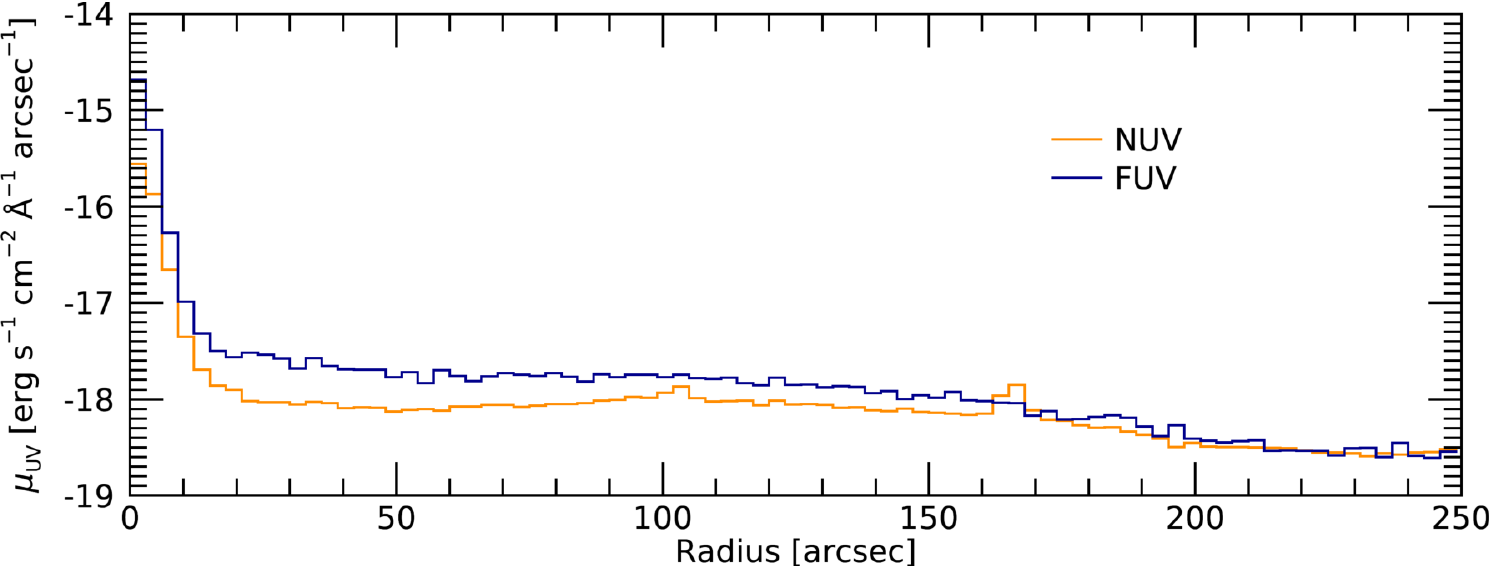}
\caption{Example of the extended ($\sim$200\,{\arcsec}) PN JnEr\,1 observed by
	GALEX \citep[left,][]{bianchi2018} in FUV (blue) 
and NUV (yellow), and by
PS1 (right) in $g$ (blue), $r$ (green), and $y$ (red).
North is up and east is to the left.
In the GALEX imaging, the CSPN appears very bright and resolved 
but the ``best'' measurement 
includes nebular emission: according to the pipeline the 
best magnitude is obtained from a Kron elliptical aperture with
a Kron radius of 5.25{\arcsec}, and semi-major and 
semi-minor axis of 11.63{\arcsec} and 9.62{\arcsec} 
respectively. In PS1 it
was measured as a point-like source \citep[with 
\texttt{PS1\_psfMag\_r} $-$ \texttt{PS1\_kronMag\_r} $<$ 0.04 
according to][definition of point-like sources]{chambers2016}.
A radial profile for the two GALEX
bands (bottom) is also shown.
Aperture measurements are included in the catalog and 
should be used for the central star.}
\label{fig:galex_optical_resolved}
\end{figure*}

As mentioned in Sect.~\ref{subsec:galex_match}, the GALEX 
database provides various measurements of the flux for
 each source, such as
Kron-like elliptical aperture,
isophotal, and circular apertures.
The ``best'' measurement (the \texttt{fuv\_mag} and \texttt{nuv\_mag} 
columns in the database) corresponds to the best fit of
the source shape as determined by the pipeline.
If the CSPN is relatively bright compared with the surrounding
 PN, the ``best'' magnitude may be extracted by treating the
object as a point-like source.  
Figure~\ref{fig:galex_optical_resolved} shows the
extended PN\,G164.8+31.1 (JnEr\,1), and its FUV and NUV 
radial profiles as an example.
More examples, showing a wide range of UV flux radial profiles 
in the GALEX imaging of PNe, are shown by \citet{bianchi2018}.

\begin{figure}
\centering
\includegraphics[width=1.\columnwidth]{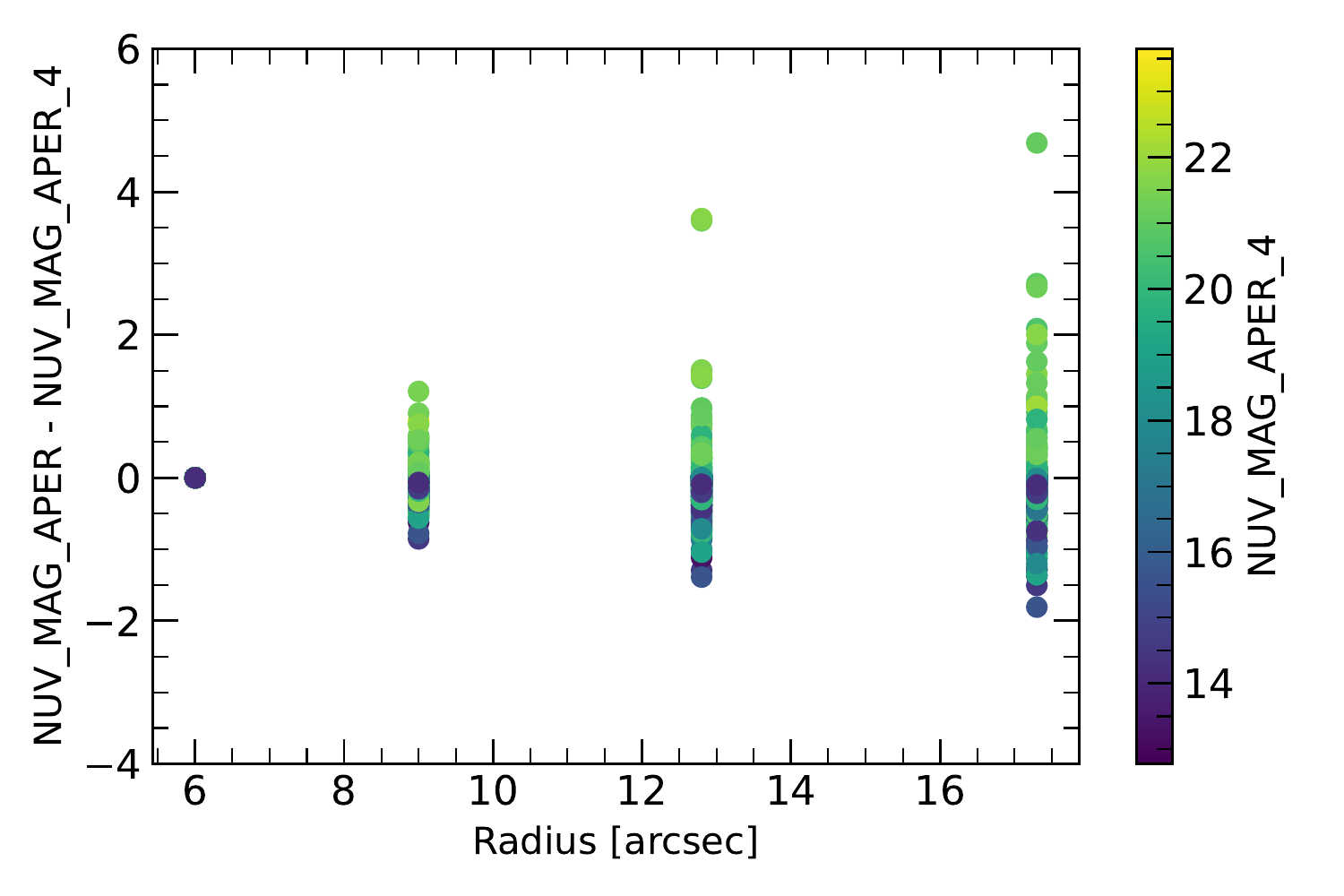} \\
\includegraphics[width=1.\columnwidth]{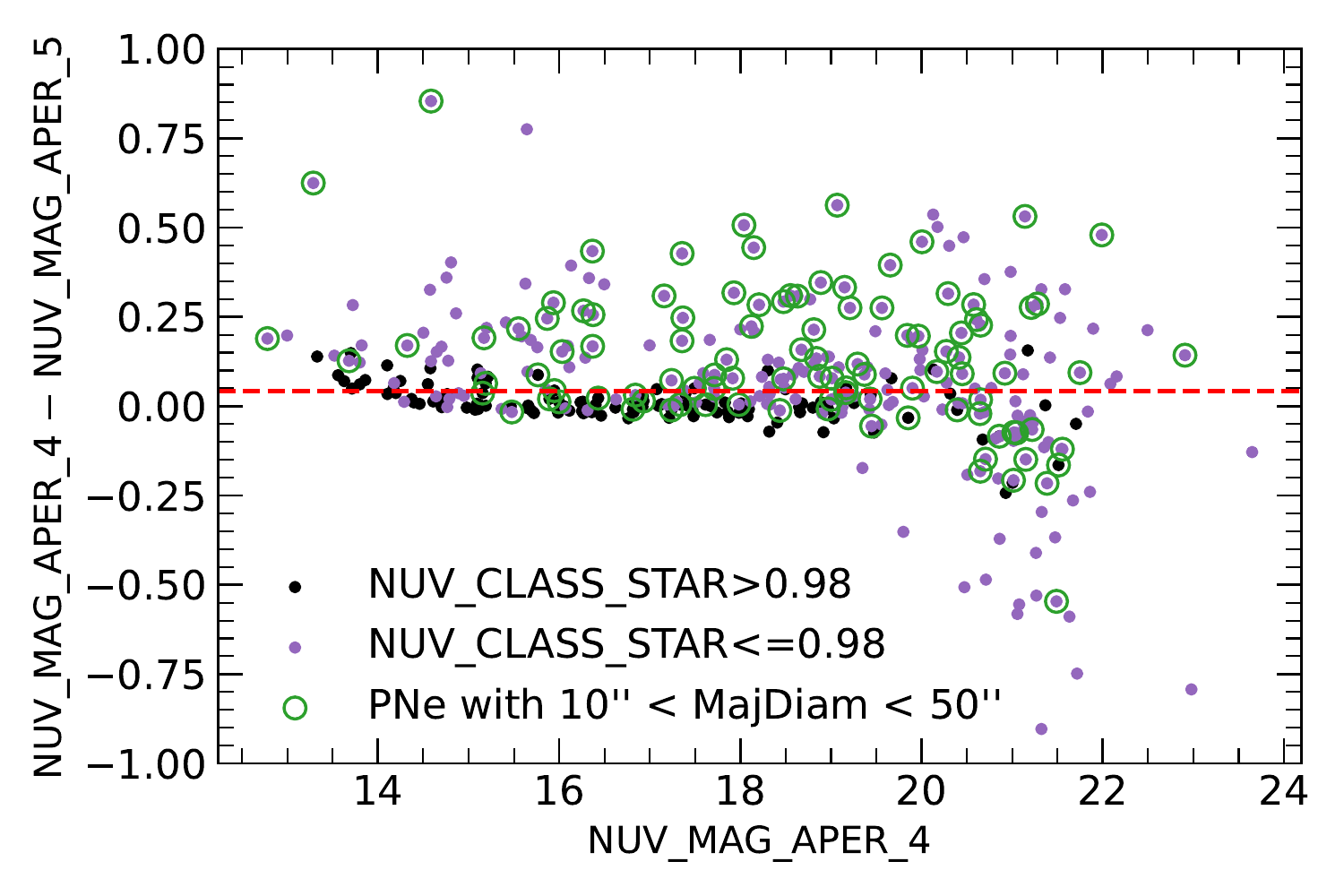}
\caption{Top panel: Curve of growth of GALEX \texttt{NUV\_MAG\_APER\_4} 
(6{\arcsec}) compared with magnitudes measured in larger GALEX apertures 
(NUV\_MAG\_APER\_5, NUV\_MAG\_APER\_6, and NUV\_MAG\_APER\_7; 9{\arcsec},
 12{\farcs}8, and 17{\farcs}3). Aperture correction from
 \citet{morrissey2007} has been applied. Bottom panel: Difference
 between the GALEX 6{\arcsec} and 9{\arcsec} radius aperture magnitudes. 
A difference larger than 0.036 (red solid line) indicate that the object 
is extended. For values below 0.036 we consider the object to be
 point-like (see text). \label{fig:nuv_cog}}
\end{figure}

In order to separate the flux of the CSPN from that of the 
PN in GALEX imaging,  we compared
the magnitudes measured in different apertures provided by 
the pipeline \citep[see][and Section~\ref{subsec:galex_match}
 for aperture radius]{morrissey2007}.
Aperture corrections are described in figure~4 of \citet{morrissey2007}
and discussed in \citet[][their figure~3]{delavega2018}.
Figure~\ref{fig:nuv_cog} shows the difference between the 
GALEX \texttt{NUV\_MAG\_APER\_4} (6{\arcsec} radius) with respect to the GALEX larger
apertures (\texttt{NUV\_MAG\_APER\_5}, \texttt{NUV\_MAG\_APER\_6}, and \texttt{NUV\_MAG\_APER\_7}).
For a point-like object, the difference between aperture magnitudes larger
than the instrument PSF is nearly zero.
In this work, all objects with 
\texttt{NUV\_MAG\_APER\_4}$-$\texttt{NUV\_MAG\_APER\_5}$>S$
were cataloged as extended.
A value of $S=0.036\pm0.003$~mag was obtained by a linear-regression fitting to all objects with
\texttt{NUV\_CLASS\_STAR} $>$ 0.95 (point-like objects according to the GALEX pipeline)
for \texttt{NUV\_MAG\_APER\_4} between 14 and 19~mag weighted by their uncertainty.
We used \texttt{NUV\_MAG\_APER\_4} to separate the resolved 
objects (Fig.~\ref{fig:nuv_cog}, bottom), as this aperture is 
more than twice the resolution of GALEX and, by comparing 
known PN diameters from PNcat (Figure~\ref{fig:nuv_cog}), the 
difference \texttt{NUV\_MAG\_APER\_4} $-$ \texttt{NUV\_MAG\_APER\_5} $>$ 0.036 
appeared to be a good indicator of an extended PN. A total of 170 extended PNe were 
deemed extended in GALEX imaging by this criterion.

We found 16 and 170 of the extended PNe from GALEX in 
the SDSS and PS1 catalogs, respectively.
SDSS classifies extended objects based on the difference 
\texttt{psfMag\_r} $-$ \texttt{r} $>$ 0.145 
(\texttt{r\_sdss\_diff}), whereas in the case of PS1, we 
identified extended PNe as described by \citet{farrow2014}, 
with \texttt{rMeanPSFMag}$-$\texttt{rMeanKronMag}$>$0.05 
(\texttt{r\_ps1\_diff}).

Some PNe that are not classified as resolved in both UV and 
optical bands are not necessarily compact PNe. PNe with 
an extended envelope of low surface brightness may only show the CSPN
in GALEX images depending on the exposure.

\begin{figure}
\centering
\includegraphics[width=1.\columnwidth]{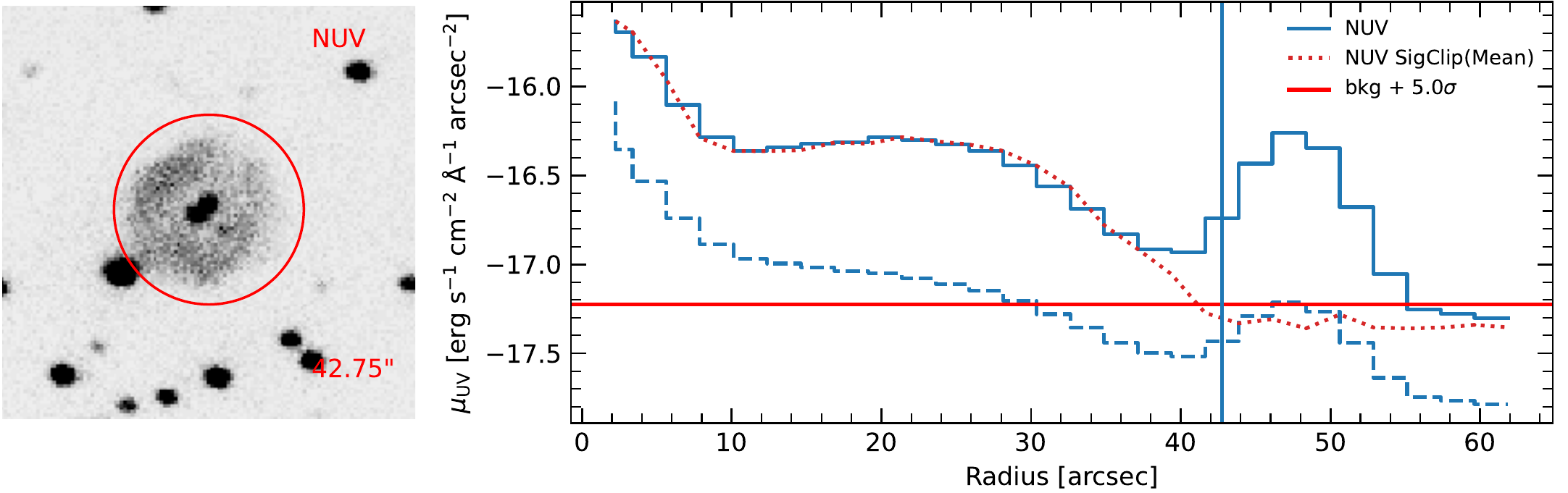}
\includegraphics[width=1.\columnwidth]{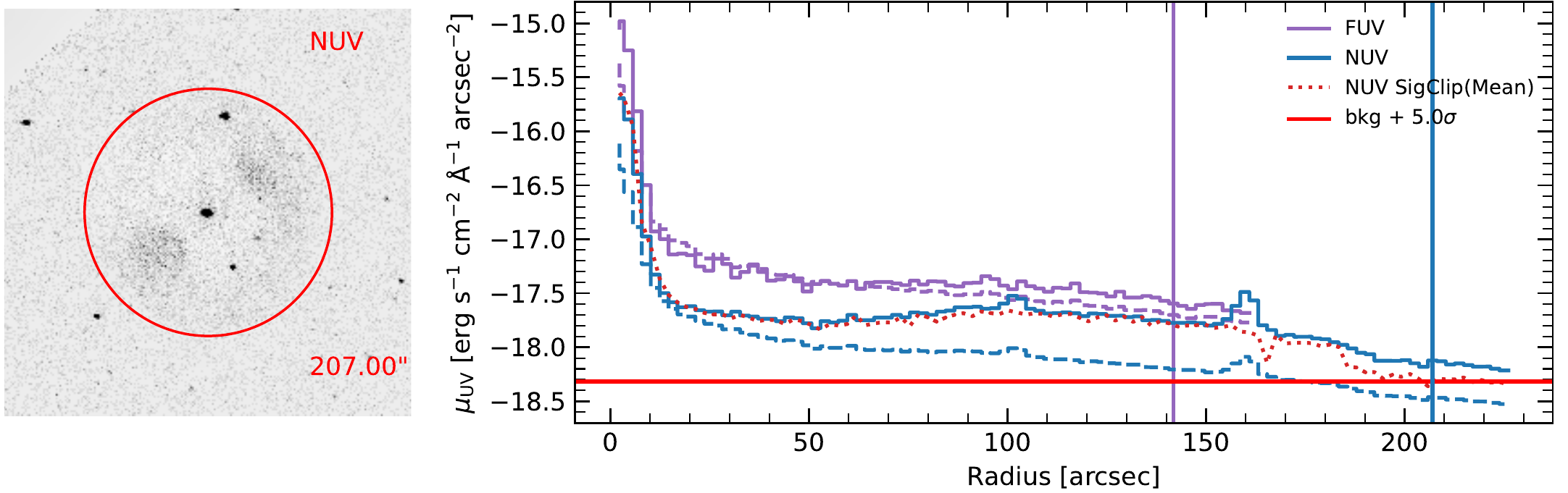}
\caption{Examples of extended PNe. GALEX NUV images of  
G\,231.8+04.1 (top-left panel) and G\,164.8+31.1 (bottom-left panel). 
The right-hand panels show the profile of the NUV flux and 
its error (blue stair solid- and dashed lines, respectively), 
the sigma clipped NUV surface flux (red dotted line), the 
background estimation (red solid line), and the calculated 
NUV radius of each PN (vertical blue line; vertical purple line for FUV radius).
Images have different size to optimally display the PN; the
 extent is labeled in the figure.}
\label{im:cog_example}
\end{figure}

The size of each extended PN in GALEX imaging was estimated by implementing 
a flux profile analysis similar to that shown in the bottom 
panel of Figure~\ref{fig:galex_optical_resolved}.  The 
analysis was carried out using an ad hoc python script. 
We downloaded the GALEX images using the \textit{astroquery} 
 package \citep{astroquery} and analyzed them with the 
\textit{astropy}  \citep{astropy:2013,astropy:2018} and the
\textit{photutils} \citep{larry_bradley_2021_4624996} packages.
For each downloaded GALEX image we estimated a global 
background mean flux, $bkg_{\rm mean}$, 
using a sigma-clipping method (to avoid the flux from 
field stars).
We then  estimated the PN flux profile,  centered on its 
coordinates, by calculating the sigma-clip mean flux, 
$\mu_{\rm mean}$,  in a number, $n$, of concentric annuli 
of 1{\arcsec} width.
The number of apertures, $n$, was calculated in such a 
way that $bkg_{\rm mean} + 5\,\sigma \geq \mu_{\rm mean}$ 
detected the border of the extended emission; we tested 
various scaling coefficients for the $bkg_{\rm mean}$ 
flux to find the best criteria to estimate the sizes of the PNe. 
Figure~\ref{im:cog_example} shows images of G\,231.8+04.1 
(top panel) and G\,164.8+31.1 (bottom panel) in the GALEX 
NUV band (left panels) as examples.
The figure also shows the flux profiles, $\mu$, calculated 
by integrating over the aperture annuli and dividing by 
their area, and the estimated error (solid and dashed lines
respectively), the $\mu_{\rm mean}$ (red dotted line),  
the background estimate (red solid line), and the measured 
PN radius (vertical solid line).
The $\mu_{\rm mean}$ profile is used to estimate the PN 
radius (e.g.,  compare $\mu$ and $\mu_{\rm mean}$ in 
right-panels of Figure~\ref{im:cog_example}).
A total of 24, 48, and 98 PNe have a NUV radius larger than 
50{\arcsec}, between 50{\arcsec} and 20{\arcsec}, and smaller 
than 20{\arcsec} respectively, when using this method; the 
sizes are added to our catalog as \texttt{FUV\_radius} and 
\texttt{NUV\_radius} for FUV and NUV respectively 
(Table~\ref{tab:gpncat_unique}).

\begin{figure}
\centering
\includegraphics[width=1.\columnwidth]{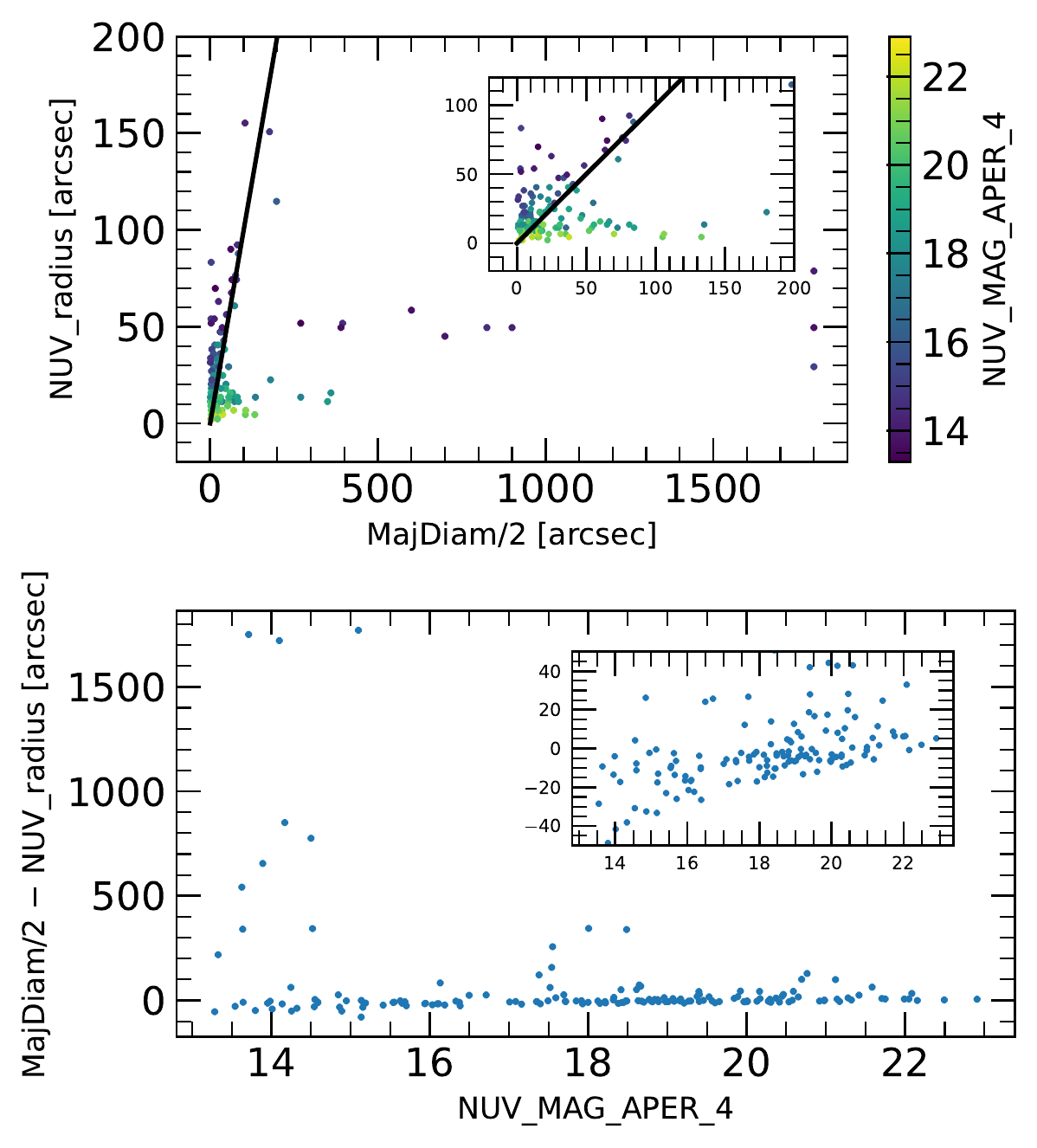}
\caption{Comparison of PNe sizes estimated in optical 
images, as extracted from PNcat, and those calculated
with the GALEX NUV radial profile analysis (top panel; 
the solid line is a 1:1 relation for comparison). The bottom
 panel shows the difference between NUV
radius and the \texttt{MajDiam}/2 from PNcat. Inset figures 
enhance a reduced range of data.}
\label{im:sizes_comparisons}
\end{figure}

Figure~\ref{im:sizes_comparisons} shows a comparison between 
the estimated PN radius in NUV and that obtained from optical 
images, as extracted from PNcat, for the 170 extended
 PNe detected in GALEX images. 
 There is generally a good correlation between the sizes 
that we estimated from GALEX and those from the literature, 
as seen in the top-panel of Figure~\ref{im:sizes_comparisons} 
for magnitudes between 14$<$NUV$<$20. The NUV radius estimated 
for PNe with NUV $\lesssim$ 13.85~mag, close to the saturation 
limit (or with a saturated nearby star), is usually 
overestimated because saturation also affects the shape and
 position of the source. In contrast, the NUV sizes
 estimated for PNe with NUV $\gtrsim$ 20~mag are underestimated 
because the extended emission is deemed to be in the background of
 the GALEX images. Note that different sizes are expected between 
UV and optical images owing to differences in the ionization structure
 of the PN, which depends on the effective temperature of the CSPN.
A CSPN with $T_\mathrm{eff}$=35\,000~K, for example, is expected to have a 
PN larger size in optical images than in UV images because of the 
lack of emission lines and continuum emission in the UV range 
(see bottom panel of Figure~\ref{im:opt_neb_example}).

The total number of extended PNe found in GALEX images represents 
$\sim$50\% of the GUVPNcat sample (compared with $\sim$70\% of PNe with 
diameter $>$10{\arcsec} obtained using the \texttt{MajDiam} 
in PNcat).  It is important to mention that the GUVPNcat 
catalog is a small sample, $\sim$10\%, of all known PNe.

\subsection{CSPNe photometry}
\label{subsec:cspne_photmetry}

Here we describe how we separated the emission of the CSPN 
from the nebular emission
in the GALEX FUV and NUV images and the optical SDSS and 
Pan-STARRS images. We have selected PNe with
values of \texttt{NUV\_DIFF\_45}$>0.036$ (see Table~\ref{tab:gpncat})
to examine extended objects.
We also restricted the sample to PNe with GALEX FUV and 
NUV measurement errors less than 0.1\,mag.

To extract the CSPN flux, an analysis of the images from GALEX, SDSS, and Pan-STARRS.
CSPN photometry was performed in three steps.
\begin{enumerate}
	\item We downloaded the images from GALEX,\footnote{We 
used the astroquery python package to download the GALEX science 
images. \url{https://astroquery.readthedocs.io/en/latest/}}
	SDSS,\footnote{\url{https://dr12.sdss.org/fields}}
	and PS1,\footnote{\url{https://ps1images.stsci.edu/cgi-bin/ps1cutouts}} 
for each PN field.
	PhotUtils was used to find the brightest unsaturated objects
	in each field and to perform aperture photometry using 
multiple aperture radii. In the case of
	GALEX images,  we transformed count rates to AB magnitudes 
using the equations of \citet{morrissey2007}, with
	\begin{equation}
	m_{\rm FUV}=-2.5\log(f_{\rm FUV}) + 18.82,
	\label{eq:FUV}
	\end{equation}
	\begin{equation}
	m_{\rm NUV}=-2.5\log(f_{\rm NUV}) + 20.08,
	\label{eq:NUV}
	\end{equation}
	whereas for SDSS we used the transformation equations 
described in
	\citet{stoughton2002}
	\begin{equation}
	m_{\rm SDSS}=-2.5\log(f_{\rm SDSS})+25.5,
	\label{eq:SDSS}
	\end{equation}
	and for PS1, as we used the cutout images already 
converted to standard
	linear flux scale,\footnote{Visit: 
\url{https://outerspace.stsci.edu/display/PANSTARRS/PS1+Image+Cutout+Service} 
for more
		information related to flux conversions.}, we used
	\begin{equation}
	m_{\rm PS1}=-2.5\log(f_{\rm PS1}/{\rm exptime})+25.0,
	\label{eq:PS1}
	\end{equation}
	where $f$ is the flux integrated in each aperture (see Table~\ref{tab:apertures_atchival}).
	Aperture photometry was applied only to point-like sources in the field.
    It is important to mention that the GALEX aperture 
photometry could include the flux of unresolved nearby
stars, that are resolved in  optical surveys, due to GALEX $\sim$5{\arcsec}
resolution. These cases can be avoided by selecting matches with the flag 
\texttt{distancerank} = 0.
	
	\item For each image we calculated, using multiple aperture photometry,
 the aperture correction 
employing a curve-of-growth analysis for each bright star in the 
field. 
We used the median value of
	the aperture correction derived for all stars as our 
adopted value. Errors were estimated according
	to the dispersion in each field's aperture correction.
	
	\item Aperture photometry was performed on each CSPN in 
the extended PNe sample. Coordinates from
	GUVPNcatxSDSSDR16xPS1MDS were used to find the CSPN in each 
instrument's image. We corrected the CSPN flux from the nebular
	contribution by calculating the mean nebular emission in 
an annulus around the CSPN (between \texttt{r\_in} and \texttt{r\_out}).
	The aperture for the CSPN  measurement was set to a fixed radius
	of 6~{\arcsec} to match aperture 4 in GALEX (Table~\ref{tab:apertures_atchival}),
	and the integrated count rates
	were converted to the AB magnitude system according 
to equations~\ref{eq:FUV}--\ref{eq:PS1},  and applying
	the corresponding aperture correction (AC),
	\begin{equation}
	m_{\rm AB}=m_{\rm X} - AC
	\label{eq:aper_corr_catalogs}
	\end{equation}
	where X corresponds to FUV, NUV, SDSS bands, or PS1 bands. 
Errors were estimated as
	\begin{equation}
	\sigma^2=\sigma_{AC}^2+\sigma_{m}^2.
	\label{eq:sigma_calc}
	\end{equation}
\end{enumerate}

The resulting CSPN fluxes,  for 8 and 50 PNe resolved in SDSS 
and Pan-STARRS respectively, from the GUVPNcatxSDSSDR16xPS1MDS 
catalog, are presented in Tables~\ref{tab:corrected_ap_photometry_sdss} 
and \ref{tab:corrected_ap_photometry_ps1}, along with their UV 
and optical sizes, \texttt{NUV\_radius} (from this work) and  \texttt{MajDiam} (from HASH), and 
the PN coordinates from PNcat.

\begin{deluxetable}{lrrr}
\tablecaption{Aperture radii used to measure the aperture correction
of each field. \label{tab:apertures_atchival}}
\tablehead{
\colhead{No.}	&	\multicolumn{3}{c}{Radius ({\arcsec})} 	\\
\colhead{(Aperture)} & \colhead{GALEX}	&	\colhead{SDSS}	&	\colhead{PS1}		
}
\startdata
1	&	1.5	&	0.4 &	0.3\\
2	&	2.3	&	0.6 &	0.4\\
3	&	3.8	&	1.0 &	0.7\\
4	&	6.0	&	1.6 &	1.0\\
5	&	9.0	&	3.0 &	1.3\\
6	&	12.8	&	2.3&	1.9\\
7	&	17.3	&	3.64&	2.3\\
\enddata
\end{deluxetable}

\begin{deluxetable*}{lllrrlllllll}
\tablecaption{Measurements for the SDSS objects with 
\texttt{NUV\_DIFF\_45}$>$0.036 from the  GUVPNcatxSDSSDR16xPS1MDS 
catalog. \label{tab:corrected_ap_photometry_sdss}}
\tabletypesize{\tiny}
        \tablehead{
                \colhead{PNG} &   \colhead{RAJ2000}&    \colhead{DEJ2000}&  \colhead{Optical}& \colhead{NUV}&    \colhead{FUV} &    \colhead{NUV} & \colhead{u} & \colhead{g} & \colhead{r} & \colhead{i} & \colhead{z} \\[-0.35cm]
                \colhead{}	&	\colhead{}	&	\colhead{}	&	\colhead{radius}	&	\colhead{radius}	&	\multicolumn{7}{c}{CSPN photometry} \\
                \cline{6-12}
                \colhead{}	&	\colhead{}	&	\colhead{}	&	\colhead{(\arcsec)}	&	\colhead{(\arcsec)}	&	\multicolumn{7}{c}{(AB mag)}
         }
\startdata
003.3+66.1 & 14:16:22.0 &  +13:52:24.1 &     25.0 &      \nodata & 16.12$\pm$0.10 & 16.79$\pm$0.02 &   17.72$\pm$0.01 &   18.19$\pm$0.00 &   18.67$\pm$0.01 &   18.86$\pm$0.01 &   18.98$\pm$0.07 \\
009.6+14.8 & 17:14:04.3 & $-$12:54:37.7 &     11.4 &    33.8 & 15.30$\pm$0.30 & 15.85$\pm$0.04 &   16.09$\pm$0.04 &   15.72$\pm$0.04 &   15.83$\pm$0.04 &   16.10$\pm$0.04 &   16.06$\pm$0.05 \\
061.9+41.3 & 16:40:18.2 &  +38:42:19.9 &      0.7 &    31.5 & 14.09$\pm$0.00 & 14.52$\pm$0.03 &   14.36$\pm$0.04 &   13.80$\pm$0.04 &   13.93$\pm$0.04 &   15.04$\pm$0.04 &   15.05$\pm$0.04 \\
144.3$-$15.5 & 02:45:23.7 &  +42:33:04.9 &     10.0 &    24.8 & 18.00$\pm$0.05 & 18.67$\pm$0.05 &   19.37$\pm$0.10 &   19.87$\pm$0.08 &   20.07$\pm$0.10 &   19.92$\pm$0.12 &   19.87$\pm$0.12 \\
164.8+31.1 & 07:57:51.6 &  +53:25:17.0 &    197.0 &   207.0 & 14.67$\pm$0.06 & 15.60$\pm$0.05 &   16.48$\pm$0.04 &   17.13$\pm$0.04 &   17.45$\pm$0.05 &   17.80$\pm$0.06 &   18.19$\pm$0.07 \\
170.3+15.8 & 06:34:07.4 &  +44:46:38.1 &     10.0 &    36.0 & 15.03$\pm$0.02 & 15.63$\pm$0.06 &   16.29$\pm$0.05 &   16.65$\pm$0.05 &   17.05$\pm$0.05 &   17.35$\pm$0.06 &   17.61$\pm$0.06 \\
211.4+18.4 & 07:55:11.3 &  +09:33:09.2 &     52.5 &      \nodata & 15.61$\pm$0.09 & 16.20$\pm$0.03 &   17.06$\pm$0.01 &   17.54$\pm$0.03 &   18.12$\pm$0.01 &   18.51$\pm$0.10 &   18.85$\pm$0.06 \\
219.1+31.2 & 08:54:13.2 &  +08:53:53.0 &    485.0 &   423.0 & 13.51$\pm$0.10 & 14.16$\pm$0.04 &   14.76$\pm$0.02 &   15.19$\pm$0.05 &   15.75$\pm$0.00 &   16.09$\pm$0.01 &   16.35$\pm$0.01 \\
\enddata
\tablecomments{The Optical radius is obtained from the 
\texttt{MajDiam} tag provided by PNcat whereas the NUV 
radius is the one calculated in section~\ref{subsec:extended_pne_galex}. 
Other useful tags for different measurements are included in 
the electronic form of this table (see Table~\ref{tab:gpncat_unique}).
The reference coordinates of the PN are from PNcat; the 
measurements were centered on the GALEX matched source position 
for FUV and NUV (Table~\ref{tab:gpncat}), and on the SDSS matched 
source position (Table~\ref{tab:pncat_sdss}) for the 
\textit{u g r i z} photometric measurements.
}
\end{deluxetable*}

\begin{deluxetable*}{lllrrlllllll}
\tablecaption{Measurements for the PS1 objects with 
\texttt{NUV\_DIFF\_45}$>$0.036 from the  GUVPNcatxSDSSDR16xPS1MDS 
catalog. \label{tab:corrected_ap_photometry_ps1}}
\tabletypesize{\tiny}
\tablehead{
       \colhead{PNG} &  \colhead{RAJ2000} &   \colhead{DEJ2000} &  \colhead{Optical} & \colhead{NUV} &     \colhead{FUV}  &   \colhead{NUV} &  \colhead{g} & \colhead{r} & \colhead{i} & \colhead{z} & \colhead{y} \\[-0.35cm]
       \colhead{} & \colhead{}	&	\colhead{}	&	\colhead{radius}	&	\colhead{radius}	&	\multicolumn{7}{c}{CSPN photometry} \\
       \cline{6-12}
       \colhead{} & \colhead{}	&	\colhead{}	&	\colhead{(\arcsec)}	&	\colhead{(\arcsec)}	&	\multicolumn{7}{c}{(AB mag)}
}
\startdata
003.3+66.1 &  14:16:22.0 &  +13:52:24.1 &    25.0 &      \nodata & 16.12$\pm$0.10 & 16.79$\pm$0.02 &  17.92$\pm$0.01 &  18.44$\pm$0.01 &  18.71$\pm$0.01 &  18.79$\pm$0.02 &  18.91$\pm$0.02 \\
007.5+07.4 &  17:35:10.2 & $-$18:34:20.4 &     4.5 &    11.2 & 19.23$\pm$0.06 & 19.07$\pm$0.08 &  17.37$\pm$0.01 &    \nodata &  15.19$\pm$0.02 &  16.90$\pm$0.02 &  17.34$\pm$0.02 \\
008.4+08.8 &  17:32:05.8 & $-$17:06:51.8 &  499.5 &    20.2 & 15.33$\pm$0.03 & 15.68$\pm$0.05 &  15.08$\pm$0.01 &  15.22$\pm$0.01 &  15.46$\pm$0.02 &  15.61$\pm$0.01 &  15.62$\pm$0.01 \\
009.6+10.5 &  17:29:02.0 & $-$15:13:05.2 &    10.1 &    18.0 & 16.92$\pm$0.06 & 17.36$\pm$0.07 &  16.33$\pm$0.01 &  16.26$\pm$0.01 &  16.40$\pm$0.01 &  16.57$\pm$0.01 &  16.62$\pm$0.01 \\
009.6+14.8 &  17:14:04.3 & $-$12:54:37.7 &    11.4 &    33.7 & 15.30$\pm$0.30 & 15.85$\pm$0.04 &  16.25$\pm$0.01 &  16.24$\pm$0.01 &  16.26$\pm$0.01 &  16.38$\pm$0.01 &  16.59$\pm$0.02 \\
012.0$-$11.9 &  18:57:46.4 &  $-$23:49:39.5 &     2.5 &    54.0 & 14.54$\pm$0.05 & 15.01$\pm$0.07 &  15.24$\pm$0.01 &  15.62$\pm$0.01 &  15.94$\pm$0.01 &  16.18$\pm$0.01 &  16.33$\pm$0.01 \\
012.5$-$09.8 &  18:50:26.0 & $-$22:34:22.6 &     2.4 &      \nodata & 17.15$\pm$0.05 & 17.15$\pm$0.05 &  14.12$\pm$0.02 &  14.89$\pm$0.01 &  15.83$\pm$0.01 &  16.02$\pm$0.01 &  15.48$\pm$0.01 \\
014.7$-$11.8 &  19:02:17.6 &  $-$21:26:51.3 &    24.0 &    27.0 & 17.37$\pm$0.05 & 18.02$\pm$0.04 &  18.47$\pm$0.01 &  18.36$\pm$0.01 &  18.32$\pm$0.01 &  18.28$\pm$0.01 &  18.23$\pm$0.01 \\
017.3$-$21.9 &  19:46:34.2 &  $-$23:08:13.1 &    76.0 &    76.5 & 14.55$\pm$0.00 & 15.13$\pm$0.00 &  15.18$\pm$0.05 &  15.14$\pm$0.04 &  15.17$\pm$0.04 &  15.96$\pm$0.03 &  15.47$\pm$0.04 \\
019.4$-$13.6 &  19:17:04.1 &  $-$18:01:35.8 &    16.5 &    22.5 & 18.63$\pm$0.03 & 18.66$\pm$0.06 &  21.09$\pm$0.04 &  20.56$\pm$0.03 &  21.38$\pm$0.03 &  21.15$\pm$0.05 &  23.56$\pm$0.05 \\
025.0$-$11.6 &  19:19:17.8 & $-$12:14:36.0 &    47.0 &    20.2 & 16.89$\pm$0.10 & 17.63$\pm$0.01 &  18.39$\pm$0.01 &  18.78$\pm$0.01 &  19.17$\pm$0.01 &  19.28$\pm$0.02 &  19.79$\pm$0.03 \\
025.4$-$16.4 &  19:37:43.8 & $-$13:51:20.0 &    20.0 &    22.5 & 18.23$\pm$0.01 & 18.71$\pm$0.06 &  19.83$\pm$0.03 &  19.94$\pm$0.03 &  20.09$\pm$0.03 &  20.07$\pm$0.03 &  19.91$\pm$0.04 \\
025.9$-$10.9 &  19:18:19.5 & $-$11:06:15.4 &     3.1 &    13.5 & 18.61$\pm$0.06 & 18.45$\pm$0.06 &  15.09$\pm$0.03 &  15.38$\pm$0.03 &  16.83$\pm$0.04 &  17.15$\pm$0.02 &  16.48$\pm$0.04 \\
028.0+10.2 &  18:06:00.8 &  +00:22:38.6 &    25.3 &      \nodata & 16.94$\pm$0.10 & 17.54$\pm$0.14 &  16.99$\pm$0.10 &  17.33$\pm$0.07 &  17.53$\pm$0.07 &  17.69$\pm$0.06 &  17.74$\pm$0.07 \\
033.0$-$05.3 &  19:10:25.8 & $-$02:20:23.5 &    28.4 &      \nodata & 19.76$\pm$0.06 & 20.26$\pm$0.06 &  19.37$\pm$0.02 &  19.44$\pm$0.08 &  19.57$\pm$0.04 &  19.55$\pm$0.07 &  19.72$\pm$0.09 \\
034.1$-$10.5 &  19:31:07.2 & $-$03:42:31.5 &    43.0 &      \nodata & 16.41$\pm$0.01 & 17.26$\pm$0.00 &  16.31$\pm$0.02 &  16.35$\pm$0.02 &  16.49$\pm$0.03 &  16.42$\pm$0.03 &  16.58$\pm$0.04 \\
038.1$-$25.4 &  20:31:33.2 &  $-$07:05:18.1 &    22.6 &    31.5 & 17.86$\pm$0.05 & 18.83$\pm$0.05 &  18.14$\pm$0.01 &  17.59$\pm$0.01 &  17.35$\pm$0.01 &  17.24$\pm$0.01 &  17.15$\pm$0.01 \\
042.5$-$14.5 &  20:00:39.2 &   +01:43:40.9 &    14.0 &    40.5 & 16.02$\pm$0.05 & 16.87$\pm$0.05 &  17.70$\pm$0.01 &  18.19$\pm$0.02 &  17.51$\pm$0.01 &  17.03$\pm$0.01 &  16.82$\pm$0.01 \\
043.5$-$13.4 &  19:58:27.1 &  +03:02:59.2 &    37.0 &    40.5 & 18.01$\pm$0.08 & 18.64$\pm$0.07 &  19.37$\pm$0.03 &  19.72$\pm$0.02 &  20.09$\pm$0.02 &  20.29$\pm$0.04 &  21.25$\pm$0.24 \\
045.0$-$12.4 &  19:57:59.3 &  +04:47:31.0 &    46.0 &    18.0 & 18.70$\pm$0.07 & 19.36$\pm$0.07 &  21.21$\pm$0.02 &  20.30$\pm$0.01 &  24.90$\pm$0.03 &    \nodata &    \nodata \\
047.0+42.4 &  16:27:33.7 &  +27:54:33.4 &    81.0 &    92.2 & 13.88$\pm$0.01 & 14.66$\pm$0.02 &  15.19$\pm$0.03 &  15.62$\pm$0.01 &  16.04$\pm$0.01 &  16.19$\pm$0.01 &  16.35$\pm$0.01 \\
050.4+05.2 &  19:04:32.3 &   +17:57:07.7 &    18.5 &    20.2 & 17.43$\pm$0.10 & 18.11$\pm$0.07 &  17.78$\pm$0.01 &  17.87$\pm$0.01 &  18.04$\pm$0.01 &  18.14$\pm$0.01 &  18.29$\pm$0.02 \\
051.5+06.1 &  19:03:37.4 &  +19:21:22.6 &    28.0 &    11.2 & 19.04$\pm$0.03 & 19.71$\pm$0.05 &  17.78$\pm$0.02 &  17.46$\pm$0.02 &  17.16$\pm$0.04 &  16.91$\pm$0.04 &  16.88$\pm$0.03 \\
055.4+16.0 &  18:31:18.3 &   +26:56:12.9 &    48.5 &    56.2 & 13.84$\pm$0.04 & 14.47$\pm$0.04 &  14.92$\pm$0.01 &  15.35$\pm$0.01 &  15.51$\pm$0.01 &  15.61$\pm$0.01 &  15.62$\pm$0.01 \\
059.7$-$18.7 &  20:50:02.1 &  +13:33:29.6 &    77.0 &      \nodata & 14.08$\pm$0.07 & 14.87$\pm$0.04 &  15.63$\pm$0.02 &  16.15$\pm$0.02 &  16.42$\pm$0.02 &  16.81$\pm$0.03 &  16.92$\pm$0.02 \\
061.9+41.3 &  16:40:18.2 &  +38:42:19.9 &     0.7 &    31.5 & 14.09$\pm$0.00 & 14.52$\pm$0.03 &  13.83$\pm$0.01 &  13.97$\pm$0.00 &  15.02$\pm$0.01 &  15.17$\pm$0.01 &  14.75$\pm$0.00 \\
066.5$-$14.8 &  20:53:03.9 &   +21:00:10.9 &    72.5 &    11.2 & 17.00$\pm$0.05 & 17.57$\pm$0.08 &  18.38$\pm$0.01 &  18.98$\pm$0.01 &  19.53$\pm$0.01 &  19.63$\pm$0.01 &    \nodata \\
076.8$-$08.1 &  20:58:10.9 &   +33:08:33.1 &    66.5 &    15.7 & 17.91$\pm$0.06 & 18.29$\pm$0.07 &    \nodata &  18.10$\pm$0.01 &  18.49$\pm$0.01 &    \nodata &  17.52$\pm$0.01 \\
077.6+14.7 &  19:19:10.2 &  +46:14:52.0 &   101.5 &      \nodata & 15.27$\pm$0.04 & 15.83$\pm$0.02 &  17.10$\pm$0.01 &  17.56$\pm$0.00 &  17.95$\pm$0.01 &  18.25$\pm$0.01 &  18.45$\pm$0.01 \\
078.5+18.7 &  18:59:19.8 &  +48:27:55.5 &    16.5 &    22.5 & 17.39$\pm$0.03 & 17.78$\pm$0.06 &  19.39$\pm$0.01 &  19.78$\pm$0.02 &  20.17$\pm$0.01 &  20.39$\pm$0.02 &  20.74$\pm$0.03 \\
084.0+09.5 &  20:04:00.1 &  +49:19:06.6 &     8.0 &    11.2 & 19.22$\pm$0.11 & 19.19$\pm$0.09 &  20.25$\pm$0.03 &  21.84$\pm$0.04 &  21.64$\pm$0.03 &    \nodata &  21.15$\pm$0.04 \\
104.1+07.9 &  21:46:08.6 &  +63:47:29.5 &    43.0 &    38.2 &   \nodata & 19.16$\pm$0.07 &  18.47$\pm$0.02 &  18.23$\pm$0.01 &  18.37$\pm$0.02 &  18.19$\pm$0.02 &  18.01$\pm$0.04 \\
104.2$-$29.6 &  23:35:53.3 &  +30:28:06.4 &   177.0 &   150.7 & 14.01$\pm$0.05 & 14.90$\pm$0.04 &  15.41$\pm$0.02 &  15.98$\pm$0.02 &  16.48$\pm$0.02 &  16.71$\pm$0.01 &  16.89$\pm$0.04 \\
110.6$-$12.9 &  23:39:10.8 &  +48:12:29.1 &    16.5 &    22.5 & 19.07$\pm$0.06 & 20.07$\pm$0.05 &  20.17$\pm$0.03 &  20.46$\pm$0.04 &  20.61$\pm$0.04 &  20.71$\pm$0.06 &  20.65$\pm$0.10 \\
117.5+18.9 &  22:42:24.9 &  +80:26:31.3 &    17.0 &    33.7 & 17.05$\pm$0.12 & 17.79$\pm$0.04 &  18.87$\pm$0.01 &  19.19$\pm$0.01 &  19.47$\pm$0.01 &  19.81$\pm$0.02 &  20.20$\pm$0.04 \\
122.1$-$04.9 &  00:45:34.7 &  +57:57:34.9 &    18.2 &     9.0 &   \nodata & 20.18$\pm$0.08 &  19.30$\pm$0.03 &  19.35$\pm$0.04 &  19.75$\pm$0.03 &  19.83$\pm$0.04 &  19.48$\pm$0.05 \\
138.8+02.8 &  03:10:19.3 &  +61:19:00.9 &    23.0 &    24.7 &   \nodata & 18.96$\pm$0.10 &  15.83$\pm$0.02 &  15.94$\pm$0.03 &  15.88$\pm$0.04 &  15.99$\pm$0.03 &  15.82$\pm$0.04 \\
144.3$-$15.5 &  02:45:23.7 &  +42:33:04.9 &    10.0 &    24.7 & 18.00$\pm$0.05 & 18.67$\pm$0.05 &  19.91$\pm$0.01 &  19.98$\pm$0.01 &  19.95$\pm$0.01 &  19.96$\pm$0.02 &  19.95$\pm$0.03 \\
153.7+22.8 &  06:43:55.5 &  +61:47:24.7 &    74.0 &      \nodata & 17.23$\pm$0.00 & 17.94$\pm$0.04 &  18.06$\pm$0.02 &  18.50$\pm$0.02 &  18.81$\pm$0.02 &  18.97$\pm$0.03 &  18.92$\pm$0.04 \\
164.8+31.1 &  07:57:51.6 &  +53:25:17.0&   197.0 &   207.0 & 14.67$\pm$0.06 & 15.60$\pm$0.05 &  16.85$\pm$0.01 &  17.35$\pm$0.01 &  17.77$\pm$0.02 &  18.06$\pm$0.03 &  18.30$\pm$0.03 \\
167.4$-$09.1 &  04:36:37.2 &  +33:39:29.9 &     1.0 &    13.5 & 17.94$\pm$0.06 & 18.26$\pm$0.07 &  14.94$\pm$0.00 &  14.07$\pm$0.01 &  14.56$\pm$0.01 &  14.47$\pm$0.01 &  14.17$\pm$0.01 \\
170.3+15.8 &  06:34:07.4 &  +44:46:38.1 &    10.0 &    36.0 & 15.03$\pm$0.02 & 15.63$\pm$0.06 &  16.75$\pm$0.02 &  17.05$\pm$0.01 &  17.37$\pm$0.01 &  17.61$\pm$0.02 &  17.79$\pm$0.01 \\
171.3$-$25.8 &  03:53:36.4 &  +19:29:38.9 &    27.0 &    24.7 & 18.05$\pm$0.02 & 18.55$\pm$0.04 &  18.01$\pm$0.01 &  18.07$\pm$0.01 &  18.31$\pm$0.02 &  18.27$\pm$0.01 &  18.35$\pm$0.01 \\
191.0+18.3 &  07:22:03.2 &  +27:13:33.5 &    32.0 &    18.0 & 17.71$\pm$0.01 & 18.36$\pm$0.06 &  17.14$\pm$0.01 &  15.32$\pm$0.01 &  15.31$\pm$0.02 &  15.52$\pm$0.01 &  15.65$\pm$0.02 \\
205.1+14.2 &  07:29:02.7 &  +13:14:48.8 &   375.0 &      \nodata & 14.15$\pm$0.03 & 14.74$\pm$0.08 &  15.39$\pm$0.02 &  15.90$\pm$0.02 &  16.33$\pm$0.02 &  16.70$\pm$0.01 &  16.79$\pm$0.02 \\
211.4+18.4 &  07:55:11.3 &  +09:33:09.2 &    52.5 &      \nodata & 15.61$\pm$0.09 & 16.20$\pm$0.03 &  17.29$\pm$0.04 &  17.93$\pm$0.09 &  18.32$\pm$0.05 &  18.71$\pm$0.05 &  18.95$\pm$0.06 \\
217.1+14.7 &  07:51:37.6 &  +03:00:20.2 &   198.0 &   114.7 & 15.12$\pm$0.04 & 15.79$\pm$0.06 &  16.93$\pm$0.02 &  17.41$\pm$0.02 &  17.81$\pm$0.03 &  18.10$\pm$0.03 &  18.31$\pm$0.04 \\
219.1+31.2 &  08:54:13.2 &  +08:53:53.0 &   485.0 &   423.0 & 13.51$\pm$0.10 & 14.16$\pm$0.04 &  14.98$\pm$0.01 &  15.51$\pm$0.03 &  15.80$\pm$0.03 &  16.14$\pm$0.02 &  16.15$\pm$0.02 \\
253.5+10.7 &  08:57:46.0 & $-$28:57:35.9 &    55.0 &    29.2 & 16.57$\pm$0.01 & 16.83$\pm$0.04 &  17.20$\pm$0.01 &  17.42$\pm$0.01 &  16.70$\pm$0.01 &  17.49$\pm$0.01 &  16.80$\pm$0.01 \\
270.1+24.8 &  10:34:30.6 & $-$29:11:15.2 &    27.0 &    29.2 & 14.39$\pm$0.01 & 15.02$\pm$0.04 &  16.32$\pm$0.02 &  16.81$\pm$0.02 &  17.16$\pm$0.01 &  17.62$\pm$0.01 &    \nodata \\
\enddata
\tablecomments{The Optical radius is obtained from the 
\texttt{MajDiam} tag provided by PNcat whereas the NUV 
radius is the one calculated in section~\ref{subsec:extended_pne_galex}. 
Other useful tags for different measurements are included 
in the electronic form of this table (see Table~\ref{tab:gpncat_unique}).
The reference coordinates of the PN are from PNcat; the 
measurements were centered on the GALEX matched source 
position for FUV and NUV (Table~\ref{tab:gpncat}), and on 
the Pan-STARRS matched source position (Table~\ref{tab:pncat_ps1}) 
for the \textit{g r i z y} photometric measurements.
}
\end{deluxetable*}

\subsection{New Observational Data for Southern GALEX PNe}
\label{sec:observational_data}

Since SDSS and PS1 cover only the northern hemisphere, 
leaving out 392 southern GALEX sources, we obtained observations
of 89 PNe in the southern hemisphere using the telescopes 
from Las Cumbres Observatory Global Telescope Network \citep[LCOGT;][]{brown2013}.
We selected PNe with \texttt{NUV\_DIFF\_45}$>0.036$ in
 our catalog. Out of the 89 PNe observed with LCOGT, measurements
of only 13 CSPNe have been obtained; of the 76 remaining
 PNe observed, the data for 20 of them were only background noise, 
25 resulted in S/N$<$5, and 
for 31 the CSPN was not resolved.

The CCD camera SBIG mounted on the 0.4\,m telescopes 
of the LCOGT network
was used to obtain images in the SLOAN \textit{g, r}, and 
\textit{i} bands. The STL-6303 detector was used. The detector
has 1534$\times$1024 pixels with a pixel scale of 
1{\farcs}142\,pix$^{-1}$ (with 2$\times$2 binning).
The resulting field of view is 15$\times$10{\arcmin}.
Data were reduced by using a dedicated LCOGT
reduction pipeline called 
BANZAI.\footnote{\url{https://github.com/LCOGT/banzai}}  Astrometry
was carried out using the 
\texttt{Astrometry.net}\footnote{\url{nova.astrometry.net}} web tool.
Raw and reduced images can be accessed through the 
LCO Data Archive\footnote{LCO Data Archive at: \url{https://archive.lco.global}}
by searching for proposal IDs IAC2017AB$-$004 and IAC2018A$-$007.

Fluxes were extracted using the python PhotUtils package.
Aperture photometry 
was performed on each field using the photometric scale of the 
American Association of Variable Star Observers 
Photometric All-Sky Survey \citep[APASS;][]{henden2015}, which provides
AB magnitudes. This procedure, using local standard stars (field stars), 
directly calibrates the SLOAN $g$, $r$, and $i$ bands
common to both LCOGT and APASS.

We performed differential aperture photometry
to calibrate our images to the APASS photometric scale.
The aperture magnitude, $m$, of a source is related to 
the measured instrumental magnitude according to
\begin{equation}
m = m_{\rm inst} + ZP + kX
\end{equation}
where $ZP$ is the instrumental zero point (defined as the 
magnitude of an object that produces one count
per second on the CCD), $k$ is the atmospheric extinction,
 and X is the airmass in the middle of the observation.
The $ZP$ and $kX$ coefficients are equal for all the 
stars in the frame. As a result, the difference
in magnitude between two sources, 1 and 2, is given by
\begin{equation}
\begin{aligned}
m_{1} - m_{2} &= m_{\rm inst\,1} + ZP + kX - (m_{\rm inst\,2} + ZP + kX) \\
	&= m_{\rm inst\,1} - m_{\rm inst\,2}
\end{aligned}
\end{equation}
and finally,
\begin{equation}
\label{eq:diff_phot_zp}
m_{1} = m_{\rm inst\,1} + (m_{2} - m_{\rm inst\,2}) = m_{\rm inst\,1} + zp
\end{equation}
with $zp= (m_{2} - m_{\rm inst\,2})$.

Aperture photometry, with an aperture radius of 
3{\farcs}5, was performed on each observed field 
star to obtain $zp$. 
We then performed aperture photometry on the CSPN using the same aperture.
A local annulus (of size twice the aperture used) 
was employed to subtract the
nebular emission from the CSPN. We then used 
equation~\ref{eq:diff_phot_zp} to calibrate
the CSPN magnitude in the AB system. The results are 
shown in Table~\ref{tab:results_obs_photometry}.

\begin{deluxetable*}{lllrrrrrrr}
\tablecaption{Measurements of the GUVPNcat sample observed at LCOGT. \label{tab:results_obs_photometry}}
\tabletypesize{\scriptsize}
\tablehead{
\colhead{PNG} & \colhead{RAJ2000} & \colhead{DECJ2000}	&	\colhead{Optical}	&	\colhead{NUV}	& \colhead{FUV}& \colhead{NUV} &    \colhead{g} &  \colhead{r} &  \colhead{i} \\[-0.3cm]
\colhead{}	&	\colhead{} & \colhead{} & \colhead{radius}& \colhead{radius}& \multicolumn{5}{c}{CSPN photometry} \\
\cline{6-10}
\colhead{}	&	\colhead{} & \colhead{} & \colhead{(\arcsec)}& \colhead{(\arcsec)}& \multicolumn{5}{c}{(AB magnitude)}
}
\startdata
243.8$-$37.1 &   05:03:02 & $-$39:45:44	&21	&	49	&   13.42$\pm$0.01 & 14.19$\pm$0.01 & 15.43$\pm$0.04 &  15.77$\pm$0.04 &  16.18$\pm$0.05 \\
274.3+09.1 &   10:05:46 &$-$44:21:33	&42	&	45	&	14.76$\pm$0.01 & \nodata & 16.62$\pm$0.06 &  17.09$\pm$0.06 &  17.36$\pm$0.07 \\
283.6+25.3 &   11:26:44 &$-$34:22:11	&200	&	99	&   14.59$\pm$0.01 & 15.29$\pm$0.05 & 16.01$\pm$0.09 &  15.82$\pm$0.09 &  15.64$\pm$0.17 \\
286.8$-$29.5 &   05:57:02& $-$75:40:23	&61	&	50	&   14.03$\pm$0.08 &  14.74$\pm$0.04 & 15.95$\pm$0.04 &  16.34$\pm$0.04 &  16.64$\pm$0.04 \\
291.4+19.2 &   11:52:29 &$-$42:17:39	&30	&	35	&    15.20$\pm$0.13 & 15.13$\pm$0.05 & 16.66$\pm$0.04 &  16.80$\pm$0.04 &  17.04$\pm$0.07 \\
308.2+07.7 &   13:28:05 &$-$54:41:58	&	19&	11	&  \nodata  & 17.54$\pm$0.04 & 17.20$\pm$0.07 &  17.34$\pm$0.07 &  17.50$\pm$0.07 \\
309.1$-$04.3 &   13:53:57& $-$66:30:51	&	10.7&	45	&  \nodata  & 14.20$\pm$0.01 &  9.98$\pm$0.07 &   9.69$\pm$0.07 &  10.81$\pm$0.13 \\
316.1+08.4 &   14:18:09 &$-$52:10:40	& 14	&	47	&  \nodata  & 14.09$\pm$0.01 & 12.44$\pm$0.08 &  11.98$\pm$0.08 &  12.42$\pm$0.07 \\
326.0$-$06.5 &   16:15:42& $-$59:54:01	&	1.8&	43	&  \nodata  & 14.38$\pm$0.01 & 12.89$\pm$0.08 &  12.33$\pm$0.08 &  12.77$\pm$0.05 \\
329.0+01.9 &   15:51:41 &$-$51:31:28	&72	&	16	&  \nodata  & 15.77$\pm$0.02 & 14.13$\pm$0.04 &  13.76$\pm$0.04 &  13.56$\pm$0.05 \\
331.3+16.8 &   15:12:51 &$-$38:07:34	&	7&	38	&    13.83$\pm$0.01 & 14.10$\pm$0.01 & 11.35$\pm$0.16 &  12.30$\pm$0.16 &  13.49$\pm$0.10 \\
349.3$-$01.1 &   17:22:16 &$-$38:29:03	&	48&	32	&  \nodata  & 16.79$\pm$0.04 & 15.57$\pm$0.13 &  15.24$\pm$0.13 &  15.17$\pm$0.14 \\
358.9$-$00.7 &   17:45:58 & $-$30:12:01	&	9&	\nodata	&  \nodata  & 16.58$\pm$0.01 & 12.32$\pm$0.09 &  10.58$\pm$0.09 &  10.86$\pm$0.16 \\
\enddata
\tablecomments{
The Optical radius is obtained from the \texttt{MajDiam} tag 
provided by PNcat whereas the NUV radius is calculated 
in section~\ref{subsec:extended_pne_galex}.
}
\end{deluxetable*}

\section{Discussion. Color--color diagram analysis of PNe and CSPNe.}
\label{subsec:pne_cc_diagram}

\begin{figure*}
\centering
\includegraphics[width=0.49\textwidth]{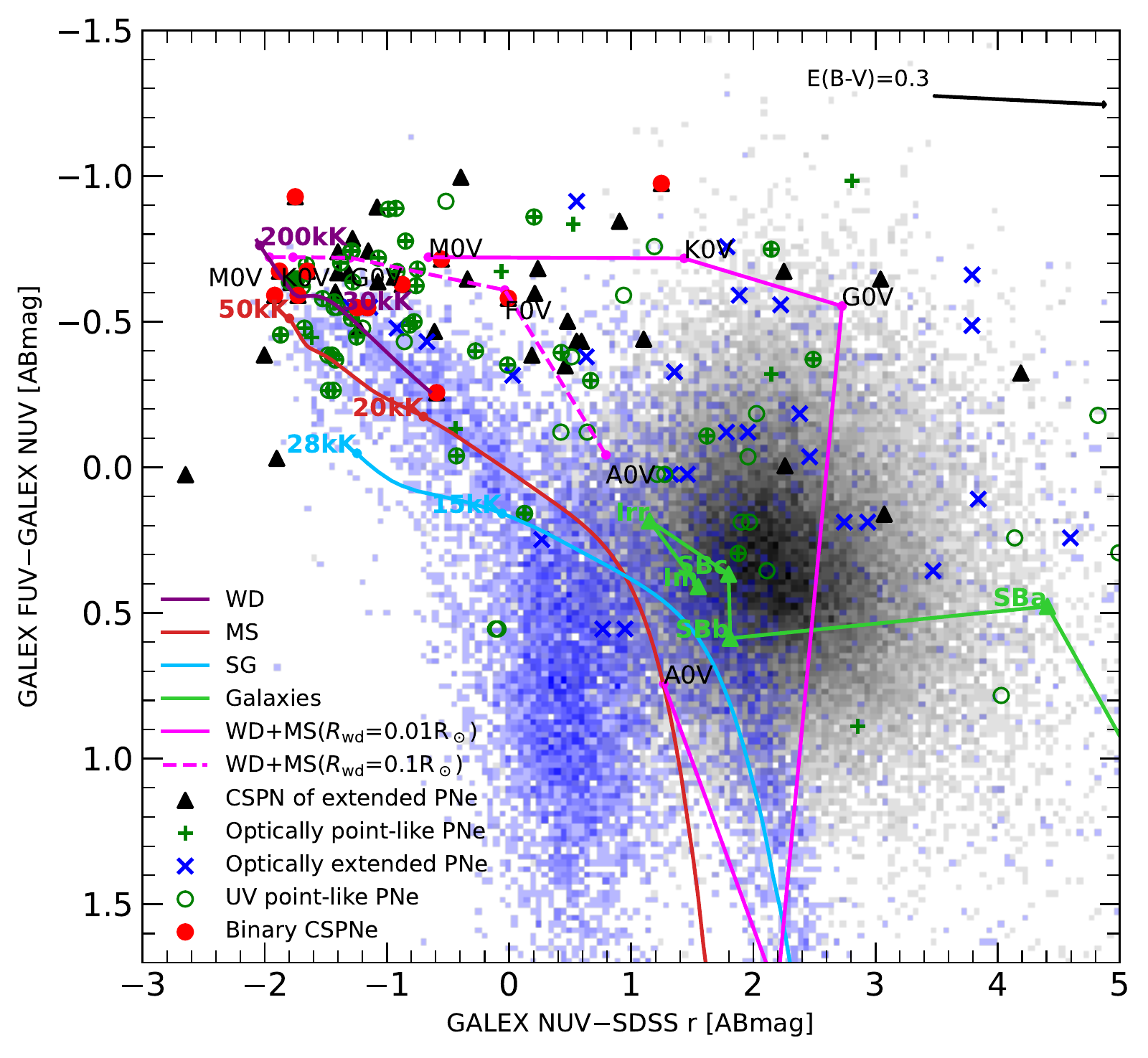}
\includegraphics[width=0.49\textwidth]{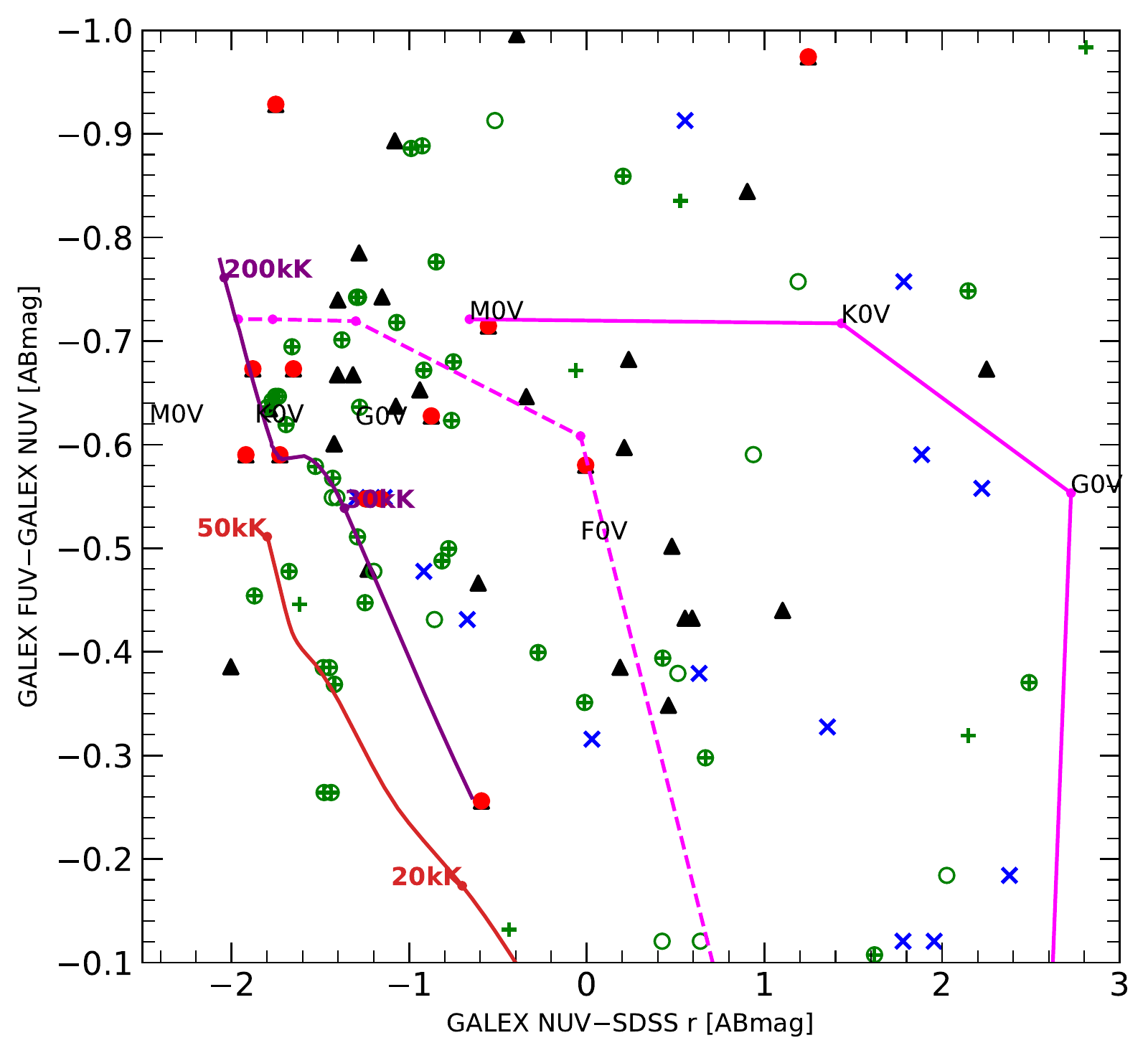}
\includegraphics[width=0.49\textwidth]{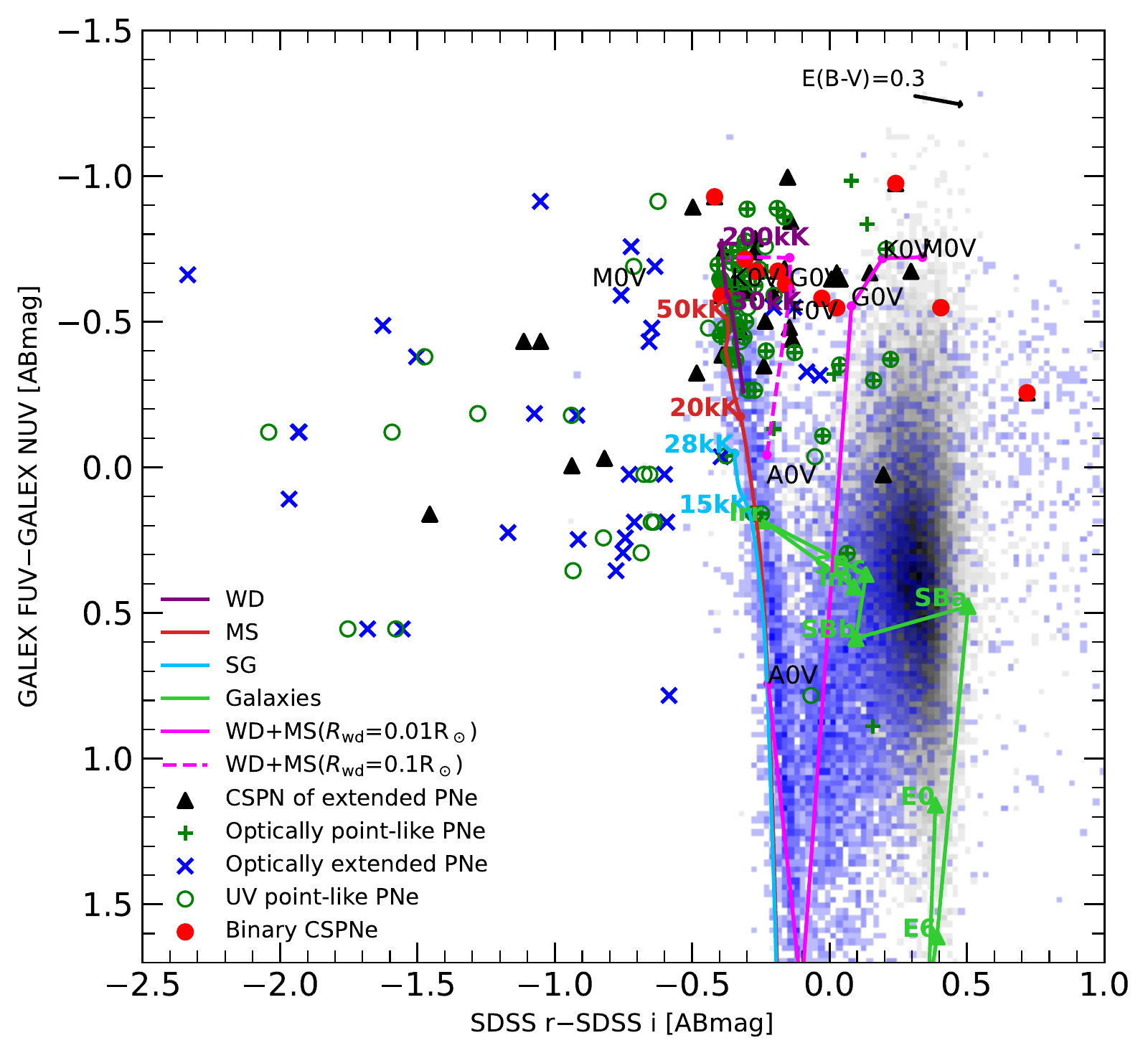}
\includegraphics[width=0.49\textwidth]{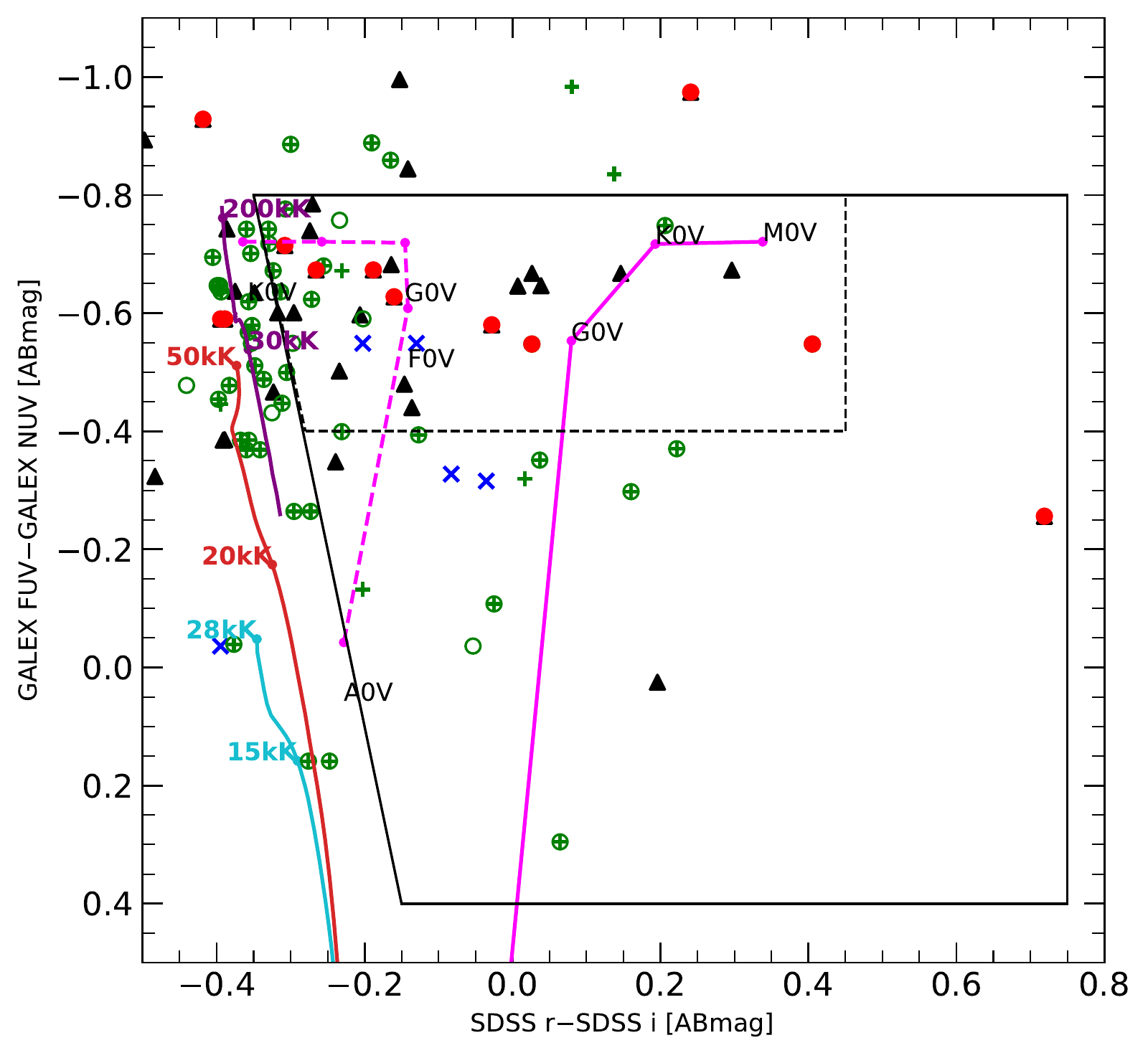}
\caption{Left-panels: Color-color diagrams for optically 
extended PNe (blue crosses), optically point-like PNe (green
 pluses), UV point-like PNe (green circles), CSPN of extended
 PNe (black up-triangles), and known binary CSPNe (red dots; 
see text). The CSPNe photometry in Tables~\ref{tab:corrected_ap_photometry_sdss} 
and \ref{tab:corrected_ap_photometry_ps1} is used for the CSPN of extended PNe. 
Purple/black density maps are point-like/extended 
sources, respectively, from the matched
GALEXxSDSS sample from \citet{bianchi2020arxiv}; from their work 
we also took the model colors shown in the plots. The purple 
sequence of stellar model colors (labeled  ``WD'') is
for $\log g = 7.0$. Main-sequence (red) and supergiant (cyan) 
sequences are for stellar model colors with
solar $\log g = 5.0$ and $\log g = 3.0$, respectively, and binaries 
composed of a WD+MS are for WD stars of $R_\mathrm{WD}=0.01$~R$_\mathrm{\sun}$ 
(pink solid line) and $R_\mathrm{WD}=0.1$~R$_\mathrm{\sun}$ (pink dashed line) for different MS spectral types. 
Galaxy templates of representative types (E0--Irr,Im; green 
solid line) are also shown.
A reddening vector for $E(B-V)=0.3$~mag is shown in the upper right
corner on each panel. Right panels: same as the left-panels 
but with a reduced range to enhance the binary CSPNe region 
(black- and dashed-line regions).}
\label{im:cc_fuvnuv_nuvr}
\end{figure*}

The observed UV--optical colors of the PNe in 
GUVPNcatxSDSSDR16xPS1MDS are shown in color--color diagrams  
in Figure~\ref{im:cc_fuvnuv_nuvr}.
Model colors of stellar objects, taken from 
\citet{bianchi2007,bianchi2011a}, are also shown to guide the 
eye in interpreting the distribution of PNe. 
We refer to \citet{bianchi2007,bianchi2011a,bianchi2020arxiv} for
other similar figures and a description of the model grids.
A separation  was made between extended and point-like PNe using 
\texttt{r\_sdss\_diff}$>$0.145 and \texttt{r\_ps1\_diff}$>$0.05
for SDSS and Pan-STARRS respectively.
For point-like sources, assuming that the majority are 
compact PNe, we used the GALEX best magnitude, the 
\texttt{psfMag} from SDSS, and \texttt{MeanPSFMag} from PS1.
These magnitudes are the integrated flux of the 
best fit  to a point-source shape from each database. 
It is important to mention that the best GALEX magnitudes, 
\texttt{fuv\_mag} and \texttt{nuv\_mag}, are in most cases equal 
to \texttt{FUV\_MAG\_APER\_4} and \texttt{NUV\_MAG\_APER\_4} 
(applying aperture correction) respectively,  for sources 
with \texttt{NUV\_DIFF\_45} $< =$ 0.036, whereas they differ 
for objects classified as extended. Best magnitudes result 
from the fit with an elliptical aperture to extended sources,  
whereas GALEX APER\_4 is a fixed circular aperture with a
6{\arcsec} radius and hence misses flux outside the aperture.
For PNe with \texttt{NUV\_DIFF\_45}$>$0.036, we used the CSPN photometry
of PNe as estimated in section~\ref{subsec:cspne_photmetry}. 
In all diagrams we used $i$- and $r$-band photometry from 
either SDSS or Pan-STARRS, as these bands 
are practically identical in both
catalogs \citep[see Figure~6 and 7 from][]{tonry2012}.  Only 
unsaturated PNe in the bands used are plotted
in Figure~\ref{im:cc_fuvnuv_nuvr}, by selecting PNe with 
\texttt{fuv\_mag}$>$13.73~mag and \texttt{nuv\_mag}$>$13.85~mag,
and \texttt{rMeanKronMag},  \texttt{petroMag\_r}, 
\texttt{iMeanKronMag}, and \texttt{petroMag\_i} $>$ 13~mag.

In Figure~\ref{im:cc_fuvnuv_nuvr} we show UV point-like PNe, 
with \texttt{NUV\_DIFF\_45} $< =$ 0.036 (green circles), optically 
extended PNe with \texttt{r\_sdss\_diff} $>$ 0.145 and 
\texttt{r\_ps1\_diff} $>$ 0.05 (blue crosses), CSPNe photometry 
for the extended PNe (black up-triangles; see 
section~\ref{subsec:cspne_photmetry}), and point-like PNe as 
selected from optical magnitudes (green pluses).
Also, point-like and extended sources, as classified by the SDSS
\texttt{class} tag, are shown as
blue and black density maps respectively for all sources 
in the \textit{GALEX Ultraviolet source Catalog} 
\citep[GUVcat;][]{bianchi2017}
matched with SDSS DR16 \citep{bianchi2020arxiv}. Model color 
sequences are shown in different colors, as explained in the
figure legend. The effect of interstellar dust is shown as
 an arrow in the upper-right corner of each panel
for typical Galactic extinction with $R_\text{V}=3.1$ and 
$E(B-V)=0.3$~mag,
using the extinction law of \citet{ccm89}.

In order to correctly interpret the loci of our types of objects, we must also examine
whether other classes of astrophysical objects overlap in color space. Therefore, we
also plot galaxy templates (green line; 
Figure~\ref{im:cc_fuvnuv_nuvr}) of E0, E4, E6, SBa, SBb, SBc, 
Irr, and Im types taken from the atlas of \citet{Brown2014}.

\subsection{Separating extended and compact PNe}

As can be seen in Figure~\ref{im:cc_fuvnuv_nuvr}, 
point-like (in optical images) PNe, and CSPN of the UV 
extended PNe are mainly located along the stellar loci, 
which is an indication that the CSPN is not contaminated 
by the nebular flux in the optical bands (bottom panel).
In contrast, the UV point-like (unresolved) PNe are dispersed 
across the UV color-color diagrams. In the bottom left-panel 
of Figure~\ref{im:cc_fuvnuv_nuvr}, the SDSS~$r-$SDSS~$i<-$0.4 
color of UV point-like PNe is strongly contaminated by 
nebular continuum and emission lines (e.g., H$\alpha$ and
 [\ion{N}{2}]; see Figure~\ref{im:opt_neb_example}), 
whereas the SDSS~$r-$SDSS~$i>=-$0.4 color of UV point-like 
PNe matches the position of the optically point-like PNe 
(i.e., the flux mainly originated from the CSPN). In the upper 
left-panel of Figure~\ref{im:cc_fuvnuv_nuvr}, optically 
extended PNe and some of the UV point-like PNe are found to 
be in the same color space as Galaxies, binary stars, and 
QSOs \citep[see][]{bianchi2007,bianchi2009,bianchi2011b}.
However, it is relatively easy to distinguish between 
extended PNe and CSPNe using these color combinations 
(see bottom-panel of Figure~\ref{im:cc_fuvnuv_nuvr}); 
this implies that it is also possible to separate compact 
PNe from extended PNe for which the nebular flux is included 
in the aperture by using the  r$-$i$<-$0.4 and $-1<$FUV$-$NUV$<$1 colors. 
The SED of the selected objects can also be analyzed using the
catalogs presented here; the single matches are 
obtained by selecting \texttt{distancerank} = 0 in UV and
 optical wavelengths. Objects with \texttt{distancerank} = 1 
can also be investigated by accounting for multiple matches 
around them.

The catalog can be used as starting point to search for 
compact PNe in GALEX in combination with optical surveys 
such as SDSS and Pan-STARRS by implementing the color cuts described above.
Similar studies of compact PNe have been done previously 
using photoionization models representing differing 
ionization/evolutionary stages of PNe and then comparing 
the synthetic SEDs with photometric surveys 
\citep[e.g.,][]{vejar2019,GutierrezSoto2020}. An analysis 
of compact and extended PNe using photoionization models 
will be provided in a forthcoming paper (G{\'o}mez-Mu{\~n}oz 
et al. 2023 in preparation).

\subsection{Identifying binary CSPNe}

An interesting feature of both color-color diagrams in 
Figure~\ref{im:cc_fuvnuv_nuvr} is that most CSPNe have 
optical colors redder than  the WD locus (purple line). 
This could indicate, as shown before by 
\citet{bianchi2007,bianchi2011a,bianchi2020arxiv}, that 
the CSPN has a cool companion \citep[e.g.,][]{douchin2015,barker2018} 
or that the CSPNe is reddened by high amounts of extinction.
Galactic extinction hardly affects FUV$-$NUV 
\citep[e.g.,][]{bianchi2011a,bianchi2011b,bianchi2017} 
but severely affects UV-optical colors, as also 
shown by the reddening arrows in the panels. 
The figure also shows the known binary CSPNe (red dots; 
taken from the compilation of Dr. David 
Jones\footnote{\url{http://www.drdjones.net/bcspn/}\label{drjonesweb}})
 along with representative binary models from 
\citet{bianchi2020arxiv} composed by a WD+main-sequence 
stars, for WDs of $T_\text{eff}=100\,000$~K with 
$R_\text{WD}=0.1$~R$_{\sun}$ (pink dashed-line) and 
$R_\text{WD}=0.01$~R$_{\sun}$ (pink solid-line) with 
less evolved companion stars (from M0V to A0V spectral types). 
Binary CSPN candidates have colors r$-$i$>-$0.4, and could be identified 
in the region (FUV$-$NUV)$\leq 6$(r$-$i)$+$1.3 and 
$-$0.8$<$(FUV$-$NUV)$<$0.4 and r$-$i$<$0.75 (black solid 
line) as shown in the bottom right-panel of 
Figure~\ref{im:cc_fuvnuv_nuvr}, provided that reddening 
is accounted for, although this color-color range does not include all types of binaries.
The position of WD + main-sequence binaries 
depends on the color combination, as seen in \citet{bianchi2020arxiv}.
About $\sim$20\% of the GUVPNcatxSDSSDR16xPS1MDS CSPNe included in
the region (FUV$-$NUV)$\leq 6$(r$-$i)$+$1.3 and $-$0.8$<$(FUV$-$NUV)$<$0.4 
and r$-$i$<$0.75 (black line contour) are known binaries (red dots),
and $\sim$33\% in the region
(FUV$-$NUV)$\leq 6$(r$-$i)$+$1.3 \& $-$0.8$<$(FUV$-$NUV)$<$$-$0.4 
\& r$-$i$<$0.45 (black dashed contour).
This was estimated by taking into account only the 
CSPN of the UV extended PNe (black triangles) and the optically 
point-like PNe (green pluses).
Note, however, that there are only $\sim$150 
binary CSPNe (from Dr. Dave Jones's compilation$^{\ref{drjonesweb}}$) 
in PNcat which represent $\sim$5\% of all known PNe, and that
binaries with main-sequence companions of types A and earlier might not be
separated by these color combinations.

The catalog can be used to extract binary CSPN candidates 
from the regions described above and to analyze the 
individual UV-optical SEDs \citep[see][who analysed the SED 
of binary hot-WD candidates using GALEX and IPHAS photometric 
data]{GomezMunoz2022}. Information on multiple measurements of the same
source can be obtained by selecting the \texttt{grankdist} = 1 
for GALEX and \texttt{nDetections}$>$5 in the case of Pan-STARRS.

\section{Summary and Conclusions}
\label{sec:summary_conclusions}

We constructed a photometric catalog of PNe in 
the footprint of the GALEX GR6/GR7 UV surveys and
in the SDSS DR16 and PS1 MDS optical databases, covering 
a spectral range from GALEX with
FUV (1344$-$1786{\AA}, $\lambda_\text{eff}=1538.6${\AA}) 
and NUV (1771$-$2831{\AA}, $\lambda_\text{eff}=2315.7${\AA}) bands,
to optical SDSS $u$ (3048$-$1028{\AA}, $\lambda_\text{eff}=3594.9${\AA}), 
$g$ (3783$-$5549{\AA}, $\lambda_\text{eff}=4640.4${\AA}),
$r$ (5415$-$6989{\AA}, $\lambda_\text{eff}=6122.3${\AA}), 
$i$ (6689$-$8389{\AA}, $\lambda_\text{eff}=7439.5${\AA}),
and $z$ (7960$-$10\,830{\AA}, $\lambda_\text{eff}=8897.1${\AA}) 
bands, and Pan-STARRS optical
$g_\text{PS1}$ (3943$-$5593{\AA}, $\lambda_\text{eff}=4775.6${\AA}),
$r_\text{PS1}$ (5386$-$7036{\AA}, $\lambda_\text{eff}=6129.5${\AA}), 
$i_\text{PS1}$ (6778$-$8304{\AA}, $\lambda_\text{eff}=7484.6${\AA}),
$z_\text{PS1}$ (8028$-$9346{\AA}, $\lambda_\text{eff}=8657.8${\AA}), 
$y_\text{PS1}$ (9100$-$10\,838{\AA}, $\lambda_\text{eff}=9603.1${\AA}) bands.
Out of the 3865 PNe in HASH, including confirmed and candidate PNe, 
61 are in both GALEX and SDSS, 388 in GALEX and PS1 MDS (with an 
overlap between SDSS and PS1 of 36 PNe), and a total of 13 PNe were observed by us
using the LCOGT facility in the  $g'$ (3968.1--5581.5{\AA}, 
$\lambda_\text{eff}=4647.9${\AA}), $r'$ (5434.7--7085.8{\AA}, 
$\lambda_\text{eff}=6150.2${\AA}), and $i'$ (6739.9--8350.9{\AA}, 
$\lambda_\text{eff}=7470.5${\AA}) bands.

With a match radius of 5{\arcsec} between PNcat (from the 
entire HASH database) and GALEX, SDSS, and Pan-STARRS we found:

\begin{enumerate}
\item \textit{GUVPNcat}. Contains 1605 GALEX matches of 
671 unique PNe in PNcat. Tags indicating to multiple 
measurements of the same source and multiple matches are 
 \texttt{grankdist} and \texttt{distancerank}, 
respectively. Out of the 1605 matches, 392 PNe have only 
one observation (\texttt{grankdist}=0), 370 PNe have 
multiple observations of the same source (\texttt{grankdist}=1), 
and 11 have a secondary observation with better exposure time 
(\texttt{grankdist} = $-$1). Of the 671 unique PNe in PNcat, 
only 79 have multiple GALEX matches within the match radius 
(\texttt{distancerank} = 1). Objects with \texttt{nuv\_artifact} = 32
 were removed from this compilation.

\item \textit{PNcatxSDSSDR16}. Contains 108 SDSS matches of 
83 PNe in PNcat. Information of multiple matches is included 
in this catalog (\texttt{distancerank} tag); 66 have only one
 SDSS counterpart (\texttt{distancerank} = 0) and 17 have more 
than one SDSS counterpart (\texttt{distancerank} = 1). Only pimary SDSS observations were 
included in this compilation.

\item \textit{PNcatxPS1MDS}. Contains 3301 Pan-STARRS matches to 
1819 PNe in PNcat. Of the 1819 PNcat objects in PS1 MDS, 927 have 
only one counterpart (\texttt{distancerank} = 0) and 892 PNcat 
objects have more than one PS1 MDS counterpart (\texttt{distancerank} = 1).

\item \textit{GUVPNcatxSDSSDR16xPS1MDS}. The comprehensive resulting 
catalog contains 375 unique PNe. The separate matched catalogs, PNcat matched with 
GALEX (GUVPNcat), SDSS (PNcatSDSSDR16), and PS1 (PNcatPS1MDS) 
contain information on multiple measurements of the 
same source (GALEX) and multiple matches within a radius of 
5{\arcsec}, useful when analyzing the SED of a 
PN when photometry of observations with \texttt{distancerank} $>$ 0 
(and \texttt{grankdst} $>$ 0 in GALEX) is used. This information 
can be retrieved by linking GUVPNcatxSDSSDR16xPS1MDS to the other
 matched catalogs. The presence of multiple matches
 is indicated in condensed form in GUVPNcatxSDSSxPS1MDS
 (see Appendix~\ref{ap:description_catalog} - Table~\ref{tab:gpncat_unique}).
 The catalog also includes distances as 
obtained from the surface brightness-ratio \citep{Frew2016} 
relation and from GaiaEDR3 \citep{gonzalessantamaria2021,Chornay2021}.
\end{enumerate}

We analyzed the different aperture magnitudes provided 
by the GALEX pipeline and found that, by comparing known 
PN diameters from previous catalogs (also included in 
GUVPNcatxSDSSDR16xPS1MDS), the difference between aperture 
magnitudes with 6{\arcsec} and 4{\arcsec} radius, 
\texttt{NUV\_DIFF\_45} $>$ 0.036$\pm$0.003, is a good 
indicator of a resolved PN in GALEX. We classify 170 PNe 
as extended in GALEX imaging, whereas 20 and 225 
are extended in SDSS and Pan-STARRS, respectively (values 
of \texttt{r\_sdss\_diff} $>$ 0.145 and \texttt{r\_ps1\_diff} $>$ 0.05, 
for SDSS and Pan-STARRS, respectively). 
We calculated the UV radius for the PNe with 
\texttt{NUV\_DIFF\_45} $>$ 0.036 by implementing a flux 
profile analysis of the extended emission.
The UV radius was determined  by measuring the clipped 
flux at a limit of 5$\sigma$ level above background. 
A total of 24, 48, and 98 PNe have an NUV radius larger than 
50{\arcsec}, between 50{\arcsec} and 20{\arcsec}, and smaller
 than 20{\arcsec}, respectively.
For 50 resolved PNe in GALEX imaging, we extracted the CSPN 
flux by subtracting the nebular emission measured in an annular aperture. 

Special attention should be given when, for a GALEX source, 
multiple SDSS or Pan-STARRS matches are found.  As the optical 
databases used here have higher spatial resolution than GALEX, 
it is possible that more than one optical counterpart to a UV 
source is found within the match radius used.
In these cases, the UV flux could be a composite of these optical 
counterparts, which are resolved in SDSS and
Pan-STARRS.
The magnitudes of the multiple optical 
counterparts must be compared to assess possible biases 
in the analysis of the spectral energy distribution 
\citep[see][]{bianchi2011b,bianchi2020arxiv}. For this 
reason, the \texttt{distancerank} tag is provided.

\textit{Compact and extended PNe}. 
We have compared the colors of PNe and CSPNe with model color 
grids of different astrophysical objects.
The comparison of UV and optical colors (Figure~\ref{im:cc_fuvnuv_nuvr}) 
shows that PNe can be identified among other astrophysical 
objects (right-panel) because of the sensitivity of the $r-i$ color 
 to the ionization of the nebula.
As discussed in Section~\ref{subsec:extended_pne_galex}, the 
$r$-band includes H$\alpha$ and [\ion{N}{2}] nebular emission 
lines which vary with the ionization of the PNe; the $i$ band 
is less affected by nebular emission.
Therefore, it is possible to find candidate compact PNe or 
extended PNe for which the nebular flux is included in the 
aperture with the CSPN flux using the colors r$-$i$<-$0.4 and
 $-1<$FUV$-$NUV$<$1. This could be a starting point in the 
search for compact versus extended PNe unresolved in GALEX matched 
with optical corollary surveys \citep[e.g.,][]{vejar2019,GutierrezSoto2020}.

\textit{Binary CSPNe}.
We also show representative loci of stellar binaries composed 
of a WD with a main-sequence companion of representative spectral 
types \citep[pink solid lines in Figure~\ref{im:cc_fuvnuv_nuvr};
 from][]{bianchi2020arxiv}.
The color combinations analyzed in this paper are also useful 
for the identification of some types of candidate binary CSPNe.
Some such binaries, however, occupy the same color space as other 
astrophysical objects, such as QSOs with enhanced Ly$\alpha$
\citep[see ][]{bianchi2009,bianchi2011b,bianchi2020arxiv}. 
Binary CSPNe in the parameter range shown in \citet{bianchi2020arxiv} 
examples have r$-$i$>-$0.4 colors and could be identified in the 
(FUV$-$NUV)$\leq 6$(r$-$i)$+$1.3 \& $-$0.8$<$(FUV$-$NUV)$<$0.4  
\& $r-i<0.75$ region, as shown in the bottom right-panel of Figure~\ref{im:cc_fuvnuv_nuvr}. 
\\

{ \small
AM acknowledges support from the ACIISI, Gobierno de Canarias
and the European Regional Development Fund (ERDF) under grant 
 PROID2020010051, as well as
from the State Research Agency (AEI) of the Spanish Ministry 
of Science and Innovation (MICINN) under grant
PID2020-115758GB-I00.
LB acknowledges partial support from NASA ADAP grant 80NSSC19K0527.

This research is based on observations made with the GALEX, 
SDSS, and Pan-STARRS, obtained from the MAST data archive at
 the Space Telescope Science Institute, which is operated by 
the Association of Universities for Research in 
Astronomy, Inc., under NASA contract NAS 5$-$26555. The 
Pan-STARRS1 Surveys (PS1) and the PS1 public science archive 
have been made possible through contributions by the Institute 
for Astronomy, the University of Hawaii, 
the Pan-STARRS Project Office, the Max-Planck Society and its 
participating institutes under Grant No. NNX08AR22G issued 
through the Planetary Science Division of the NASA Science 
Mission Directorate, the National Science 
Foundation Grant No. AST-1238877, the University of Maryland, 
Eotvos Lorand University (ELTE), the Los Alamos National 
Laboratory, and the Gordon and Betty Moore Foundation. Funding
 for the Sloan Digital Sky 
Survey IV has been provided by the 
Alfred P. Sloan Foundation, the U.S. 
Department of Energy Office of 
Science, and the Participating 
Institutions. 
This work also makes use of observations from the Las Cumbres 
Observatory global telescope network. This research has made 
use of the HASH PN database at hashpn.space.

This research made use of Astropy,\footnote{http://www.astropy.org} 
a community-developed core Python package for Astronomy \citep{astropy:2013, astropy:2018}. 
We also made use of Photutils, an Astropy package for the
detection and photometry of astronomical sources \citep{larry_bradley_2021_4624996}.

We thank the anonymous referee for valuable comments.
}

\vspace{5mm}
\facilities{LCOGT:0.4m, AAVSO}

\software{astropy \citep{astropy:2013,astropy:2018},
astroquery \citep{astroquery},
photutils \citep{larry_bradley_2021_4624996},
TOPCAT \citep{topcat},
}


\appendix

\section{Description of the matched catalogs}
\label{ap:description_catalog}

Tables~\ref{tab:pncat}--\ref{tab:gpncat_unique} give 
the description of each column tag included in
the matched catalogs presented in this paper.

\begin{deluxetable}{rll}
\tablecaption{Information from the HASH database included in 
PNcat and all the matched catalogs. \label{tab:pncat}}
\tablehead{
\colhead{No.}	&	\colhead{Tag}	&	\colhead{Description}
}
\startdata
1       &       PNG     &       Name given in the PN G nomenclature (PN GLLL.l+BB.b). \\
2       &       Name    &       Common name of the PN. \\
3       &       RAJ2000 &       Right Ascension (J2000). (HMS) \\
4       &       DECJ2000        &       Declination (J2000). (DMS) \\
5       &       DRAJ2000        &       Right Ascension (J2000). (deg) \\
6       &       DDECJ2000       &       Declination (J2000). (deg) \\
7       &       PNstat  &       Status of the PN: L=Likely, P=Probable, and T=True PN. \\
8       &       Catalogue       &       The source catalog of the PN. \\
9       &       MajDiam &       Major diameter of the PN. (arcsec) \\
10      &       MinDiam &       Minor diameter of the PN. (arcsec) \\
11      &       mainClass       &       PN main morphological type. \\
12      &       subClass        &       PN sub-morphological type. \\
\enddata
\tablerefs{(a) \citet{weidmann2011}, (b) \citet{kerber2003}, 
(c) \citet{parker2016}, (d) \citet{stanghellini2010}, 
(e) \citet{bailerjones2021}, (f) \citet{gonzalessantamaria2021}}
\end{deluxetable}

\clearpage

\startlongtable
\begin{deluxetable}{rll}
\tablecaption{GUVPNcat column information. The catalog also 
includes the columns described in Table~\ref{tab:pncat}. \label{tab:gpncat}}
\tablehead{
\colhead{No.}	&	\colhead{Tag}	&	\colhead{Description}
}
\startdata
1--12    &   &   as in Table~\ref{tab:pncat}. \\
13      &       \textbf{GALEX\_IAUName}  &       GALEX IAU Name of the source, from the GALEX coordinates. \\
14      &       GALEX\_objID    &       GALEX identifier of the source. \\
15      &       photoExtractID  &       Pointer to GALEX photoExtract parent image. \\
16      &       GALEX\_RA       &       Right Ascension for the GALEX source. (deg) \\
17      &       GALEX\_DEC      &       Declination for the  GALEX source. (deg) \\
18      &       E(B$-$V)   &       \parbox{0.6\textwidth}{E(B$-$V) from the GALEX database, obtained interpolating the source’s Galactic coordinates onto the reddening maps of \citet{schlegel1998}.} \\
19      &       glon    &       Galactic longitude for the GALEX source. (deg) \\
20      &       glat    &       Galactic latitude for the GALEX source. (deg) \\
21      &       fov\_radius     &       \parbox{0.6\columnwidth}{Source distance from the center of the field of view (from visitphotoobjall table) in the GALEX image. (deg)} \\
22      &       nuv\_mag        &       NUV magnitude. (ABmag) \\
23      &       nuv\_magerr     &       NUV  magnitude error. (ABmag) \\
24      &       fuv\_mag        &       FUV magnitude. (ABmag) \\
25      &       fuv\_magerr     &       FUV  magnitude error. (ABmag) \\
26      &       nuv\_artifact   &       NUV artifact flag. \\
27      &       fuv\_artifact   &       FUV artifact flag. \\
28      &       fuv\_weight     &       FUV effective exposure time. (sec) \\
29      &       nuv\_weight     &       NUV effective exposure time. (sec) \\
30      &       NUV\_MAG\_ISO   &       NUV ISO magnitude. (mag) \\
31      &       NUV\_MAGERR\_ISO        &       NUV ISO magnitude error. (mag) \\
32      &       NUV\_MAG\_APER\_1       &       NUV aperture magnitude (aperture radius of 1.5 arcsec). (mag) \\
33      &       NUV\_MAG\_APER\_2       &       NUV aperture magnitude (aperture radius of 2.3 arcsec). (mag) \\
34      &       NUV\_MAG\_APER\_3       &       NUV aperture magnitude (aperture radius of 3.8 arcsec). (mag) \\
35      &       NUV\_MAG\_APER\_4       &       NUV aperture magnitude (aperture radius of 6.0 arcsec). (mag) \\
36      &       NUV\_MAG\_APER\_5       &       NUV aperture magnitude (aperture radius of 9.0 arcsec). (mag) \\
37      &       NUV\_MAG\_APER\_6       &       NUV aperture magnitude (aperture radius of 12.8 arcsec). (mag) \\
38      &       NUV\_MAG\_APER\_7       &       NUV aperture magnitude (aperture radius of 17.3 arcsec). (mag) \\
39      &       NUV\_MAGERR\_APER\_1    &       NUV aperture magnitude error (aperture radius of 1.5 arcsec). (mag) \\
40      &       NUV\_MAGERR\_APER\_2    &       NUV aperture magnitude error (aperture radius of 2.3 arcsec). (mag) \\
41      &       NUV\_MAGERR\_APER\_3    &       NUV aperture magnitude error (aperture radius of 3.8 arcsec). (mag) \\
42      &       NUV\_MAGERR\_APER\_4    &       NUV aperture magnitude error (aperture radius of 6.0 arcsec). (mag) \\
43      &       NUV\_MAGERR\_APER\_5    &       NUV aperture magnitude error (aperture radius of 9.0 arcsec). (mag) \\
44      &       NUV\_MAGERR\_APER\_6    &       NUV aperture magnitude error (aperture radius of 12.8 arcsec). (mag) \\
45      &       NUV\_MAGERR\_APER\_7    &       NUV aperture magnitude error (aperture radius of 17.3 arcsec). (mag) \\
46      &       NUV\_MAG\_AUTO  &       NUV AUTO magnitude.  (mag)\\
47      &       NUV\_MAGERR\_AUTO       &       NUV AUTO magnitude error.  (mag)\\
48      &       NUV\_KRON\_RADIUS       &       NUV Kron apertures. (pix) \\
49      &       NUV\_A\_IMAGE   &       NUV profile rms along major axis. (pix) \\
50      &       NUV\_B\_IMAGE   &       NUV profile rms along minor axis. (pix) \\
51      &       NUV\_THETA\_IMAGE       &       NUV position angle. (deg) \\
52      &       NUV\_ELLIPTICITY        &       NUV ellipticity. (pix) \\
53      &       NUV\_FWHM\_IMAGE        &       NUV FWHM assuming a Gaussian core. (pix) \\
54      &       FUV\_MAG\_ISO   &       FUV ISO magnitude.  (mag)\\
55      &       FUV\_MAGERR\_ISO        &       FUV ISO magnitude error.  (mag)\\
56      &       FUV\_MAG\_APER\_1       &       FUV aperture magnitude (aperture radius of 1.5 arcsec).  (mag) \\
57      &       FUV\_MAG\_APER\_2       &       FUV aperture magnitude (aperture radius of 2.3 arcsec).  (mag) \\
58      &       FUV\_MAG\_APER\_3       &       FUV aperture magnitude (aperture radius of 3.8 arcsec).  (mag) \\
59      &       FUV\_MAG\_APER\_4       &       FUV aperture magnitude (aperture radius of 6.0 arcsec).  (mag) \\
60      &       FUV\_MAG\_APER\_5       &       FUV aperture magnitude (aperture radius of 9.0 arcsec).  (mag) \\
61      &       FUV\_MAG\_APER\_6       &       FUV aperture magnitude (aperture radius of 12.8 arcsec).  (mag) \\
62      &       FUV\_MAG\_APER\_7       &       FUV aperture magnitude (aperture radius of 17.3 arcsec).  (mag) \\
63      &       FUV\_MAGERR\_APER\_1    &       FUV aperture magnitude error (aperture radius of 1.5 arcsec).  (mag) \\
64      &       FUV\_MAGERR\_APER\_2    &       FUV aperture magnitude error (aperture radius of 2.3 arcsec).  (mag) \\
65      &       FUV\_MAGERR\_APER\_3    &       FUV aperture magnitude error (aperture radius of 3.8 arcsec).  (mag) \\
66      &       FUV\_MAGERR\_APER\_4    &       FUV aperture magnitude error (aperture radius of 6.0 arcsec).  (mag) \\
67      &       FUV\_MAGERR\_APER\_5    &       FUV aperture magnitude error (aperture radius of 9.0 arcsec).  (mag) \\
68      &       FUV\_MAGERR\_APER\_6    &       FUV aperture magnitude error (aperture radius of 12.8 arcsec).  (mag) \\
69      &       FUV\_MAGERR\_APER\_7    &       FUV aperture magnitude error (aperture radius of 17.3 arcsec).  (mag) \\
70      &       FUV\_MAG\_AUTO  &       FUV AUTO magnitude.  (mag) \\
71      &       FUV\_MAGERR\_AUTO       &       FUV AUTO magnitude error. (mag) \\
72      &       FUV\_KRON\_RADIUS       &       FUV Kron apertures. (pix) \\
73      &       FUV\_A\_IMAGE   &       FUV profile rms along major axis. (pix) \\
74      &       FUV\_B\_IMAGE   &       FUV profile rms along minor axis. (pix) \\
75      &       FUV\_THETA\_IMAGE       &       FUV Position angle. (deg) \\
76      &       FUV\_ELLIPTICITY        &       FUV ellipticity. (pix) \\
77      &       FUV\_FWHM\_IMAGE        &       FWHM assuming a Gaussian core. (pix) \\
78      &       \textbf{grankdist}   &       Rank for multiple measurements.$^{a}$ \\
79      &       \textbf{ngrankdist}  &       Number of multiple measurements of the same source.$^{a}$ \\
80      &       \textbf{primggroupid} &       Concatenated GALEX objid of multiple measurements of the same source.$^{a}$ \\
81      &       \textbf{primgid} &       GALEX objid of the primary source measurement.$^{a}$ \\
82      &       \textbf{FUVavg}  &       FUV averaged magnitude weighted-averaged for multiple measurements (if applicable). (mag) \\
83      &       \textbf{NUVavg}  &       NUV averaged magnitude weighted-averaged for multiple measurements (if applicable). (mag) \\
84      &       \textbf{FUVavgerr}       &       FUV averaged magnitude error weighted-averaged for multiple measurements (if applicable). (mag) \\
85      &       \textbf{NUVavgerr}       &       NUV averaged magnitude error weighted-averaged for multiple measurements (if applicable). (mag) \\
86      &       \textbf{distancerank}    &       Multiple matches.$^{b}$ \\
87      &       \textbf{NUV\_DIFF\_45}       &   Difference between NUV\_MAG\_APER\_4$-$NUV\_MAG\_APER\_5 magnitudes. (mag) \\
88      &       \textbf{HASHGALEXdist}       &   Distance between HASH position and GALEX matched source. (arcsec) \\
\enddata
\tablecomments{Tags from 14 to 77 are taken from the 
visitphotoobjall table  of the GALEX database. Additional tags 
created for the analysis of this paper are in bold font.}
\tablerefs{(a) See appendix~A of \citet{bianchi2017}, (b) \citet{bianchi2020arxiv}}
\end{deluxetable}
\clearpage

\startlongtable
\begin{deluxetable}{rll}
\tablecaption{PNcatxSDSSDR16 column information. \label{tab:pncat_sdss}}
\tablehead{
\colhead{No.}	&	\colhead{Tag}	&	\colhead{Description}
}
\startdata
1--12    &   &   as in Table~\ref{tab:pncat}. \\
13	&	SDSS\_objID	&	SDSS object ID of the matched source. \\ 
14	&	SDSS\_RA	&	Right Ascension for the SDSS source. (deg) \\ 
15	&	SDSS\_DEC	&	Declination for the SDSS source. (deg) \\ 
16	&	mode	&	SDSS observation mode. \\ 
17	&	SDSS\_type	&	SDSS object type. \\ 
18	&	psfMag\_u	&	SDSS u PSF magnitude. (mag)\\ 
19	&	psfMag\_g	&	SDSS g PSF magnitude. (mag)\\ 
20	&	psfMag\_r	&	SDSS r PSF magnitude. (mag)\\ 
21	&	psfMag\_i	&	SDSS i PSF magnitude. (mag)\\ 
22	&	psfMag\_z	&	SDSS z PSF magnitude. (mag)\\ 
23	&	psfMagErr\_u	&	SDSS u PSF magnitude error. (mag)\\ 
24	&	psfMagErr\_g	&	SDSS g PSF magnitude error. (mag)\\ 
25	&	psfMagErr\_r	&	SDSS r PSF magnitude error. (mag)\\ 
26	&	psfMagErr\_i	&	SDSS i PSF magnitude error. (mag)\\ 
27	&	psfMagErr\_z	&	SDSS z PSF magnitude error. (mag)\\ 
28	&	petroMag\_u	&	SDSS u petrosian magnitude. (mag)\\ 
29	&	petroMag\_g	&	SDSS g petrosian magnitude. (mag)\\ 
30	&	petroMag\_r	&	SDSS r petrosian magnitude. (mag)\\ 
31	&	petroMag\_i	&	SDSS i petrosian magnitude. (mag)\\ 
32	&	petroMag\_z	&	SDSS z petrosian magnitude. (mag)\\ 
33	&	petroMagErr\_u	&	SDSS u PSF petrosian error. (mag)\\ 
34	&	petroMagErr\_g	&	SDSS g PSF petrosian error. (mag)\\ 
35	&	petroMagErr\_r	&	SDSS r PSF petrosian error. (mag)\\ 
36	&	petroMagErr\_i	&	SDSS i PSF petrosian error. (mag)\\ 
37	&	petroMagErr\_z	&	SDSS z PSF petrosian error. (mag)\\ 
38	&	flags\_u	&	Object detection flag for SDSS u magnitude. \\ 
39	&	flags\_g	&	Object detection flag for SDSS g magnitude. \\ 
40	&	flags\_r	&	Object detection flag for SDSS r magnitude. \\ 
41	&	flags\_i	&	Object detection flag for SDSS i magnitude. \\ 
42	&	flags\_z	&	Object detection flag for SDSS z magnitude. \\ 
43	&	u	&	SDSS model u magnitude. (mag)\\ 
44	&	g	&	SDSS model g magnitude. (mag)\\ 
45	&	r	&	SDSS model r magnitude. (mag)\\ 
46	&	i	&	SDSS model i magnitude. (mag)\\ 
47	&	z	&	SDSS model z magnitude. (mag)\\ 
48	&	err\_u	&	SDSS model u magnitude error. (mag)\\ 
49	&	err\_g	&	SDSS model g magnitude error. (mag)\\ 
50	&	err\_r	&	SDSS model r magnitude error. (mag)\\ 
51	&	err\_i	&	SDSS model i magnitude error. (mag)\\ 
52	&	err\_z	&	SDSS model z magnitude error. (mag)\\ 
53	&	{\bf HASHSDSSdist}	&	Distance between HASH position and SDSS matched source. (arcsec) \\ 
54	&	{\bf distancerank\_SDSS}	&	\parbox{0.6\columnwidth}{Rank (by distance) of the multiple matches within the match radius of 5{\arcsec} around HASH source coordinates.$^{a}$} \\ 
55  &   {\bf r\_SDSS\_diff} &   Difference between SDSS~r-band PSF and model magnitudes. (mag)\\
\enddata
\tablecomments{Tags 13 to 52 are from SDSS database. Additional 
tags created for the analysis of this work are in bold font.}
\tablerefs{(a) see \citet{bianchi2020arxiv} for definition.}
\end{deluxetable}

\startlongtable
\begin{deluxetable}{rll}
\tablecaption{PNcatxPS1MDS column information. \label{tab:pncat_ps1}}
\tablehead{
\colhead{No.}	&	\colhead{Tag}	&	\colhead{Description}
}
\startdata
1--12   &   & as in Table~\ref{tab:pncat}. \\
13	&	objName	&	PS1 object name designation. \\ 
14	&	PS1\_objID	&	PS1 object ididentifier. \\ 
15	&	objInfoFlag	&	Information flag bitmask indicating details of the photometry. \\ 
16	&	qualityFlag	&	Subset of objInfoFlag indicating whether this object is real or likely false positive. \\
17	&	PS1\_RA	&	PS1 Right Ascension for the source. (deg)\\ 
18	&	PS1\_DEC	&	PS1 Declination for the source. (deg)\\ 
19	&	nDetections	&	Number of single epoch detections in all filters. \\ 
20	&	gMeanPSFMag	&	PS1 g mean PSF magnitude. (mag)\\ 
21	&	gMeanPSFMagErr	&	PS1 g mean PSF magnitude error. (mag)\\ 
22	&	gMeanKronMag	&	PS1 g mean Kron magnitude. (mag)\\ 
23	&	gMeanKronMagErr	&	PS1 g mean Kron magnitude error. (mag)\\ 
24	&	rMeanPSFMag	&	PS1 r mean PSF magnitude. (mag)\\ 
25	&	rMeanPSFMagErr	&	PS1 r mean PSF magnitude error. (mag)\\ 
26	&	rMeanKronMag	&	PS1 r mean Kron magnitude. (mag)\\ 
27	&	rMeanKronMagErr	&	PS1 r mean Kron magnitude error. (mag)\\ 
28	&	iMeanPSFMag	&	PS1 i mean PSF magnitude. (mag)\\ 
29	&	iMeanPSFMagErr	&	PS1 i mean PSF magnitude error. (mag)\\ 
30	&	iMeanKronMag	&	PS1 i mean Kron magnitude. (mag)\\ 
31	&	iMeanKronMagErr	&	PS1 i mean Kron magnitude error. (mag)\\ 
32	&	zMeanPSFMag	&	PS1 z mean PSF magnitude. (mag)\\ 
33	&	zMeanPSFMagErr	&	PS1 z mean PSF magnitude error. (mag)\\ 
34	&	zMeanKronMag	&	PS1 z mean Kron magnitude. (mag)\\ 
35	&	zMeanKronMagErr	&	PS1 z mean Kron magnitude error. (mag)\\ 
36	&	yMeanPSFMag	&	PS1 y mean PSF magnitude. (mag)\\ 
37	&	yMeanPSFMagErr	&	PS1 y mean PSF magnitude error. (mag)\\ 
38	&	yMeanKronMag	&	PS1 y mean Kron magnitude. (mag)\\ 
39	&	yMeanKronMagErr	&	PS1 y mean Kron magnitude error. (mag)\\ 
40	&	{\bf HASHPS1dist}	&	Distance between HASH position and Pan-STARRS matched source. (arcsec) \\ 
41	&	{\bf distancerank\_PS1}	&	\parbox{0.6\columnwidth}{Rank (by distance) of the multiple matches within the match radius of 5{\arcsec} around HASH source coordinates.$^{a}$} \\ 
42  &   {\bf r\_PS1\_diff}  &   Difference between PS1~r-band PSF and Kron magnitudes. (mag)\\
\enddata
\tablecomments{Tags 13 to 39 are from Pan-STARRS1 MeanObjView 
and StackObjectAttributes  tables. Additional tags created for 
the analysis of this paper are in bold font.}
\tablerefs{(a) \citet{bianchi2020arxiv}}
\end{deluxetable}

\startlongtable
\begin{deluxetable}{rll}
\tablecaption{GUVPNcatxSDSSDR16xPS1MDS catalog columns information. 
The magnitude measurements of the extracted CSPN flux described 
in Section~\ref{subsec:cspne_photmetry}, as well as the FUV and
 NUV PN radius (see Section~\ref{subsec:extended_pne_galex}) are
 also included (147--186); for those PNe not re-measured by us,
 the value of these columns is -99. \label{tab:gpncat_unique}}
\tablehead{
\colhead{No.}   &   \colhead{Tag}&  \colhead{Description}
}
\startdata
1--12   &   & as in Table~\ref{tab:pncat}. \\
13--88  &   & columns 13--88 from Table~\ref{tab:gpncat}. \\
89--131  &   & columns 13--55 from Table~\ref{tab:pncat_sdss}. \\
132--161  &   & columns 13--42 from Table~\ref{tab:pncat_ps1}. \\
162     &       FUV\_radius       &       FUV PN radius. (arcsec) \\
163     &       NUV\_radius       &       NUV PN radius. (arcsec) \\
164     &       mNUV    &       Extracted NUV CSPN photometry. (mag) \\
165     &       e\_mNUV &       Extracted NUV CSPN photmetry error. (mag) \\
166     &       mFUV    &       Extracted FUV CSPN photometry. (mag) \\
167     &       e\_mFUV &       Extracted FUV CSPN photmetry error. (mag) \\
168     &       mg\_PS1 &       Extracted PS1 g CSPN photometry. (mag) \\
169     &       mr\_PS1 &       Extracted PS1 r CSPN photometry. (mag) \\
170     &       mi\_PS1 &       Extracted PS1 i CSPN photometry. (mag) \\
171     &       mz\_PS1 &       Extracted PS1 z CSPN photometry. (mag) \\
172     &       my\_PS1 &       Extracted PS1 y CSPN photometry. (mag) \\
173     &       e\_mg\_PS1      &       Extracted PS1 g CSPN photometry error. (mag) \\
174     &       e\_mr\_PS1      &       Extracted PS1 r CSPN photometry error. (mag) \\
175     &       e\_mi\_PS1      &       Extracted PS1 i CSPN photometry error. (mag) \\
176     &       e\_mz\_PS1      &       Extracted PS1 z CSPN photometry error. (mag) \\
177     &       e\_my\_PS1      &       Extracted PS1 y CSPN photometry error. (mag) \\
178     &       recalc\_ps1     &       Flag to indicate if CSPN flux was extracted in PS1 images (=1). \\
179     &       mu\_SDSS        &       Extracted SDSS u CSPN photometry. (mag) \\
180     &       mg\_SDSS        &       Extracted SDSS g CSPN photometry. (mag) \\
181     &       mr\_SDSS        &       Extracted SDSS r CSPN photometry. (mag) \\
182     &       mi\_SDSS        &       Extracted SDSS i CSPN photometry. (mag) \\
183     &       mz\_SDSS        &       Extracted SDSS z CSPN photometry. (mag) \\
184     &       e\_mu\_SDSS     &       Extracted SDSS u CSPN photometry error. (mag) \\
185     &       e\_mg\_SDSS     &       Extracted SDSS g CSPN photometry error. (mag) \\
186     &       e\_mr\_SDSS     &       Extracted SDSS r CSPN photometry error. (mag) \\
187     &       e\_mi\_SDSS     &       Extracted SDSS i CSPN photometry error. (mag) \\
188     &       e\_mz\_SDSS     &       Extracted SDSS z CSPN photometry error. (mag) \\
189     &       recalc\_sdss    &       Flag to indicate if CSPN flux was extracted in SDSS images (=1). \\
190     &       binaryFlag      &       Flag to indicate if the PN has a binary nucleus. \\
191     &       r\_in\_SDSS     &       Inner annulus radius of extracted CSPN photometry. (arcsec) \\
192     &       r\_out\_SDSS    &       Outter annulus radius of extracted CSPN photometry. (arcsec) \\
193     &       r\_in\_PS1      &       Inner annulus radius of extracted CSPN photometry.. (arcsec) \\
194     &       r\_out\_PS1     &       Outter annulus radius of extracted CSPN photometry. (arcsec) \\
195     &       r\_aper\_SDSS   &       Aperture radius of extracted CSPN photometry. (arcsec) \\
196     &       r\_aper\_PS1    &       Aperture radius of extracted CSPN photometry. (arcsec) \\
197     &       ac\_u\_SDSS     &       Aperture correction for SDSS u. (mag) \\
198     &       ac\_g\_SDSS     &       Aperture correction for SDSS g. (mag) \\
199     &       ac\_r\_SDSS     &       Aperture correction for SDSS r. (mag) \\
200     &       ac\_i\_SDSS     &       Aperture correction for SDSS i. (mag) \\
201     &       ac\_z\_SDSS     &       Aperture correction for SDSS z. (mag) \\
202     &       ac\_g\_PS1      &       Aperture correction for PS1 g. (mag) \\
203     &       ac\_r\_PS1      &       Aperture correction for PS1 r. (mag) \\
204     &       ac\_i\_PS1      &       Aperture correction for PS1 i. (mag) \\
205     &       ac\_z\_PS1      &       Aperture correction for PS1 z. (mag) \\
206     &       ac\_y\_PS1      &       Aperture correction for PS1 y. (mag) \\
207     &       ac\_FUV &       Aperture correction for GALEX FUV. (mag) \\
208     &       ac\_NUV &       Aperture correction for GALEX NUV. (mag) \\
209     &       r\_in\_GALEX    &       Inner annulus radius of extracted CSPN photometry. (arcsec) \\
210     &       r\_out\_GALEX   &       Outter annulus radius of extracted CSPN photometry.. (arcsec) \\
211     &   lcogt   &   Flag to indicate if PN was observed by LCOGT. \\
212     &   g\_lcogt    &   Extracted LCOGT g CSPN photometry. (mag)\\
213     &   e\_g\_lcogt &   Extracted LCOGT g CSPN photometry error. (mag) \\
214     &   r\_lcogt    &   Extracted LCOGT r CSPN photometry. (mag) \\
215     &   e\_r\_lcogt &   Extracted LCOGT r CSPN photmetry error. (mag) \\
216     &   i\_lcogt    &   Extracted LCOGT i CSPN photometry. (mag) \\
217     & e\_i\_lcogt   &   Extracted LCOGT i CSPN photometry error. (mag) \\
218     &   Dmean   &   Mean statistical distance from \citet{Frew2016}. (kpc)\\
219     &   e\_Dmean &   Mean statistical distance error from \citet{Frew2016}. (kpc) \\
220     &   GaiaEDR3\_C &   Gaia EDR3 object identifier.$^{a}$ \\
221     &   rcomb   &   \parbox{0.6\textwidth}{Median of the combined (parallax and statistical) distance posterior from \citet{Chornay2021}. (pc)} \\
222     &   b\_rcomb &   16th percentile of the combined (parallax and statistical) distance posterior.$^{a}$ (pc)\\
223     &   B\_rcomb &   84th percentile of the combined (parallax and statistical) distance posterior.$^{a}$ (pc)\\
224     &   GaiaEDR3\_GS &   Gaia EDR3 object identifier.$^{b}$ \\
225     &   d   &   \parbox{0.6\textwidth}{Estimated distance from the Sun$^{c}$ from the compilation of \citet{gonzalessantamaria2021}. (pc)} \\
226     &   d\_min  &   Estimated minimum distance.$^{c}$ (pc)\\
227     &   d\_max  &   Estimated maximum distance.$^{c}$ (pc) \\
228     &   sep\_GS &  Distance between best GALEX position and best SDSS matched source. (arcsec) \\
229     &   sep\_GP &  Distance between best GALEX position and best Pan-STARRS matched source. (arcsec) \\
\enddata
\tablerefs{(a) \citet{Chornay2021}, (b) \citet{gonzalessantamaria2021}, (c) \citet{bailerjones2021}}
\end{deluxetable}




\bibliography{main}{}

\begin{thebibliography}{}
\expandafter\ifx\csname natexlab\endcsname\relax\def\natexlab#1{#1}\fi
\providecommand{\url}[1]{\href{#1}{#1}}
\providecommand{\dodoi}[1]{doi:~\href{http://doi.org/#1}{\nolinkurl{#1}}}
\providecommand{\doeprint}[1]{\href{http://ascl.net/#1}{\nolinkurl{http://ascl.net/#1}}}
\providecommand{\doarXiv}[1]{\href{https://arxiv.org/abs/#1}{\nolinkurl{https://arxiv.org/abs/#1}}}

\bibitem[{{Acker} {et~al.}(1992){Acker}, {Marcout}, {Ochsenbein}, {Stenholm},
  {Tylenda}, \& {Schohn}}]{acker1992}
{Acker}, A., {Marcout}, J., {Ochsenbein}, F., {et~al.} 1992, {The
  Strasbourg-ESO Catalogue of Galactic Planetary Nebulae. Parts I, II.}
  ({European Southern Observatory, Garching (Germany)})

\bibitem[{{Ahumada} {et~al.}(2020){Ahumada}, {Prieto}, {Almeida}, {Anders},
  {Anderson}, {Andrews}, {Anguiano}, {Arcodia}, {Armengaud}, {Aubert}, {Avila},
  {Avila-Reese}, {Badenes}, {Balland}, {Barger}, {Barrera-Ballesteros}, {Basu},
  {Bautista}, {Beaton}, {Beers}, {Benavides}, {Bender}, {Bernardi}, {Bershady},
  {Beutler}, {Bidin}, {Bird}, {Bizyaev}, {Blanc}, {Blanton}, {Boquien},
  {Borissova}, {Bovy}, {Brandt}, {Brinkmann}, {Brownstein}, {Bundy}, {Bureau},
  {Burgasser}, {Burtin}, {Cano-D{\'\i}az}, {Capasso}, {Cappellari}, {Carrera},
  {Chabanier}, {Chaplin}, {Chapman}, {Cherinka}, {Chiappini}, {Doohyun Choi},
  {Chojnowski}, {Chung}, {Clerc}, {Coffey}, {Comerford}, {Comparat}, {da
  Costa}, {Cousinou}, {Covey}, {Crane}, {Cunha}, {Ilha}, {Dai}, {Damsted},
  {Darling}, {Davidson}, {Davies}, {Dawson}, {De}, {de la Macorra}, {De Lee},
  {Queiroz}, {Deconto Machado}, {de la Torre}, {Dell'Agli}, {du Mas des
  Bourboux}, {Diamond-Stanic}, {Dillon}, {Donor}, {Drory}, {Duckworth},
  {Dwelly}, {Ebelke}, {Eftekharzadeh}, {Davis Eigenbrot}, {Elsworth},
  {Eracleous}, {Erfanianfar}, {Escoffier}, {Fan}, {Farr},
  {Fern{\'a}ndez-Trincado}, {Feuillet}, {Finoguenov}, {Fofie},
  {Fraser-McKelvie}, {Frinchaboy}, {Fromenteau}, {Fu}, {Galbany}, {Garcia},
  {Garc{\'\i}a-Hern{\'a}ndez}, {Oehmichen}, {Ge}, {Maia}, {Geisler}, {Gelfand},
  {Goddy}, {Gonzalez-Perez}, {Grabowski}, {Green}, {Grier}, {Guo}, {Guy},
  {Harding}, {Hasselquist}, {Hawken}, {Hayes}, {Hearty}, {Hekker}, {Hogg},
  {Holtzman}, {Horta}, {Hou}, {Hsieh}, {Huber}, {Hunt}, {Chitham}, {Imig},
  {Jaber}, {Angel}, {Johnson}, {Jones}, {J{\"o}nsson}, {Jullo}, {Kim},
  {Kinemuchi}, {Kirkpatrick}, {Kite}, {Klaene}, {Kneib}, {Kollmeier}, {Kong},
  {Kounkel}, {Krishnarao}, {Lacerna}, {Lan}, {Lane}, {Law}, {Le Goff}, {Leung},
  {Lewis}, {Li}, {Lian}, {Lin}, {Long}, {Longa-Pe{\~n}a}, {Lundgren}, {Lyke},
  {Ted Mackereth}, {MacLeod}, {Majewski}, {Manchado}, {Maraston}, {Martini},
  {Masseron}, {Masters}, {Mathur}, {McDermid}, {Merloni}, {Merrifield},
  {M{\'e}sz{\'a}ros}, {Miglio}, {Minniti}, {Minsley}, {Miyaji}, {Mohammad},
  {Mosser}, {Mueller}, {Muna}, {Mu{\~n}oz-Guti{\'e}rrez}, {Myers}, {Nadathur},
  {Nair}, {Nandra}, {do Nascimento}, {Nevin}, {Newman}, {Nidever}, {Nitschelm},
  {Noterdaeme}, {O'Connell}, {Olmstead}, {Oravetz}, {Oravetz}, {Osorio},
  {Pace}, {Padilla}, {Palanque-Delabrouille}, {Palicio}, {Pan}, {Pan},
  {Parker}, {Paviot}, {Peirani}, {Ram{\'r}ez}, {Penny}, {Percival},
  {Perez-Fournon}, {P{\'e}rez-R{\`a}fols}, {Petitjean}, {Pieri},
  {Pinsonneault}, {Poovelil}, {Povick}, {Prakash}, {Price-Whelan}, {Raddick},
  {Raichoor}, {Ray}, {Rembold}, {Rezaie}, {Riffel}, {Riffel}, {Rix}, {Robin},
  {Roman-Lopes}, {Rom{\'a}n-Z{\'u}{\~n}iga}, {Rose}, {Ross}, {Rossi},
  {Rowlands}, {Rubin}, {Salvato}, {S{\'a}nchez}, {S{\'a}nchez-Menguiano},
  {S{\'a}nchez-Gallego}, {Sayres}, {Schaefer}, {Schiavon}, {Schimoia},
  {Schlafly}, {Schlegel}, {Schneider}, {Schultheis}, {Schwope}, {Seo},
  {Serenelli}, {Shafieloo}, {Shamsi}, {Shao}, {Shen}, {Shetrone}, {Shirley},
  {Aguirre}, {Simon}, {Skrutskie}, {Slosar}, {Smethurst}, {Sobeck}, {Sodi},
  {Souto}, {Stark}, {Stassun}, {Steinmetz}, {Stello}, {Stermer},
  {Storchi-Bergmann}, {Streblyanska}, {Stringfellow}, {Stutz}, {Su{\'a}rez},
  {Sun}, {Taghizadeh-Popp}, {Talbot}, {Tayar}, {Thakar}, {Theriault}, {Thomas},
  {Thomas}, {Tinker}, {Tojeiro}, {Toledo}, {Tremonti}, {Troup}, {Tuttle},
  {Unda-Sanzana}, {Valentini}, {Vargas-Gonz{\'a}lez}, {Vargas-Maga{\~n}a},
  {V{\'a}zquez-Mata}, {Vivek}, {Wake}, {Wang}, {Weaver}, {Weijmans}, {Wild},
  {Wilson}, {Wilson}, {Wolthuis}, {Wood-Vasey}, {Yan}, {Yang}, {Y{\`e}che},
  {Zamora}, {Zarrouk}, {Zasowski}, {Zhang}, {Zhao}, {Zhao}, {Zheng}, {Zheng},
  {Zhu}, \& {Zou}}]{Ahumada2020}
{Ahumada}, R., {Prieto}, C.~A., {Almeida}, A., {et~al.} 2020, \apjs, 249, 3,
  \dodoi{10.3847/1538-4365/ab929e}

\bibitem[{{Astropy Collaboration} {et~al.}(2013){Astropy Collaboration},
  {Robitaille}, {Tollerud}, {Greenfield}, {Droettboom}, {Bray}, {Aldcroft},
  {Davis}, {Ginsburg}, {Price-Whelan}, {Kerzendorf}, {Conley}, {Crighton},
  {Barbary}, {Muna}, {Ferguson}, {Grollier}, {Parikh}, {Nair}, {Unther},
  {Deil}, {Woillez}, {Conseil}, {Kramer}, {Turner}, {Singer}, {Fox}, {Weaver},
  {Zabalza}, {Edwards}, {Azalee Bostroem}, {Burke}, {Casey}, {Crawford},
  {Dencheva}, {Ely}, {Jenness}, {Labrie}, {Lim}, {Pierfederici}, {Pontzen},
  {Ptak}, {Refsdal}, {Servillat}, \& {Streicher}}]{astropy:2013}
{Astropy Collaboration}, {Robitaille}, T.~P., {Tollerud}, E.~J., {et~al.} 2013,
  \aap, 558, A33, \dodoi{10.1051/0004-6361/201322068}

\bibitem[{{Astropy Collaboration} {et~al.}(2018){Astropy Collaboration},
  {Price-Whelan}, {Sip{\H{o}}cz}, {G{\"u}nther}, {Lim}, {Crawford}, {Conseil},
  {Shupe}, {Craig}, {Dencheva}, {Ginsburg}, {Vand erPlas}, {Bradley},
  {P{\'e}rez-Su{\'a}rez}, {de Val-Borro}, {Aldcroft}, {Cruz}, {Robitaille},
  {Tollerud}, {Ardelean}, {Babej}, {Bach}, {Bachetti}, {Bakanov}, {Bamford},
  {Barentsen}, {Barmby}, {Baumbach}, {Berry}, {Biscani}, {Boquien}, {Bostroem},
  {Bouma}, {Brammer}, {Bray}, {Breytenbach}, {Buddelmeijer}, {Burke},
  {Calderone}, {Cano Rodr{\'\i}guez}, {Cara}, {Cardoso}, {Cheedella}, {Copin},
  {Corrales}, {Crichton}, {D'Avella}, {Deil}, {Depagne}, {Dietrich}, {Donath},
  {Droettboom}, {Earl}, {Erben}, {Fabbro}, {Ferreira}, {Finethy}, {Fox},
  {Garrison}, {Gibbons}, {Goldstein}, {Gommers}, {Greco}, {Greenfield},
  {Groener}, {Grollier}, {Hagen}, {Hirst}, {Homeier}, {Horton}, {Hosseinzadeh},
  {Hu}, {Hunkeler}, {Ivezi{\'c}}, {Jain}, {Jenness}, {Kanarek}, {Kendrew},
  {Kern}, {Kerzendorf}, {Khvalko}, {King}, {Kirkby}, {Kulkarni}, {Kumar},
  {Lee}, {Lenz}, {Littlefair}, {Ma}, {Macleod}, {Mastropietro}, {McCully},
  {Montagnac}, {Morris}, {Mueller}, {Mumford}, {Muna}, {Murphy}, {Nelson},
  {Nguyen}, {Ninan}, {N{\"o}the}, {Ogaz}, {Oh}, {Parejko}, {Parley}, {Pascual},
  {Patil}, {Patil}, {Plunkett}, {Prochaska}, {Rastogi}, {Reddy Janga},
  {Sabater}, {Sakurikar}, {Seifert}, {Sherbert}, {Sherwood-Taylor}, {Shih},
  {Sick}, {Silbiger}, {Singanamalla}, {Singer}, {Sladen}, {Sooley},
  {Sornarajah}, {Streicher}, {Teuben}, {Thomas}, {Tremblay}, {Turner},
  {Terr{\'o}n}, {van Kerkwijk}, {de la Vega}, {Watkins}, {Weaver}, {Whitmore},
  {Woillez}, {Zabalza}, \& {Astropy Contributors}}]{astropy:2018}
{Astropy Collaboration}, {Price-Whelan}, A.~M., {Sip{\H{o}}cz}, B.~M., {et~al.}
  2018, \aj, 156, 123, \dodoi{10.3847/1538-3881/aabc4f}

\bibitem[{{Bailer-Jones} {et~al.}(2021){Bailer-Jones}, {Rybizki}, {Fouesneau},
  {Demleitner}, \& {Andrae}}]{bailerjones2021}
{Bailer-Jones}, C.~A.~L., {Rybizki}, J., {Fouesneau}, M., {Demleitner}, M., \&
  {Andrae}, R. 2021, \aj, 161, 147, \dodoi{10.3847/1538-3881/abd806}

\bibitem[{{Balick}(1987)}]{balick1987}
{Balick}, B. 1987, \aj, 94, 671, \dodoi{10.1086/114504}

\bibitem[{{Barker} {et~al.}(2018){Barker}, {Zijlstra}, {De Marco}, {Frew},
  {Drew}, {Corradi}, {Eisl{\"o}ffel}, \& {Parker}}]{barker2018}
{Barker}, H., {Zijlstra}, A., {De Marco}, O., {et~al.} 2018, \mnras, 475, 4504,
  \dodoi{10.1093/mnras/stx3240}

\bibitem[{{Bianchi}(2009)}]{bianchi2009b}
{Bianchi}, L. 2009, \apss, 320, 11, \dodoi{10.1007/s10509-008-9761-3}

\bibitem[{{Bianchi}(2014)}]{bianchi2014}
---. 2014, \apss, 354, 103, \dodoi{10.1007/s10509-014-1935-6}

\bibitem[{{Bianchi} {et~al.}(2019){Bianchi}, {de la Vega}, {Shiao}, \&
  {Souter}}]{bianchi2019}
{Bianchi}, L., {de la Vega}, A., {Shiao}, B., \& {Souter}, B.~J. 2019, \apjs,
  241, 14, \dodoi{10.3847/1538-4365/aafee8}

\bibitem[{{Bianchi} {et~al.}(2011{\natexlab{a}}){Bianchi}, {Efremova},
  {Herald}, {Girardi}, {Zabot}, {Marigo}, \& {Martin}}]{bianchi2011b}
{Bianchi}, L., {Efremova}, B., {Herald}, J., {et~al.} 2011{\natexlab{a}},
  \mnras, 411, 2770, \dodoi{10.1111/j.1365-2966.2010.17890.x}

\bibitem[{{Bianchi} {et~al.}(2011{\natexlab{b}}){Bianchi}, {Herald},
  {Efremova}, {Girardi}, {Zabot}, {Marigo}, {Conti}, \& {Shiao}}]{bianchi2011a}
{Bianchi}, L., {Herald}, J., {Efremova}, B., {et~al.} 2011{\natexlab{b}},
  \apss, 335, 161, \dodoi{10.1007/s10509-010-0581-x}

\bibitem[{{Bianchi} {et~al.}(2009){Bianchi}, {Hutchings}, {Efremova}, {Herald},
  {Bressan}, \& {Martin}}]{bianchi2009}
{Bianchi}, L., {Hutchings}, J.~B., {Efremova}, B., {et~al.} 2009, \aj, 137,
  3761, \dodoi{10.1088/0004-6256/137/4/3761}

\bibitem[{{Bianchi} {et~al.}(2018){Bianchi}, {Keller}, {Bohlin}, {Barstow}, \&
  {Casewell}}]{bianchi2018z}
{Bianchi}, L., {Keller}, G.~R., {Bohlin}, R., {Barstow}, M., \& {Casewell}, S.
  2018, \apss, 363, 166, \dodoi{10.1007/s10509-018-3369-z}

\bibitem[{{Bianchi} \& {Shiao}(2020)}]{bianchi2020arxiv}
{Bianchi}, L., \& {Shiao}, B. 2020, \apjs, 250, 36,
  \dodoi{10.3847/1538-4365/aba2d7}

\bibitem[{{Bianchi} {et~al.}(2017){Bianchi}, {Shiao}, \&
  {Thilker}}]{bianchi2017}
{Bianchi}, L., {Shiao}, B., \& {Thilker}, D. 2017, \apjs, 230, 24,
  \dodoi{10.3847/1538-4365/aa7053}

\bibitem[{{Bianchi} \& {Thilker}(2018)}]{bianchi2018}
{Bianchi}, L., \& {Thilker}, D. 2018, \apss, 363, 85,
  \dodoi{10.1007/s10509-018-3300-7}

\bibitem[{{Bianchi} {et~al.}(2007){Bianchi}, {Rodriguez-Merino}, {Viton},
  {Laget}, {Efremova}, {Herald}, {Conti}, {Shiao}, {Gil de Paz}, {Salim},
  {Thakar}, {Friedman}, {Rey}, {Thilker}, {Barlow}, {Budav{\'a}ri}, {Donas},
  {Forster}, {Heckman}, {Lee}, {Madore}, {Martin}, {Milliard}, {Morrissey},
  {Neff}, {Rich}, {Schiminovich}, {Seibert}, {Small}, {Szalay}, {Wyder},
  {Welsh}, \& {Yi}}]{bianchi2007}
{Bianchi}, L., {Rodriguez-Merino}, L., {Viton}, M., {et~al.} 2007, \apjs, 173,
  659, \dodoi{10.1086/516648}

\bibitem[{{Bloecker}(1995)}]{bloecker1995}
{Bloecker}, T. 1995, \aap, 299, 755

\bibitem[{{Bond}(2009)}]{bond2009}
{Bond}, H.~E. 2009, in Journal of Physics Conference Series, Vol. 172, Journal
  of Physics Conference Series, 012029, \dodoi{10.1088/1742-6596/172/1/012029}

\bibitem[{Bradley {et~al.}(2021)Bradley, Sipőcz, Robitaille, Tollerud,
  Vinícius, Deil, Barbary, Wilson, Busko, Donath, Günther, Cara, Conseil,
  Bostroem, Droettboom, Bray, krachyon, Lim, Bratholm, Barentsen, Craig, Rathi,
  Pascual, Perren, Georgiev, de~Val-Borro, Kerzendorf, Bach, Quint, \&
  Souchereau}]{larry_bradley_2021_4624996}
Bradley, L., Sipőcz, B., Robitaille, T., {et~al.} 2021, astropy/photutils:
  1.1.0, 1.1.0,  Zenodo, \dodoi{10.5281/zenodo.4624996}

\bibitem[{{Brown} {et~al.}(2014){Brown}, {Moustakas}, {Smith}, {da Cunha},
  {Jarrett}, {Imanishi}, {Armus}, {Brandl}, \& {Peek}}]{Brown2014}
{Brown}, M. J.~I., {Moustakas}, J., {Smith}, J. D.~T., {et~al.} 2014, \apjs,
  212, 18, \dodoi{10.1088/0067-0049/212/2/18}

\bibitem[{{Brown} {et~al.}(2013){Brown}, {Baliber}, {Bianco}, {Bowman},
  {Burleson}, {Conway}, {Crellin}, {Depagne}, {De Vera}, {Dilday}, {Dragomir},
  {Dubberley}, {Eastman}, {Elphick}, {Falarski}, {Foale}, {Ford}, {Fulton},
  {Garza}, {Gomez}, {Graham}, {Greene}, {Haldeman}, {Hawkins}, {Haworth},
  {Haynes}, {Hidas}, {Hjelstrom}, {Howell}, {Hygelund}, {Lister}, {Lobdill},
  {Martinez}, {Mullins}, {Norbury}, {Parrent}, {Paulson}, {Petry}, {Pickles},
  {Posner}, {Rosing}, {Ross}, {Sand}, {Saunders}, {Shobbrook}, {Shporer},
  {Street}, {Thomas}, {Tsapras}, {Tufts}, {Valenti}, {Vander Horst}, {Walker},
  {White}, \& {Willis}}]{brown2013}
{Brown}, T.~M., {Baliber}, N., {Bianco}, F.~B., {et~al.} 2013, \pasp, 125,
  1031, \dodoi{10.1086/673168}

\bibitem[{{Cardelli} {et~al.}(1989){Cardelli}, {Clayton}, \& {Mathis}}]{ccm89}
{Cardelli}, J.~A., {Clayton}, G.~C., \& {Mathis}, J.~S. 1989, \apj, 345, 245,
  \dodoi{10.1086/167900}

\bibitem[{{Chambers} {et~al.}(2016){Chambers}, {Magnier}, {Metcalfe},
  {Flewelling}, {Huber}, {Waters}, {Denneau}, {Draper}, {Farrow}, {Finkbeiner},
  {Holmberg}, {Koppenhoefer}, {Price}, {Rest}, {Saglia}, {Schlafly}, {Smartt},
  {Sweeney}, {Wainscoat}, {Burgett}, {Chastel}, {Grav}, {Heasley}, {Hodapp},
  {Jedicke}, {Kaiser}, {Kudritzki}, {Luppino}, {Lupton}, {Monet}, {Morgan},
  {Onaka}, {Shiao}, {Stubbs}, {Tonry}, {White}, {Ba{\~n}ados}, {Bell},
  {Bender}, {Bernard}, {Boegner}, {Boffi}, {Botticella}, {Calamida},
  {Casertano}, {Chen}, {Chen}, {Cole}, {Deacon}, {Frenk}, {Fitzsimmons},
  {Gezari}, {Gibbs}, {Goessl}, {Goggia}, {Gourgue}, {Goldman}, {Grant},
  {Grebel}, {Hambly}, {Hasinger}, {Heavens}, {Heckman}, {Henderson}, {Henning},
  {Holman}, {Hopp}, {Ip}, {Isani}, {Jackson}, {Keyes}, {Koekemoer}, {Kotak},
  {Le}, {Liska}, {Long}, {Lucey}, {Liu}, {Martin}, {Masci}, {McLean}, {Mindel},
  {Misra}, {Morganson}, {Murphy}, {Obaika}, {Narayan}, {Nieto-Santisteban},
  {Norberg}, {Peacock}, {Pier}, {Postman}, {Primak}, {Rae}, {Rai}, {Riess},
  {Riffeser}, {Rix}, {R{\"o}ser}, {Russel}, {Rutz}, {Schilbach}, {Schultz},
  {Scolnic}, {Strolger}, {Szalay}, {Seitz}, {Small}, {Smith}, {Soderblom},
  {Taylor}, {Thomson}, {Taylor}, {Thakar}, {Thiel}, {Thilker}, {Unger},
  {Urata}, {Valenti}, {Wagner}, {Walder}, {Walter}, {Watters}, {Werner},
  {Wood-Vasey}, \& {Wyse}}]{chambers2016}
{Chambers}, K.~C., {Magnier}, E.~A., {Metcalfe}, N., {et~al.} 2016, arXiv
  e-prints, arXiv:1612.05560.
\newblock \doarXiv{1612.05560}

\bibitem[{{Chornay} \& {Walton}(2021)}]{Chornay2021}
{Chornay}, N., \& {Walton}, N.~A. 2021, \aap, 656, A110,
  \dodoi{10.1051/0004-6361/202142008}

\bibitem[{{de la Vega} \& {Bianchi}(2018)}]{delavega2018}
{de la Vega}, A., \& {Bianchi}, L. 2018, \apjs, 238, 25,
  \dodoi{10.3847/1538-4365/aaddf5}

\bibitem[{{Douchin} {et~al.}(2015){Douchin}, {De Marco}, {Frew}, {Jacoby},
  {Jasniewicz}, {Fitzgerald}, {Passy}, {Harmer}, {Hillwig}, \&
  {Moe}}]{douchin2015}
{Douchin}, D., {De Marco}, O., {Frew}, D.~J., {et~al.} 2015, \mnras, 448, 3132,
  \dodoi{10.1093/mnras/stu2700}

\bibitem[{{Farrow} {et~al.}(2014){Farrow}, {Cole}, {Metcalfe}, {Draper},
  {Norberg}, {Foucaud}, {Burgett}, {Chambers}, {Kaiser}, {Kudritzki},
  {Magnier}, {Price}, {Tonry}, \& {Waters}}]{farrow2014}
{Farrow}, D.~J., {Cole}, S., {Metcalfe}, N., {et~al.} 2014, \mnras, 437, 748,
  \dodoi{10.1093/mnras/stt1933}

\bibitem[{{Feibelman}(2000)}]{feibelman2000}
{Feibelman}, W.~A. 2000, \pasp, 112, 861, \dodoi{10.1086/316584}

\bibitem[{{Flewelling} {et~al.}(2020){Flewelling}, {Magnier}, {Chambers},
  {Heasley}, {Holmberg}, {Huber}, {Sweeney}, {Waters}, {Calamida}, {Casertano},
  {Chen}, {Farrow}, {Hasinger}, {Henderson}, {Long}, {Metcalfe}, {Narayan},
  {Nieto-Santisteban}, {Norberg}, {Rest}, {Saglia}, {Szalay}, {Thakar},
  {Tonry}, {Valenti}, {Werner}, {White}, {Denneau}, {Draper}, {Hodapp},
  {Jedicke}, {Kaiser}, {Kudritzki}, {Price}, {Wainscoat}, {Chastel}, {McLean},
  {Postman}, \& {Shiao}}]{flewelling2020}
{Flewelling}, H.~A., {Magnier}, E.~A., {Chambers}, K.~C., {et~al.} 2020, \apjs,
  251, 7, \dodoi{10.3847/1538-4365/abb82d}

\bibitem[{{Frew} {et~al.}(2013){Frew}, {Boji{\v{c}}i{\'c}}, \&
  {Parker}}]{frew2013}
{Frew}, D.~J., {Boji{\v{c}}i{\'c}}, I.~S., \& {Parker}, Q.~A. 2013, \mnras,
  431, 2, \dodoi{10.1093/mnras/sts393}

\bibitem[{{Frew} {et~al.}(2016){Frew}, {Parker}, \&
  {Boji{\v{c}}i{\'c}}}]{Frew2016}
{Frew}, D.~J., {Parker}, Q.~A., \& {Boji{\v{c}}i{\'c}}, I.~S. 2016, \mnras,
  455, 1459, \dodoi{10.1093/mnras/stv1516}

\bibitem[{{Gauba} {et~al.}(2001){Gauba}, {Parthasarathy}, {Nakada}, \&
  {Fujii}}]{gauba2001}
{Gauba}, G., {Parthasarathy}, M., {Nakada}, Y., \& {Fujii}, T. 2001, \aap, 373,
  572, \dodoi{10.1051/0004-6361:20010623}

\bibitem[{{Ginsburg} {et~al.}(2019){Ginsburg}, {Sip{\H o}cz}, {Brasseur},
  {Cowperthwaite}, {Craig}, {Deil}, {Guillochon}, {Guzman}, {Liedtke}, {Lian
  Lim}, {Lockhart}, {Mommert}, {Morris}, {Norman}, {Parikh}, {Persson},
  {Robitaille}, {Segovia}, {Singer}, {Tollerud}, {de Val-Borro}, {Valtchanov},
  {Woillez}, {The Astroquery collaboration}, \& {a subset of the astropy
  collaboration}}]{astroquery}
{Ginsburg}, A., {Sip{\H o}cz}, B.~M., {Brasseur}, C.~E., {et~al.} 2019, \aj,
  157, 98, \dodoi{10.3847/1538-3881/aafc33}

\bibitem[{{G{\'o}mez-Mu{\~n}oz} {et~al.}(2022){G{\'o}mez-Mu{\~n}oz}, {Sabin},
  {Raddi}, \& {Wells}}]{GomezMunoz2022}
{G{\'o}mez-Mu{\~n}oz}, M.~A., {Sabin}, L., {Raddi}, R., \& {Wells}, R.~D. 2022,
  \mnras, 514, 2434, \dodoi{10.1093/mnras/stac1403}

\bibitem[{{Gonz{\'a}lez-Santamar{\'\i}a}
  {et~al.}(2021){Gonz{\'a}lez-Santamar{\'\i}a}, {Manteiga}, {Manchado}, {Ulla},
  {Dafonte}, \& {L\'opez-Varela}}]{gonzalessantamaria2021}
{Gonz{\'a}lez-Santamar{\'\i}a}, I., {Manteiga}, M., {Manchado}, A., {et~al.}
  2021, \aap

\bibitem[{{Guerrero} {et~al.}(2010){Guerrero}, {Ramos-Larios}, \&
  {Massa}}]{guerrero2010}
{Guerrero}, M.~A., {Ramos-Larios}, G., \& {Massa}, D. 2010, \pasa, 27, 210,
  \dodoi{10.1071/AS09024}

\bibitem[{{Guerrero} {et~al.}(2018){Guerrero}, {Sabin}, {Tovmassian},
  {Santamar{\'\i}a}, {Michel}, {Ramos-Larios}, {Alarie}, {Morisset},
  {Berm{\'u}dez Bustamante}, {Gonz{\'a}lez}, \& {Wright}}]{guerrero2018}
{Guerrero}, M.~A., {Sabin}, L., {Tovmassian}, G., {et~al.} 2018, \apj, 857, 80,
  \dodoi{10.3847/1538-4357/aab669}

\bibitem[{{Guti{\'e}rrez-Soto} {et~al.}(2020){Guti{\'e}rrez-Soto},
  {Gon{\c{c}}alves}, {Akras}, {Cortesi}, {L{\'o}pez-Sanjuan}, {Guerrero},
  {Daflon}, {Borges Fernandes}, {Mendes de Oliveira}, {Ederoclite},
  {Sodr{\'e}}, {Pereira}, {Kanaan}, {Werle}, {V{\'a}zquez Rami{\'o}},
  {Alcaniz}, {Angulo}, {Cenarro}, {Crist{\'o}bal-Hornillos}, {Dupke},
  {Hern{\'a}ndez-Monteagudo}, {Mar{\'\i}n-Franch}, {Moles}, {Varela},
  {Ribeiro}, {Schoenell}, {Alvarez-Candal}, {Galbany}, {Jim{\'e}nez-Esteban},
  {Logro{\~n}o-Garc{\'\i}a}, \& {Sobral}}]{GutierrezSoto2020}
{Guti{\'e}rrez-Soto}, L.~A., {Gon{\c{c}}alves}, D.~R., {Akras}, S., {et~al.}
  2020, \aap, 633, A123, \dodoi{10.1051/0004-6361/201935700}

\bibitem[{{Henden} {et~al.}(2015){Henden}, {Levine}, {Terrell}, \&
  {Welch}}]{henden2015}
{Henden}, A.~A., {Levine}, S., {Terrell}, D., \& {Welch}, D.~L. 2015, in
  American Astronomical Society Meeting Abstracts, Vol. 225, American
  Astronomical Society Meeting Abstracts \#225, 336.16

\bibitem[{{Herald} \& {Bianchi}(2011)}]{herald2011}
{Herald}, J.~E., \& {Bianchi}, L. 2011, \mnras, 417, 2440,
  \dodoi{10.1111/j.1365-2966.2011.19319.x}

\bibitem[{{Hoogerwerf} {et~al.}(2007){Hoogerwerf}, {Szentgyorgyi}, {Raymond},
  {Brickhouse}, {Slane}, \& {Franco}}]{hoogerwerf2007}
{Hoogerwerf}, R., {Szentgyorgyi}, A., {Raymond}, J., {et~al.} 2007, \apj, 670,
  442, \dodoi{10.1086/521637}

\bibitem[{{Jones} \& {Boffin}(2017)}]{jones2017}
{Jones}, D., \& {Boffin}, H. M.~J. 2017, Nature Astronomy, 1, 0117,
  \dodoi{10.1038/s41550-017-0117}

\bibitem[{{Jones} {et~al.}(2019{\natexlab{a}}){Jones}, {Boffin}, {Sowicka},
  {Miszalski}, {Rodr{\'\i}guez-Gil}, {Santand er-Garc{\'\i}a}, \&
  {Corradi}}]{jones2019}
{Jones}, D., {Boffin}, H. M.~J., {Sowicka}, P., {et~al.} 2019{\natexlab{a}},
  \mnras, 482, L75, \dodoi{10.1093/mnrasl/sly142}

\bibitem[{{Jones} {et~al.}(2019{\natexlab{b}}){Jones}, {Pejcha}, \&
  {Corradi}}]{jones2019b}
{Jones}, D., {Pejcha}, O., \& {Corradi}, R. L.~M. 2019{\natexlab{b}}, \mnras,
  489, 2195, \dodoi{10.1093/mnras/stz2293}

\bibitem[{{Jones} {et~al.}(2017){Jones}, {Van Winckel}, {Aller}, {Exter}, \&
  {De Marco}}]{jones2017a}
{Jones}, D., {Van Winckel}, H., {Aller}, A., {Exter}, K., \& {De Marco}, O.
  2017, \aap, 600, L9, \dodoi{10.1051/0004-6361/201730700}

\bibitem[{{Kameswara Rao} {et~al.}(2018{\natexlab{a}}){Kameswara Rao}, {De
  Marco}, {Krishna}, {Murthy}, {Ray}, {Sutaria}, \& {Mohan}}]{rao2018b}
{Kameswara Rao}, N., {De Marco}, O., {Krishna}, S., {et~al.}
  2018{\natexlab{a}}, \aap, 620, A138, \dodoi{10.1051/0004-6361/201833507}

\bibitem[{{Kameswara Rao} {et~al.}(2018{\natexlab{b}}){Kameswara Rao},
  {Sutaria}, {Murthy}, {Krishna}, {Mohan}, \& {Ray}}]{rao2018a}
{Kameswara Rao}, N., {Sutaria}, F., {Murthy}, J., {et~al.} 2018{\natexlab{b}},
  \aap, 609, L1, \dodoi{10.1051/0004-6361/201732188}

\bibitem[{{Karakas} \& {Lattanzio}(2014)}]{karakas2014}
{Karakas}, A.~I., \& {Lattanzio}, J.~C. 2014, \pasa, 31, e030,
  \dodoi{10.1017/pasa.2014.21}

\bibitem[{{Keller} {et~al.}(2011){Keller}, {Herald}, {Bianchi}, {Maciel}, \&
  {Bohlin}}]{keller2011}
{Keller}, G.~R., {Herald}, J.~E., {Bianchi}, L., {Maciel}, W.~J., \& {Bohlin},
  R.~C. 2011, \mnras, 418, 705, \dodoi{10.1111/j.1365-2966.2011.19085.x}

\bibitem[{{Kerber} {et~al.}(2003){Kerber}, {Mignani}, {Guglielmetti}, \&
  {Wicenec}}]{kerber2003}
{Kerber}, F., {Mignani}, R.~P., {Guglielmetti}, F., \& {Wicenec}, A. 2003,
  \aap, 408, 1029, \dodoi{10.1051/0004-6361:20031046}

\bibitem[{{Kwok} {et~al.}(1978){Kwok}, {Purton}, \& {Fitzgerald}}]{kwok1978}
{Kwok}, S., {Purton}, C.~R., \& {Fitzgerald}, P.~M. 1978, \apjl, 219, L125,
  \dodoi{10.1086/182621}

\bibitem[{{Le D{\^u}} {et~al.}(2022){Le D{\^u}}, {Mulato}, {Parker}, {Petit},
  {Ritter}, {Drechsler}, {Strottner}, {Patchick}, {Prestgard}, {Garde},
  {Outters}, \& {Raffaelli}}]{LeDu2022}
{Le D{\^u}}, P., {Mulato}, L., {Parker}, Q.~A., {et~al.} 2022, \aap, 666, A152,
  \dodoi{10.1051/0004-6361/202243393}

\bibitem[{{Manchado}(2004)}]{manchado2004}
{Manchado}, A. 2004, in Astronomical Society of the Pacific Conference Series,
  Vol. 313, Asymmetrical Planetary Nebulae III: Winds, Structure and the
  Thunderbird, ed. M.~{Meixner}, J.~H. {Kastner}, B.~{Balick}, \& N.~{Soker}, 3

\bibitem[{{Martin} {et~al.}(2005){Martin}, {Fanson}, {Schiminovich},
  {Morrissey}, {Friedman}, {Barlow}, {Conrow}, {Grange}, {Jelinsky},
  {Milliard}, {Siegmund}, {Bianchi}, {Byun}, {Donas}, {Forster}, {Heckman},
  {Lee}, {Madore}, {Malina}, {Neff}, {Rich}, {Small}, {Surber}, {Szalay},
  {Welsh}, \& {Wyder}}]{martin2005}
{Martin}, D.~C., {Fanson}, J., {Schiminovich}, D., {et~al.} 2005, \apjl, 619,
  L1, \dodoi{10.1086/426387}

\bibitem[{{Miller Bertolami}(2016)}]{miller2016}
{Miller Bertolami}, M.~M. 2016, \aap, 588, A25,
  \dodoi{10.1051/0004-6361/201526577}

\bibitem[{{Miszalski} {et~al.}(2012){Miszalski}, {Boffin}, {Frew}, {Acker},
  {K{\"o}ppen}, {Moffat}, \& {Parker}}]{miszalski2012}
{Miszalski}, B., {Boffin}, H.~M.~J., {Frew}, D.~J., {et~al.} 2012, \mnras, 419,
  39, \dodoi{10.1111/j.1365-2966.2011.19667.x}

\bibitem[{{Morrissey} {et~al.}(2007){Morrissey}, {Conrow}, {Barlow}, {Small},
  {Seibert}, {Wyder}, {Budav{\'a}ri}, {Arnouts}, {Friedman}, {Forster},
  {Martin}, {Neff}, {Schiminovich}, {Bianchi}, {Donas}, {Heckman}, {Lee},
  {Madore}, {Milliard}, {Rich}, {Szalay}, {Welsh}, \& {Yi}}]{morrissey2007}
{Morrissey}, P., {Conrow}, T., {Barlow}, T.~A., {et~al.} 2007, \apjs, 173, 682,
  \dodoi{10.1086/520512}

\bibitem[{{Oke} \& {Gunn}(1983)}]{oke1983}
{Oke}, J.~B., \& {Gunn}, J.~E. 1983, \apj, 266, 713, \dodoi{10.1086/160817}

\bibitem[{{Parker} {et~al.}(2016){Parker}, {Boji{\v{c}}i{\'c}}, \&
  {Frew}}]{parker2016}
{Parker}, Q.~A., {Boji{\v{c}}i{\'c}}, I.~S., \& {Frew}, D.~J. 2016, in Journal
  of Physics Conference Series, Vol. 728, Journal of Physics Conference Series,
  032008, \dodoi{10.1088/1742-6596/728/3/032008}

\bibitem[{{Parker} {et~al.}(2006){Parker}, {Acker}, {Frew}, {Hartley},
  {Peyaud}, {Ochsenbein}, {Phillipps}, {Russeil}, {Beaulieu}, {Cohen},
  {K{\"o}ppen}, {Miszalski}, {Morgan}, {Morris}, {Pierce}, \&
  {Vaughan}}]{Parker2006}
{Parker}, Q.~A., {Acker}, A., {Frew}, D.~J., {et~al.} 2006, \mnras, 373, 79,
  \dodoi{10.1111/j.1365-2966.2006.10950.x}

\bibitem[{{Perek} \& {Kohoutek}(1967)}]{Perek1967}
{Perek}, L., \& {Kohoutek}, L. 1967, {Catalogue of Galactic Planetary Nebulae}

\bibitem[{{Sahai} {et~al.}(2011){Sahai}, {Morris}, \& {Villar}}]{Sahai2011}
{Sahai}, R., {Morris}, M.~R., \& {Villar}, G.~G. 2011, \aj, 141, 134,
  \dodoi{10.1088/0004-6256/141/4/134}

\bibitem[{{Schlegel} {et~al.}(1998){Schlegel}, {Finkbeiner}, \&
  {Davis}}]{schlegel1998}
{Schlegel}, D.~J., {Finkbeiner}, D.~P., \& {Davis}, M. 1998, \apj, 500, 525,
  \dodoi{10.1086/305772}

\bibitem[{{Stanghellini} \& {Haywood}(2010)}]{stanghellini2010}
{Stanghellini}, L., \& {Haywood}, M. 2010, \apj, 714, 1096,
  \dodoi{10.1088/0004-637X/714/2/1096}

\bibitem[{{Stoughton} {et~al.}(2002){Stoughton}, {Lupton}, {Bernardi},
  {Blanton}, {Burles}, {Castand er}, {Connolly}, {Eisenstein}, {Frieman},
  {Hennessy}, {Hindsley}, {Ivezi{\'c}}, {Kent}, {Kunszt}, {Lee}, {Meiksin},
  {Munn}, {Newberg}, {Nichol}, {Nicinski}, {Pier}, {Richards}, {Richmond},
  {Schlegel}, {Smith}, {Strauss}, {SubbaRao}, {Szalay}, {Thakar}, {Tucker},
  {Vand en Berk}, {Yanny}, {Adelman}, {Anderson}, {Anderson}, {Annis},
  {Bahcall}, {Bakken}, {Bartelmann}, {Bastian}, {Bauer}, {Berman},
  {B{\"o}hringer}, {Boroski}, {Bracker}, {Briegel}, {Briggs}, {Brinkmann},
  {Brunner}, {Carey}, {Carr}, {Chen}, {Christian}, {Colestock}, {Crocker},
  {Csabai}, {Czarapata}, {Dalcanton}, {Davidsen}, {Davis}, {Dehnen},
  {Dodelson}, {Doi}, {Dombeck}, {Donahue}, {Ellman}, {Elms}, {Evans}, {Eyer},
  {Fan}, {Federwitz}, {Friedman}, {Fukugita}, {Gal}, {Gillespie}, {Glazebrook},
  {Gray}, {Grebel}, {Greenawalt}, {Greene}, {Gunn}, {de Haas}, {Haiman},
  {Haldeman}, {Hall}, {Hamabe}, {Hansen}, {Harris}, {Harris}, {Harvanek},
  {Hawley}, {Hayes}, {Heckman}, {Helmi}, {Henden}, {Hogan}, {Hogg}, {Holmgren},
  {Holtzman}, {Huang}, {Hull}, {Ichikawa}, {Ichikawa}, {Johnston}, {Kauffmann},
  {Kim}, {Kimball}, {Kinney}, {Klaene}, {Kleinman}, {Klypin}, {Knapp},
  {Korienek}, {Krolik}, {Kron}, {Krzesi{\'n}ski}, {Lamb}, {Leger},
  {Limmongkol}, {Lindenmeyer}, {Long}, {Loomis}, {Loveday}, {MacKinnon},
  {Mannery}, {Mantsch}, {Margon}, {McGehee}, {McKay}, {McLean}, {Menou},
  {Merelli}, {Mo}, {Monet}, {Nakamura}, {Narayanan}, {Nash}, {Neilsen},
  {Newman}, {Nitta}, {Odenkirchen}, {Okada}, {Okamura}, {Ostriker}, {Owen},
  {Pauls}, {Peoples}, {Peterson}, {Petravick}, {Pope}, {Pordes}, {Postman},
  {Prosapio}, {Quinn}, {Rechenmacher}, {Rivetta}, {Rix}, {Rockosi}, {Rosner},
  {Ruthmansdorfer}, {Sandford}, {Schneider}, {Scranton}, {Sekiguchi}, {Sergey},
  {Sheth}, {Shimasaku}, {Smee}, {Snedden}, {Stebbins}, {Stubbs}, {Szapudi},
  {Szkody}, {Szokoly}, {Tabachnik}, {Tsvetanov}, {Uomoto}, {Vogeley}, {Voges},
  {Waddell}, {Walterbos}, {Wang}, {Watanabe}, {Weinberg}, {White}, {White},
  {Wilhite}, {Wolfe}, {Yasuda}, {York}, {Zehavi}, \& {Zheng}}]{stoughton2002}
{Stoughton}, C., {Lupton}, R.~H., {Bernardi}, M., {et~al.} 2002, \aj, 123, 485,
  \dodoi{10.1086/324741}

\bibitem[{{Taylor}(2005)}]{topcat}
{Taylor}, M.~B. 2005, in Astronomical Society of the Pacific Conference Series,
  Vol. 347, Astronomical Data Analysis Software and Systems XIV, ed.
  P.~{Shopbell}, M.~{Britton}, \& R.~{Ebert}, 29

\bibitem[{{Tonry} {et~al.}(2012){Tonry}, {Stubbs}, {Lykke}, {Doherty},
  {Shivvers}, {Burgett}, {Chambers}, {Hodapp}, {Kaiser}, {Kudritzki},
  {Magnier}, {Morgan}, {Price}, \& {Wainscoat}}]{tonry2012}
{Tonry}, J.~L., {Stubbs}, C.~W., {Lykke}, K.~R., {et~al.} 2012, \apj, 750, 99,
  \dodoi{10.1088/0004-637X/750/2/99}

\bibitem[{{Vassiliadis} \& {Wood}(1994)}]{vassiliadis1994}
{Vassiliadis}, E., \& {Wood}, P.~R. 1994, \apjs, 92, 125,
  \dodoi{10.1086/191962}

\bibitem[{{Vejar} {et~al.}(2019){Vejar}, {Montez}, {Morris}, \&
  {Stassun}}]{vejar2019}
{Vejar}, G., {Montez}, Rodolfo, J., {Morris}, M., \& {Stassun}, K.~G. 2019,
  \apj, 879, 38, \dodoi{10.3847/1538-4357/ab21ba}

\bibitem[{{Weidmann} \& {Gamen}(2011)}]{weidmann2011}
{Weidmann}, W.~A., \& {Gamen}, R. 2011, \aap, 526, A6,
  \dodoi{10.1051/0004-6361/200913984}

\bibitem[{{York} {et~al.}(2000){York}, {Adelman}, {Anderson}, {Anderson},
  {Annis}, {Bahcall}, {Bakken}, {Barkhouser}, {Bastian}, {Berman}, {Boroski},
  {Bracker}, {Briegel}, {Briggs}, {Brinkmann}, {Brunner}, {Burles}, {Carey},
  {Carr}, {Castander}, {Chen}, {Colestock}, {Connolly}, {Crocker}, {Csabai},
  {Czarapata}, {Davis}, {Doi}, {Dombeck}, {Eisenstein}, {Ellman}, {Elms},
  {Evans}, {Fan}, {Federwitz}, {Fiscelli}, {Friedman}, {Frieman}, {Fukugita},
  {Gillespie}, {Gunn}, {Gurbani}, {de Haas}, {Haldeman}, {Harris}, {Hayes},
  {Heckman}, {Hennessy}, {Hindsley}, {Holm}, {Holmgren}, {Huang}, {Hull},
  {Husby}, {Ichikawa}, {Ichikawa}, {Ivezi{\'c}}, {Kent}, {Kim}, {Kinney},
  {Klaene}, {Kleinman}, {Kleinman}, {Knapp}, {Korienek}, {Kron}, {Kunszt},
  {Lamb}, {Lee}, {Leger}, {Limmongkol}, {Lindenmeyer}, {Long}, {Loomis},
  {Loveday}, {Lucinio}, {Lupton}, {MacKinnon}, {Mannery}, {Mantsch}, {Margon},
  {McGehee}, {McKay}, {Meiksin}, {Merelli}, {Monet}, {Munn}, {Narayanan},
  {Nash}, {Neilsen}, {Neswold}, {Newberg}, {Nichol}, {Nicinski}, {Nonino},
  {Okada}, {Okamura}, {Ostriker}, {Owen}, {Pauls}, {Peoples}, {Peterson},
  {Petravick}, {Pier}, {Pope}, {Pordes}, {Prosapio}, {Rechenmacher}, {Quinn},
  {Richards}, {Richmond}, {Rivetta}, {Rockosi}, {Ruthmansdorfer}, {Sand ford},
  {Schlegel}, {Schneider}, {Sekiguchi}, {Sergey}, {Shimasaku}, {Siegmund},
  {Smee}, {Smith}, {Snedden}, {Stone}, {Stoughton}, {Strauss}, {Stubbs},
  {SubbaRao}, {Szalay}, {Szapudi}, {Szokoly}, {Thakar}, {Tremonti}, {Tucker},
  {Uomoto}, {Vanden Berk}, {Vogeley}, {Waddell}, {Wang}, {Watanabe},
  {Weinberg}, {Yanny}, {Yasuda}, \& {SDSS Collaboration}}]{york2000}
{York}, D.~G., {Adelman}, J., {Anderson}, John~E., J., {et~al.} 2000, \aj, 120,
  1579, \dodoi{10.1086/301513}

\end{thebibliography}
\bibliographystyle{aasjournal}

\listofchanges
\end{document}